\PassOptionsToPackage{english}{babel}

\documentclass[%
10pt,
aip,
amsmath,amssymb,
floatfix
]{revtex4-1}

\usepackage{graphicx}
\usepackage[mathlines]{lineno}
\usepackage[english]{babel}

\renewcommand{\d}[1]{\ensuremath{\operatorname{d}\!{#1}}}
\newcommand{\D}[1]{\ensuremath{\operatorname{D}\!{#1}}}
\DeclareMathOperator{\Id}{Id}
\DeclareMathOperator*{\spec}{spec}

\DeclareMathOperator*{\sign}{sign}

\def\pct{\%}

\begin{document}

\selectlanguage{English}

\title{Building a Maxey--Riley framework for surface ocean inertial
particle dynamics}

\author{F.\ J.\ Beron-Vera} \email{fberon@miami.edu}
\affiliation{Department of Atmospheric Sciences, Rosenstiel School
of Marine and Atmospheric Science, University of Miami, Miami,
Florida 33149, USA} \author{M.\ J.\ Olascoaga} \affiliation{Department
of Ocean Sciences, Rosenstiel School of Marine and Atmospheric
Science, University of Miami, Miami, Florida 33149, USA} \author{P.\
Miron} \affiliation{Department of Atmospheric Sciences, Rosenstiel
School of Marine and Atmospheric Science, University of Miami,
Miami, Florida 33149, USA}

\date{Started: March 2, 2019; this version: \today.}%

\begin{abstract}
  A framework for the study of surface ocean inertial particle
  motion is built from the Maxey--Riley set.  A new set is obtained
  by vertically averaging each term of the original set, adapted
  to account for Earth's rotation effects, across the extent of a
  sufficiently small spherical particle that floats at an assumed
  unperturbed air--sea interface with unsteady nonuniform winds and
  ocean currents above and below, respectively.  The inertial
  particle velocity is shown to exponentially decay in time to a
  velocity that lies close to an average of seawater and air
  velocities, weighted by a function of the seawater-to-particle
  density ratio.  Such a weighted average velocity turns out to
  fortuitously be of the type commonly discussed in the search-and-rescue
  literature, which alone cannot explain the observed role of
  anticyclonic mesoscale eddies as traps for marine debris or the
  formation of great garbage patches in the subtropical gyres,
  phenomena dominated by finite-size effects. A heuristic extension
  of the theory is proposed to describe the motion of nonspherical
  particles by means of a simple shape factor correction, and
  recommendations are made for incorporating wave-induced Stokes
  drift, and allowing for inhomogeneities of the carrying fluid
  density.  The new Maxey--Riley set outperforms an ocean adaptation
  that ignored wind drag effects and the first reported adaption
  that attempted to incorporate them.
\end{abstract}


\maketitle

\tableofcontents

\section{Introduction}

The study of the motion of \emph{inertial} (i.e., buoyant, finite-size)
particles was pioneered by \citet{Stokes-51} by solving the linearized
Navier--Stokes equations for the oscillatory motion of a small solid
sphere (pendulum) immersed in a fluid at rest. This was followed
by the efforts of \citet{Basset-88}, \citet{Boussinesq-85}, and
\citet{Oseen-27} to model a solid sphere settling under gravity,
also in a quiescent fluid.  \citet{Tchen-47} extended these efforts
to model motion in nonuniform unsteady flow by writing the resulting
equation, known as the BBO equation, on a frame of reference moving
with the fluid.  Several corrections to the precise form of the
forces exerted on the particle due to the solid--fluid interaction
were made along the years \cite{Corrsin-Lumely-56} until
the now widely accepted form of the forces was derived by
\citet{Maxey-Riley-83} from first principles, following an approach
introduced by \citet{Riley-71}, and independently and nearly
simultaneously by \citet{Gatignol-83}.  The resulting equation,
with a correction made by \citet{Auton-etal-88}, is commonly referred
to as the \emph{Maxey--Riley equation}.

The Maxey--Riley set is a classical mechanics second Newton's law
that provides the de-jure framework for modeling inertial particle
motion in fluid mechanics \cite{Michaelides-97, Provenzale-99,
Cartwright-etal-10}.  Conveniently given in the form of an ordinary
differential equation, it has for instance facilitated the understanding
of why buoyant particles can behave quite differently than fluid
(i.e., neutrally buoyant, infinitesimally small) particles no matter
how small \cite{Babiano-etal-00, Vilela-etal-06}.  Such an
understanding would have been very difficult to be attained by
solving the numerically expensive Navier--Stokes partial differential
equations with a moving boundary.

Understanding inertial particle motion is crucial in oceanography
for a number of reasons.  These include a need of improving the
success of search-and-rescue operations at sea \cite{Breivik-etal-13,
Bellomo-etal-15}, better understanding the drift of macroalgae
\cite{Gower-King-08, Brooks-etal-19}, or the motion of flotsam in
general such as plastic litter \cite{Law-etal-10, Cozar-etal-14},
airplane wreckage \cite{Trinanes-etal-16, Miron-etal-19b}, tsunami
debris \cite{Rypina-etal-13a, Matthews-etal-17}, and even sea-ice
pieces in a warming climate \cite{Szanyi-etal-16}.

With the well-founded expectation that the Maxey--Riley set can
provide insight into inertial particle motion in the ocean, two
ocean adaptations of the set were recently proposed (additional
applications in oceanography have been reported \cite{Nielsen-94,
Reigada-etal-03, Peng-Dabiri-09, Monroy-etal-16}, but we do not
discuss them here as these mostly deal with settling of particles
under gravity or biological problems rather than motion near the
ocean surface).  \citet{Beron-etal-15} included Earth rotation
effects, and restricting to quasigeostrophic carrying flow,
investigated the motion of inertial particles near mesoscale eddies.
These authors found that mesoscale eddies with coherent material
boundaries \cite{Haller-Beron-13, Haller-Beron-14, Haller-etal-16}
can attract or repel inertial particles depending on the buoyancy
of the particles and the polarity of the eddies.  The result was
formalized by \citet{Haller-etal-16} by providing rigorous conditions
under which finite-time attractors or repellors can be found inside
eddies.  The prediction was supported in \citet{Beron-etal-15} by
an observation in the Pacific Ocean of two submerged drifting buoys
(floats), which, deployed nearby inside a anticyclonic mesoscale
eddy, one remained looping inside the eddy while the other was
expelled away from it.  According to the theory heavy (light)
inertial particles should be attracted (repelled) by anticyclonic
eddies and vice verse by cyclonic eddies. And indeed the observation
adhered to the theoretical result since the float that remained
trapped in the eddy was seen to take a slightly descending path
while the float that escaped the eddy took a slightly ascending
path.  While some evidence was presented in \citet{Beron-etal-15}
for similar behavior at the ocean surface, the dynamics there can
be expected to be different than those below due to the wind action.
A consequence of this is the inability of the ocean adaptation of
the Maxey--Riley set by \citet{Beron-etal-15} to describe the
accumulation of marine debris into large patches in the subtropical
gyres \cite{Cozar-etal-14}.

The above motivated \citet{Beron-etal-16} to extend the theory to
account for the combined effect on a particle of ocean current and
wind drag.  With this in mind, \citet{Beron-etal-16} proceeded
heuristically by modeling the particle piece immersed in the seawater
(air) as a sphere of the fractional volume that is immersed in the
seawater (air), and assuming that it evolves according to the
Maxey--Riley set.  The subspheres were advected together and the
forces acting on each of them were calculated at the same position.
This heuristics resulted in a Maxey--Riley set, which, including
Earth's rotation and sphericity effects, predicted the formation
of great garbage patches in the subtropical gyres as a phenomenon
dominated by inertial effects, rather than Ekman convergence as
commonly argued \cite{Maximenko-Niiler-06, Brach-etal-18}.
The Maxey--Riley equation for surface ocean inertial particle
dynamics by \citet{Beron-etal-16}, just as that for subsurface ocean
inertial particle dynamics by \citet{Beron-etal-15}, predicts
accumulation of (light) particles into cyclonic eddies and repulsion
from anticyclonic eddies.  However, recent in-situ observations are
showing the contrary \cite{Brach-etal-18}, consistent with the
traditional paradigm \cite{Chelton-etal-11b} that does not account
for inertial effects, which represents a puzzle.  On the other hand,
the neutrally buoyant limit of the Maxey--Riley equation of
\citet{Beron-etal-16} does not coincide with that of the standard
Maxey--Riley set as it includes descriptors of the air component
of the carrying flow when the particle is completely immersed in
the seawater below the surface.  Furthermore, results from a dedicated
experiment involving satellite-tracked floating objects of different
buoyancies, sizes, and shapes \cite{Olascoaga-etal-19} are showing
little trajectory prediction skill for the Maxey--Riley set proposed
by \citet{Beron-etal-16}.

To improve the description of inertial particle motion at the
air--sea interface provided by the Maxey--Riley set, a new ocean
adaptation of the set is proposed here.  The new set is obtained
by vertically integrating the original set, appropriately extended
to represent Earth's rotation and sphericity effects, across a
sufficiently small spherical particle which floats at an unperturbed
air--sea interface with unsteady nonuniform winds and ocean currents
above and below, respectively.  The new set, while preserving the
important capability of the one derived by \citet{Beron-etal-16}
in predicting garbage patch formation, predicts concentration of
particles inside anticyclonic eddies consistent with observations,
thereby explaining this phenomenon as a result of inertial effects.
As the Maxey--Riley set proposed by \citet{Beron-etal-16}, the
inertial particle velocity is shown to exponentially decay in time
to a velocity that lies close to an average of seawater and air
velocities, weighted by a certain function of the seawater-to-particle
density ratio that conveys it additional margin for modeling in a
wider range of conditions.  This velocity coincidentally is of the
type extensively discussed in the search-and-rescue literature and
obtained mainly empirically or from considerations that are difficult
to justify.  In any case, the weighted average velocity alone cannot
explain the observed role of anticyclonic mesoscale eddies as traps
for marine debris or the formation of great garbage patches in the
subtropical gyres, phenomena dominated by finite-size effects.  A
heuristic extension of the Maxey--Rile theory derived here to
describe the motion of nonspherical particles is proposed, and
recommendations are made for accounting for lateral gradients and
time variations of the advecting fluid density.

The rest of the paper is organized as follows.  Section 2 starts
with the mathematical setup.  In \S 3 we present the proposed ocean
adaptation of the Maxey--Riley set after introducing and discussing
the forcing terms involved.  Limiting buoyancy behavior of the
Maxey--Riley set and its small-size asymptotic dynamics (slow
manifold reduction) are discussed in \S 4.  The ability of the model
derived to describe observed behavior is demonstrated in \S 5.
Section 6 addresses corrections of the set to account for the motion
of nonspherical particles, the incorporation of wave-induced drift,
and the inclusion of memory effects and those produced by the
carrying fluid density varying in space and time.  Section 7 presents
the conclusions of the paper. Finally, Appendix A includes the
full spherical form of the Maxey--Riley set and its slow manifold
reduction, and Appendix B presents some mathematical details.

\section{Setup}

\begin{figure}[t!]
  \centering%
  \includegraphics[width=\textwidth]{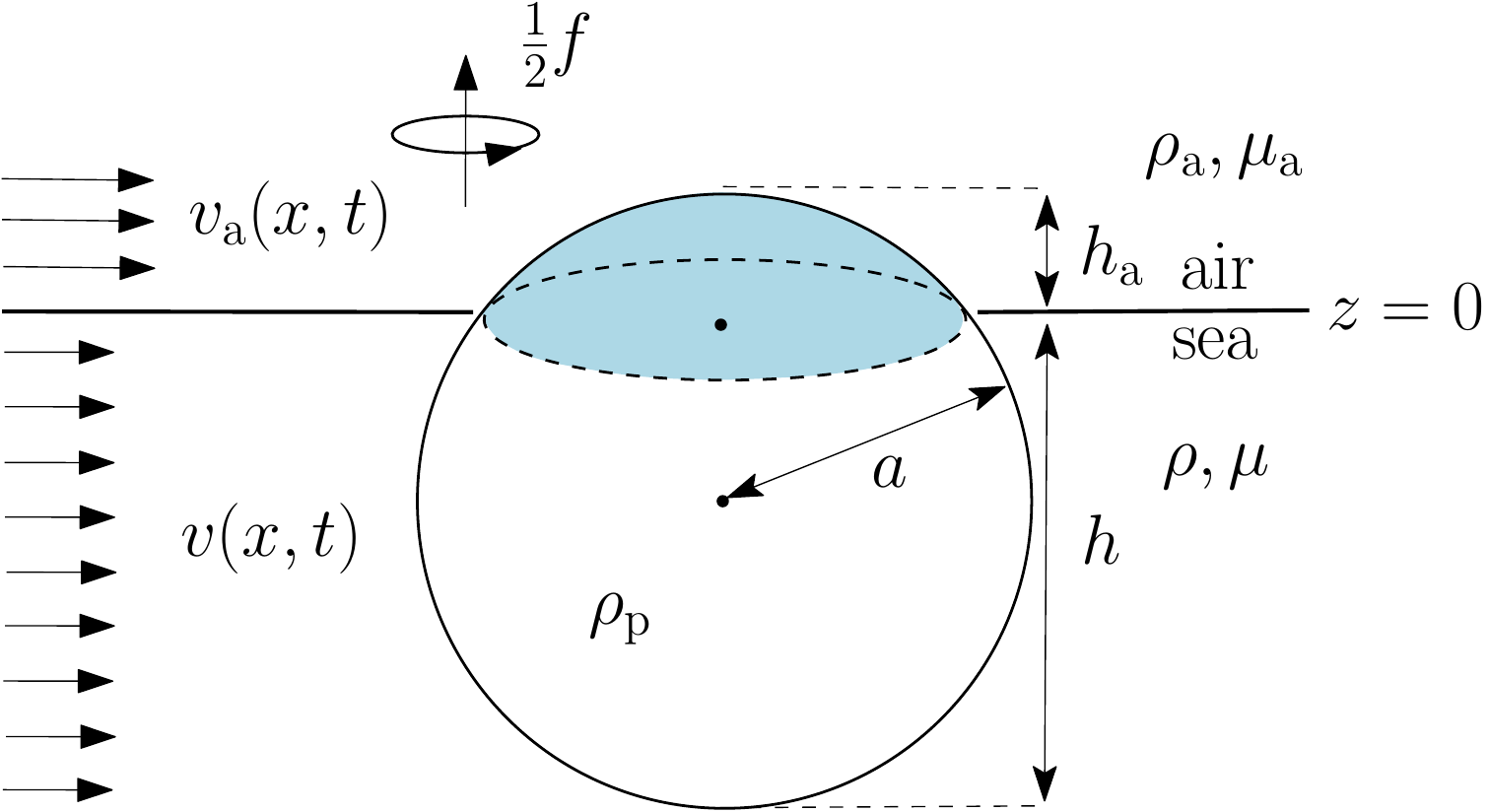}%
  \caption{Solid spherical particle that floats at an assumed flat
  interface between homogeneous seawater and air, and is subjected
  to flow, added mass, and drag forces resulting from the action
  of unsteady, horizontally sheared ocean currents and winds. See
  text for variable and parameter definitions.}
  \label{fig:setup}%
\end{figure}

Let $x = (x^1,x^2)$ with $x_1$ (resp., $x_2$) pointing eastward
(resp., northward) be position on some domain $D$ of the $\beta$
plane, i.e., $D\subset \mathbb{R}^2$ rotates with angular speed
$\frac{1}{2}f$ where $f = f_0 + \beta x^2$ is the Coriolis parameter;
let $z$ denote the vertical direction; and let $t$ stand for time,
ranging on some finite interval $I\subset \mathbb{R}$ (Figure
\ref{fig:setup}).  Consider a stack of two homogeneous fluid layers
separated by an interface, assumed to be \emph{fixed} at $z = 0$
Figure \ref{fig:setup}.  The fluid in the bottom layer represents
the seawater and has density $\rho$.  The top-layer fluid is much
lighter, representing the air; its density is $\rho_\mathrm{a} \ll
\rho$.  Let $\mu$ and $\mu_\mathrm{a}$ stand for dynamic viscosities
of seawater and air, respectively.  The seawater and air velocities
vary in horizontal position and time, and are denoted $v(x,t)$ and
$v_\mathrm{a}(x,t)$, respectively.  Consider finally a solid spherical
particle, of radius $a$ and density $\rho_\mathrm{p}$, floating at
the air--sea interface.

Let
\begin{equation}
  \delta : = \frac{\rho}{\rho_\mathrm{p}},\quad 
  \delta_\mathrm{a} : = \frac{\rho_\mathrm{a}}{\rho_\mathrm{p}}.
  \label{eq:dta}
\end{equation}
Clearly, $\delta \gg \delta_\mathrm{a}$.  Let $0 \le \sigma \le 1$ be
the fraction of submerged (in seawater) particle volume. The emerged
fraction then is $1 - \sigma$, which is sometimes referred to as
\emph{reserve buoyancy}.  Static (in the \emph{vertical}) stability
of the particle (Archimedes' principle),
\begin{equation}
  \sigma\delta + (1-\sigma)\delta_\mathrm{a} = 1,
  \label{eq:arch}
\end{equation}
is satisfied for
\begin{equation}
  \sigma = \frac{1-\delta_\mathrm{a}}{\delta-\delta_\mathrm{a}},
  \label{eq:frac1}
\end{equation}
which requires
\begin{equation}
  \delta \ge 1,\quad \delta_\mathrm{a} \le 1.
\end{equation}
We will conveniently
assume \cite{Beron-etal-16}
\begin{equation}
  \delta_\mathrm{a} \ll 1,
\end{equation}
so \eqref{eq:frac1} can be well approximated by
\begin{equation}
  \sigma = \delta^{-1}.
  \label{eq:frac2}
\end{equation}

The height ($h_\mathrm{a}$) of the emerged spherical cap can be
expressed in terms of $\delta$. This follows from equating its
volume formula expressed in terms of $a$ and $h_\mathrm{a}$ with
the volume of the emerged spherical cap.  To wit,
\begin{equation}
  \frac{\pi h_\mathrm{a}^2}{3}(3a - h_\mathrm{a}) =
  (1-\delta^{-1})\frac{4}{3}\pi a^3,
  \label{eq:cubic}
\end{equation}
whose only physically meaningful root is
\begin{equation}
  h_\mathrm{a}/a = \Phi := \frac{\mathrm{i}\sqrt{3}}{2}
  \left(\frac{1}{\varphi}-\varphi\right) - \frac{1}{2\varphi} -
  \frac{\varphi}{2} + 1
  \label{eq:Phi}
\end{equation}
where
\begin{equation}
  \varphi := \sqrt[3]{\mathrm{i}\sqrt{1-(2\delta^{-1}-1)^2} +
  2\delta^{-1} - 1}.
  \label{eq:phi}
\end{equation}
The depth ($h$) of the submerged spherical cap,
\begin{equation}
  h = (2-\Phi)a.
\end{equation}
For a neutrally buoyant particle, i.e., $\delta = 1$, $\varphi =
0$ and thus $\Phi = 0$. Consequently, as expected, $h_\mathrm{a} =
0$  (and hence $h = 2a$).  Near neutrality, $\Phi =
\smash{\frac{2\sqrt{3}}{3}}\smash{\sqrt{\delta-1}} - \smash{\frac{2}{9}}
(\delta-1) + O((\delta-1)^2)$, which reveals the real nature of the
root(s) of \eqref{eq:cubic} explicitly\footnote{While having a
positive discriminant, the cubic polynomial in \eqref{eq:cubic} is
irreducible over the reals.  Thus while its three roots are real,
they require complex numbers to be expressed in radicals
\cite{Wantzel-43}.}.  A half-emerged (equivalently, half-submerged)
particle, namely, $h_\mathrm{a} = a = h$, corresponds to $\delta =
2$.  On the other hand, $h_\mathrm{a} \to 2a$ (and thus $h \to 0$)
slowly as $\delta \to \infty$.

For future reference, the projected (in the flow direction) area
of the emerged spherical cap, $A_\mathrm{a}$, can be readily seen
to be given by
\begin{equation}
  A_\mathrm{a} = \pi\Psi a^2,\quad 
  \Psi := \pi^{-1}\cos^{-1}(1-\Phi)
  - \pi^{-1}(1-\Phi) \sqrt{1-(1-\Phi)^2},
  \label{eq:Aa}
\end{equation}
a function of $\delta$ exclusively. In turn, the immersed projected
area, denoted $A$, is equal to
\begin{equation}
  A = \pi a^2 - A_\mathrm{a} = \pi(1 - \Psi)a^2.
  \label{eq:A}
\end{equation}
When $\delta = 1$, $A_\mathrm{a} = 0$  (and hence $A = \pi a^2$),
which immediately follows from $\Psi = \smash{\frac{16\sqrt[4]{3}}{9\pi}}
(\delta-1)^{3/4} + O((\delta-1)^{5/4})$ as $\delta \to 1$.  The
situation in which $A_\mathrm{a} = \frac{1}{2}\pi a^2$ corresponds
to $\delta = 2$.  Finally, $A_\mathrm{a} \to \pi a^2$ (and thus $A
\to 0$) slowly as $\delta \to \infty$.

We close the setup with a few remarks.  Ignoring the vertical shear
of the ocean currents (resp., winds) below (resp., above) the
interface over the extent of the particle piece that is immersed
in the seawater (resp., air) is a reasonable approximation under
the assumption that the particle is small. On the other hand, that
the interface remains flat at all times is clearly a strong assumption.
Recommendations for incorporating the effects of wind-induced
(Stokes) drift are given below.  In turn, ignoring lateral gradients
and time variations of the carrying fluid density can be of
consequence, particularly near frontal regions.  Below we provide
means for incorporating their effects as well.

\section{The Maxey--Riley set}

\subsection{The original fluid mechanics formulation}

The exact motion of \emph{inertial particles} such as that in Figure
\ref{fig:setup} is controlled by the Navier--Stokes equation with
moving boundaries as such particles are extended objects in the
fluid with their own boundaries.  This approach results in complicated
partial differential equations which are extremely difficult---if
not impossible---to solve and analyze.

Here we are concerned with the approximation, formulated in terms
of an ordinary differential equation, provided by the Maxey--Riley
equation, which, as noted in the Introduction, has become the de-jure
fluid mechanics paradigm for inertial particle dynamics.

More specifically, the Maxey--Riley equation is a classical mechanics
Newton's second law with several forcing terms that describe the
motion of solid spherical particles immersed in the unsteady
nonuniform flow of a homogeneous viscous fluid.  Normalized by
particle mass, $m_\mathrm{p} = \frac{4}{3}\pi a^3\rho_\mathrm{p}$,
the relevant forcing terms for the \emph{horizontal} motion of a
sufficiently small particle are:
\begin{enumerate}
\item the \emph{flow force} exerted on the particle by the undisturbed
fluid,
\begin{equation}
  F_\mathrm{flow} =
  \frac{m_\mathrm{f}}{m_\mathrm{p}}\frac{\D{v_\mathrm{f}}}{\D{t}} 
  \label{eq:FF},
\end{equation}
where $m_\mathrm{f} = \frac{4}{3}\pi a^3\rho_\mathrm{f}$ is the
mass of the displaced fluid (of density $\rho_\mathrm{f}$), and
$\smash{\frac{\D{v_\mathrm{f}}}{\D{t}}}$ is the material derivative
of the fluid velocity ($v_\mathrm{f}$) or its total derivative taken
along the trajectory of a fluid particle, $x = X_\mathrm{f}(t)$,
i.e., $\smash{\frac{\D{v_\mathrm{f}}}{\D{t}}} =
\smash{\left[\frac{\d{}}{\d{t}}v_\mathrm{f}(x,t)\right]_{x=X_\mathrm{f}(t)}} =
\partial_t v_\mathrm{f} + (\nabla v_\mathrm{f}) v_\mathrm{f}$; 
\item the \emph{added mass force} resulting from part of the fluid moving
with the particle,
\begin{equation}
  F_\mathrm{mass} =
  \frac{\frac{1}{2}m_\mathrm{f}}{m_\mathrm{p}}\left(\frac{\D{v_\mathrm{f}}}{\D{t}}
  - \dot v_\mathrm{p}\right)
  \label{eq:AM},
\end{equation}
where $\dot v_\mathrm{p}$ is the acceleration of an inertial particle
with trajectory $x = X_\mathrm{p}(t)$, i.e., $\dot v_\mathrm{p} =
\smash{\frac{\d{}}{\d{t}}\left[v_\mathrm{p}(x,t)\right]_{x=X_\mathrm{p}(t)}} = \partial_t
v_\mathrm{p}$ where $v_\mathrm{p} = \partial_t X_\mathrm{p} = \dot
x$ is the inertial particle velocity; 
\item the \emph{lift force}, which arises when the particle
rotates as it moves in a (horizontally) sheared flow,
\begin{equation}
  F_\mathrm{lift} =
  \frac{\frac{1}{2}m_\mathrm{f}}{m_\mathrm{p}}\omega_\mathrm{f}
  (v_\mathrm{f} - v_\mathrm{p})^\perp,
  \label{eq:FL}
\end{equation}
where $\omega_\mathrm{f} = \partial_1 v^2_\mathrm{f} -
\partial_2 v^1_\mathrm{f}$ is the (vertical) vorticity of the
fluid and
\begin{equation}
  w^\perp = J w,\quad
  J := 
  \begin{pmatrix}
	 0 & -1\\
	 1 &  0
  \end{pmatrix}
\end{equation}
for any vector $w$ in $\mathbb R^2$; and  
\item the \emph{drag force} caused by the fluid viscosity,
\begin{equation}
  F_\mathrm{drag} = \frac{12\mu_\mathrm{f}
  \frac{A_\mathrm{f}}{\ell_\mathrm{f}}}{m_\mathrm{p}}
  (v_\mathrm{f} - v_\mathrm{p}),
  \label{eq:SD}
\end{equation}
where $\mu_\mathrm{f}$ is the dynamic viscosity of the fluid, and
$A_\mathrm{f}$ ($=\pi a^2$) is the projected area of the particle
and $\ell_\mathrm{f}$ ($=2a$) is the characteristic projected length,
which we have intentionally left unspecified for future appropriate
evaluation. 
\end{enumerate}

Except for the lift force \eqref{eq:FL}, due to \citet{Auton-87},
the above forces are included in the original formulation by
\citet{Maxey-Riley-83} (cf.\ also \citet{Gatignol-83}), yet with a
form of the added mass term different than \eqref{eq:AM}, which
corresponds to the correction due to \citet{Auton-etal-88}.  A
Maxey--Riley model with lift force, which has not been so far
considered in ocean dynamics despite its relevance in the presence
of unbalanced (submesoscale) motions (e.g., \citet{Beron-etal-18a}),
can be found in \citet[Chapter 4]{Montabone-02} (cf.\ similar forms
in \citet{Henderson-etal-07, Sapsis-etal-11}).

In writing \eqref{eq:AM} and \eqref{eq:SD}, terms proportional to
$\nabla^2v_\mathrm{f}$, so-called Faxen corrections, have been
ignored.  These account for the horizontal variation of the flow
field across the particle, which is negligible for a particle with
a radius much smaller than the typical length scale of the flow.
Also, the complete set of Maxey--Riley forces involves an additional
term, the Basset--Boussinesq history or memory term.  This is an
integral term that accounts for the lagging boundary layer developed
around the particle.  The memory term turns the Maxey--Riley set
into a fractional differential equation that does not generate a
dynamical system as the corresponding flow map does not satisfy a
semi-group property \cite{Farazmand-Haller-15, Langlois-etal-15}.
Numerical experimentation \cite{Daitche-Tel-11} reveals that the
Basset history term mainly tends to slow down the inertial particle
motion.  More rigorously, \citet{Langlois-etal-15} show that the
particle velocity decays algebraically, rather than exponentially
as in the absence of the memory term, in time to a limit that is
close, in the square of the particle's radius, to the carrying fluid
velocity.  The memory term cannot be neglected on sufficiently small
particle assumption grounds \cite{Daitche-Tel-14, Langlois-etal-15},
but it may be under the assumption that the time it takes a particle
to return to a region that has visited earlier is long compared to
the time scale of the flow \cite{Sudharsan-etal-16}, condition
that should not be too difficult to be satisfied in the ocean,
except, for instance, inside vortices.

\subsection{The proposed adaptation to surface ocean dynamics}

We first account for the geophysical nature of the fluid by including
the Coriolis force\footnote{In an earlier geophysical adaptation
of the Maxey--Riley equation \cite{Provenzale-99}, the centrifugal
force was included as well, but this is actually balanced out by
the gravitational force on the horizontal plane.}.  This amounts
to replacing \eqref{eq:FF} and \eqref{eq:AM} with
\begin{equation}
  F_\mathrm{flow} =
  \frac{m_\mathrm{f}}{m_\mathrm{p}}\left(\frac{\D{v_\mathrm{f}}}{\D{t}}
  + f v_\mathrm{f}^\bot\right) 
\end{equation}
and
\begin{equation}
  F_\mathrm{mass} =
  \frac{\frac{1}{2}m_\mathrm{f}}{m_\mathrm{p}}\left(\frac{\D{v_\mathrm{f}}}{\D{t}}
  + f v_\mathrm{f}^\bot - \dot v_\mathrm{p} - f
  v_\mathrm{p}^\bot\right),
\end{equation}
respectively. Geometric terms due to the planet's sphericity, which
should be included when $f$ is allowed to vary with $x^2$, making
$(x^1,x^2)$ curvilinear rather than Cartesian \cite{Pedlosky-87},
were omitted as traditionally done for simplicity, yet recognizing
that some consequences may be expected \cite{Ripa-JPO-97b}.
Nevertheless, the full spherical form of the equations derived
below, appropriate for operational use, is given in Appendix
\ref{ap:sph}.

Then, noting that fluid variables and parameters take different
values when pertaining to seawater or air, e.g.,
\begin{equation}
 v_\mathrm{f}(x,z,t) = 
  \begin{cases} 
	v_\mathrm{a}(x,t) & \text{if } z \in (0,h_\mathrm{a}], \\
   v(x,t)  & \text{if } z \in [-h,0),
  \end{cases}
\end{equation}
we write
\begin{equation}
  \dot v_\mathrm{p} + f v_\mathrm{p}^\perp = \langle F_\mathrm{flow}\rangle
  + \langle F_\mathrm{mass}\rangle + \langle F_\mathrm{lift}\rangle
  + \langle F_\mathrm{drag}\rangle, 
  \label{eq:mr}
\end{equation}
where $\langle\,\rangle$ is an average over $z\in
[-h,h_\mathrm{a}]$.

Specifically,
\begin{align}
  \langle F_\mathrm{flow}\rangle 
  = {} &
  \frac{1}{2a}\int_{-h}^{h_\mathrm{a}}
  \frac{m_\mathrm{f}}{m_\mathrm{p}}\left(\frac{\D{v_\mathrm{f}}}{\D{t}} + f
  v_\mathrm{f}^\bot\right)\d{z}\nonumber\\
  = {} &
  \frac{1}{2a}\int_{(\Phi-2)a}^{0}
  \frac{\delta^{-1}\frac{4}{3}\pi a^3\rho}{\frac{4}{3}\pi 
  a^3\rho_\mathrm{p}}\left(\frac{\D{v}}{\D{t}} + f
  v^\bot\right)\d{z}\nonumber\\
  & +
  \frac{1}{2a}\int_{0}^{\Phi a}
  \frac{(1-\delta^{-1})\frac{4}{3}\pi a^3\rho_\mathrm{a}}{\frac{4}{3}\pi 
  a^3\rho_\mathrm{p}}\left(\frac{\D{v_\mathrm{a}}}{\D{t}} + 
  f v_\mathrm{a}^\bot\right)\d{z}\nonumber\\
  \overset{\hphantom{\delta_\mathrm{a} \ll 1}}{=} {} & 
  \frac{1}{2}(2-\Phi)\left(\frac{\D{v}}{\D{t}}
  + f v^\bot\right) + 
  \frac{1}{2}(1-\delta^{-1})\Phi\delta_\mathrm{a}\left(\frac{\D{v_\mathrm{a}}}{\D{t}} + 
  f v_\mathrm{a}^\bot\right),
  \label{eq:FFavg-full}
\end{align}
where $\smash{\frac{\D{}}{\D{t}}}v$
(resp., $\smash{\frac{\D{}}{\D{t}}}v_\mathrm{a}$) is understood to be taken
along the trajectory of a seawater (resp., air) particle, obtained by solving $\dot
x = v$ (resp., $\dot x = v_\mathrm{a}$), namely, $\smash{\frac{\D{}}{\D{t}}}v
= \partial_t v + (\nabla v)v$ (resp., $\smash{\frac{\D{}}{\D{t}}}v_\mathrm{a}
= \partial_t v_\mathrm{a} + (\nabla v_\mathrm{a})v_\mathrm{a}$).
Taking into account that $\delta_\mathrm{a} \ll 1$, \eqref{eq:FFavg-full}
is well approximated by
\begin{equation}
  \langle F_\mathrm{flow}\rangle =  
  \left(1-\frac{\Phi}{2}\right)\left(\frac{\D{v}}{\D{t}} + f
  v^\bot\right).
 \label{eq:FFavg}
\end{equation}

Similarly,
\begin{align}
  \langle F_\mathrm{mass}\rangle 
  &\overset{\hphantom{\delta_\mathrm{a} \ll 1}}{=}
  \frac{1}{2a}\int_{-h}^{h_\mathrm{a}}
  \frac{\frac{1}{2}m_\mathrm{f}}{m_\mathrm{p}}\left(\frac{\D{v_\mathrm{f}}}{\D{t}}
  + f v_\mathrm{f}^\bot - \dot v_\mathrm{p} - f v_\mathrm{p}^\bot\right)\d{z}\nonumber\\
  &\overset{\delta_\mathrm{a} \ll 1}{=} 
  \frac{1}{2}\left(1-\frac{\Phi}{2}\right)\left(\frac{\D{v}}{\D{t}} + f v^\bot -
  \dot v_\mathrm{p} - f v_\mathrm{p}^\bot\right)
  \label{eq:AMavg}
\end{align}
and
\begin{align}
  \langle F_\mathrm{lift}\rangle 
  &\overset{\hphantom{\delta_\mathrm{a} \ll 1}}{=}
  \frac{1}{2a}\int_{-h}^{h_\mathrm{a}}
  \frac{\frac{1}{2}m_\mathrm{f}}{m_\mathrm{p}}\omega_\mathrm{f} 
  \left(v_\mathrm{f} - v_\mathrm{p}\right)^\perp\d{z}\nonumber\\
  &\overset{\delta_\mathrm{a} \ll 1}{=} 
  \frac{1}{2}\left(1-\frac{\Phi}{2}\right)\omega 
  \left(v - v_\mathrm{p}\right)^\perp.
  \label{eq:FLavg}
\end{align}

Now, to evaluate the drag force, appropriate projected length scales
for the submerged and emerged particle pieces must be chosen.  We
conveniently take $\ell = kh$ and $\ell_\mathrm{a} = k_\mathrm{a}
h_\mathrm{a}$ for some $k,k_\mathrm{a} > 0$.  For instance, if
$\delta = 1$ (resp., $\delta \to \infty$), namely, the particle is
completely submerged below (resp., emerged above) the sea surface,
$k = 1$ (resp., $k_\mathrm{a} = 1$) is an appropriate choice so
$\ell = 2a$ (resp., $\ell_\mathrm{a} = 2a$).  Thus, with this in
mind,
\begin{align}
  \langle F_\mathrm{drag}\rangle
  = & {} 
  \frac{1}{2a}\int_{-h}^{h_\mathrm{a}}
  \frac{12\mu_\mathrm{f}\frac{A_\mathrm{f}}{\ell_\mathrm{f}}}{m_\mathrm{p}}
  (v_\mathrm{f} - v_\mathrm{p})\d{z}\nonumber\\
  = & {}
  \frac{1}{2a}\int_{(\Phi-2)a}^{0}
  \frac{12\mu\frac{\pi(1-\Psi)a^2}{k(2-\Phi)a}}{\frac{4}{3}\pi a^3\rho_\mathrm{p}}
  (v - v_\mathrm{p})\d{z}\nonumber\\
  & + 
  \frac{1}{2a}\int_0^{\Phi a}
  \frac{12\mu_\mathrm{a}\frac{\pi\Psi a^2}{k_\mathrm{a}\Phi a}}{\frac{4}{3}\pi a^3\rho_\mathrm{p}}
  (v_\mathrm{a} - v_\mathrm{p})\d{z}\nonumber\\
  = & {} 
  \frac{9\mu k^{-1}(1-\Psi)}{2\rho_\mathrm{p}a^2} (v
  - v_\mathrm{p}) +
  \frac{9\mu_\mathrm{a}k_\mathrm{a}^{-1}\Psi}{2\rho_\mathrm{p}a^2} (v_\mathrm{a}
  - v_\mathrm{p})\nonumber\\
  = & {}
  \frac{3}{2}\left(1-\frac{\Phi}{6}\right)\frac{u - v_\mathrm{p}}{\tau},
  \label{eq:SDavg}
\end{align}
where
\begin{equation}
  u := (1-\alpha)v + \alpha v_\mathrm{a}, 
  \label{eq:u}
\end{equation}
and the parameters
\begin{equation}
  \tau := \frac{1-\frac{1}{6}\Phi}{3\left(k^{-1}(1-\Psi) +
  \gamma k_\mathrm{a}^{-1}\Psi\right)\delta
  }\cdot \frac{a^2}{\mu/\rho},\quad 
  \alpha := \frac{\gamma k_\mathrm{a}^{-1}\Psi}{k^{-1}(1-\Psi) +
  \gamma k_\mathrm{a}^{-1}\Psi},\quad
  \gamma := \frac{\mu_\mathrm{a}}{\mu}.
  \label{eq:par}
\end{equation}

Finally, plugging \eqref{eq:FFavg}--\eqref{eq:SDavg} in \eqref{eq:mr},
we obtain, after some algebraic manipulation,
\begin{equation}
  \boxed{
  \dot v_\mathrm{p} + \left(f +
  \tfrac{1}{3}R\omega\right)v_\mathrm{p}^\perp
  + \tau^{-1} v_\mathrm{p} =
  R\frac{\D{v}}{\D{t}} + 
  R\left(f + \tfrac{1}{3}\omega\right)v^\perp +
  \tau^{-1}u, 
  }
  \label{eq:MR}
\end{equation}
where
\begin{equation}
  \quad
  R : = \frac{1 - \frac{1}{2}\Phi}{1 - \frac{1}{6}\Phi},
  \label{eq:R}
\end{equation}
which is the explicit form of the Maxey--Riley set proposed in this
paper.  As a second-order ordinary differential equation in position,
to integrate this classical mechanics motion law, not only initial
position has to be specified but clearly also initial velocity.

\begin{figure}[t!]
  \centering%
  \includegraphics[width=\textwidth]{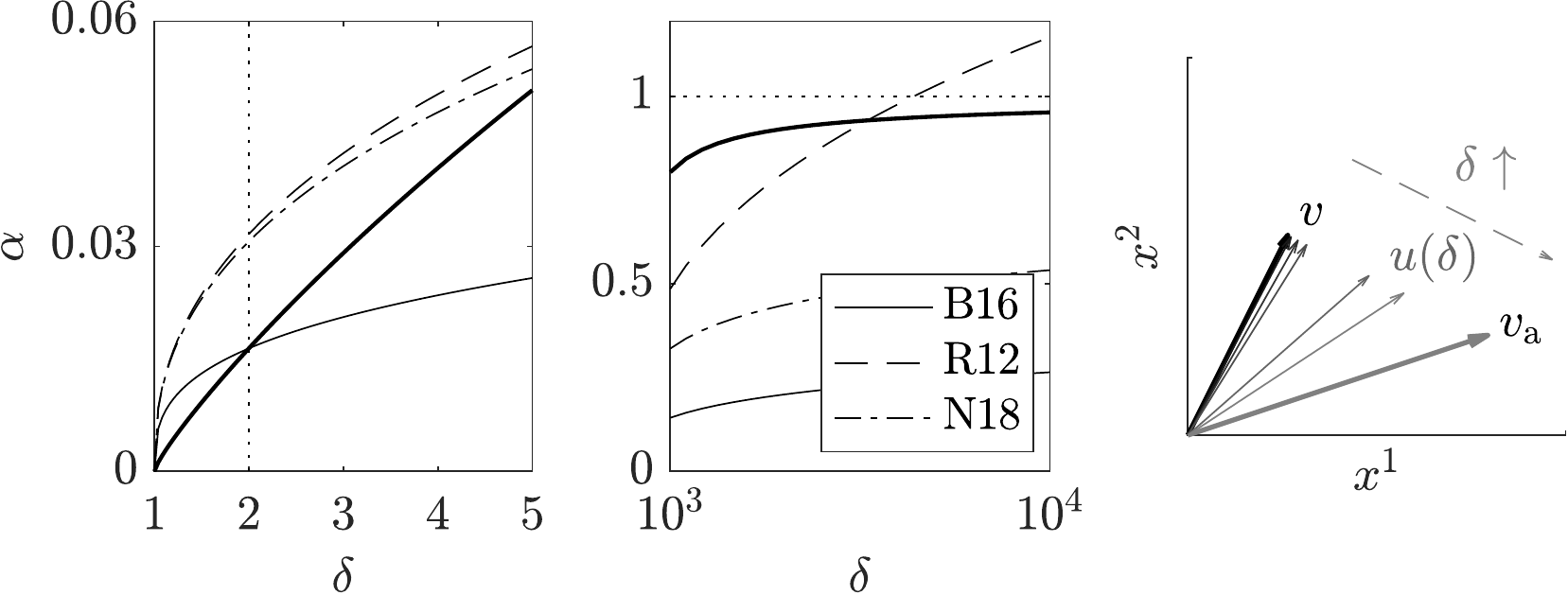}%
  \caption{(left and middle) The behavior of the leeway factor
  $\alpha$ in \eqref{eq:par} as a function of $\delta$ (with $\gamma
  = 1/60$ and $k = 1 = k_\mathrm{a}$).  B16, R12, and N18 indicate
  $\alpha(\delta)$ curves derived by \citet{Beron-etal-16},
  \citet{Rohrs-etal-12}, and \citet{Nesterov-18}, respectively.
  (right) The $\delta$-weighted velocity in \eqref{eq:u} for selected
  values of $\delta$.}
  \label{fig:u}%
\end{figure}

In \eqref{eq:par} parameter $\gamma > 0$ is less than unity ($\gamma
\approx 1/60$, typically), while parameters $\alpha$ and $\tau$
behave as follows.  First recall that $0 \le \Phi < 2$, so $0 \le
\Psi < 1$.  Then given that $k,k_\mathrm{a} > 0$, it is easy to see
that $0 \le \alpha < 1$. More specifically, $\alpha = 0$ when $\delta
= 1$ and $\alpha \to 1$ slowly as $\delta \to \infty$ (cf.\ thick
curve(s) in the left and middle panels of Figure \ref{fig:u}).
Parameter $\tau$, with units of time and representing a generalization
of the so-called \emph{Stokes time} \cite{Sozza-etal-16}, decays
as a function of $\delta$ from $\smash{\frac{a^2}{3\mu/\rho}}$
(since $k = 1$ is an appropriate choice when $\delta = 1$) to 0.
Yet it can be brought arbitrarily close to 0 for finite $\delta$
if the inertial particle radius ($a$) is small enough.  Finally,
parameter $R$ in \eqref{eq:R} decays from 1 to 0 as $\delta$ increases
from 1.

Because $\alpha \ge 0$, the convex combination $u$ in \eqref{eq:u}
can be interpreted as a \emph{$\delta$-weighted average of the seawater}
($v$) \emph{and air} ($v_\mathrm{a}$) \emph{velocities}. In fact,
$u$ coincides with $v$ in the neutrally buoyant case ($\delta = 1$)
in which the particle lies fully immersed in the seawater below the
surface, whereas $u$ approaches $v_\mathrm{a}$ as the particle
lightens (i.e., as $\delta$ departs from unity) until it becomes
fully exposed to the air above the sea surface.

The original Maxey--Riley set was derived under the assumption that
particle Reynolds number is less than unity, so the Stokes law for
drag \eqref{eq:SD} can be used.  The particle Reynolds number,
$\mathrm{Re_p} := \smash{\frac{V_\mathrm{slip}\ell_\mathrm{f}}
{\mu_\mathrm{f}/\rho_\mathrm{f}}}$ where $V_\mathrm{slip}$ is a
measure of the magnitude of the \emph{slip velocity}, i.e., that
of the particle velocity ($v_\mathrm{p}$) \emph{relative} to that
of the carrying flow ($v_\mathrm{f}$).  The asymptotic analysis of
set \eqref{eq:MR} as $\tau \to 0$ (or, equivalently, $a \to 0$ if
$\delta$ is kept finite) in the following section will reveal that
an appropriate measure of $V_\mathrm{slip}$ is that of $|v_\mathrm{p}
- u|$.  Furthermore, this asymptotic analysis will reveal that
$v_\mathrm{p}-u = O(\tau)$, so the use of the Stokes drag law will
be well justified for sufficiently small particles independent of
the magnitude of the carrying flow velocity, effectively given by
that of the $\delta$-weighted velocity $u$, and the carrying fluid
kinematic viscosity, taken as that of either the seawater or the
air, or some average thereof.

As it follows from the aforementioned asymptotic analysis, in the
sizeless particle case ($\tau = 0$), $v_\mathrm{p}$ coincides with
$u$.  The search-and-rescue literature (e.g., Breivik and Allen
\cite{Breivik-Allen-08} and references therein) often models windage
effects on the drift of objects as an additive contribution to the
ocean current.  In our notation this is $v_\mathrm{p} = u$ for some
$\alpha$, commonly referred to as a \emph{leeway factor}.  Obtained
empirically, $\alpha$ is taken as some fixed value in the range
1--5\pct{ }\cite{Duhec-etal-15, Trinanes-etal-16,
Allshouse-etal-17}.  However, formulas depending on the projected
areas of emerged and submerged pieces of the objects and their
floatation characteristics have been proposed
\cite{Rohrs-etal-12, Nesterov-18}.  These formulas,
seemingly valid for arbitrary shaped objects, are obtained by
assuming that the drag in the seawater is exactly balanced by that
in the air above, a hard to justify assumption apparently first
made by \citet{Geyer-89}.  Furthermore, these formulas consider a
quadratic (in the slip particle) drag law.  Such a law assumes that
the particle is in Newton's (rather than Stokes') regime, which is
valid for high particle Reynolds numbers \cite{Kundu-etal-12}.
Assuming that the (constant) drag coefficient is the same below and
above the sea surface as in \citet{Nesterov-18}, we show in the
left and right panels of Figure \ref{fig:u} the resulting leeway
factors as a function of $\delta$ for the case of spherical objects.
Note for instance that the formula derived by \citet{Rohrs-etal-12}
(cf.\ also \citet{Daniel-etal-02}) exceeds unity in the $\delta\to
\infty$ limit, while that of \citet{Nesterov-18} has not converged
to unity for $\delta$ values for which a particle is almost completely
exposed to the air (indeed, for $\delta = 10^3$, $\Phi = h_\mathrm{a}/2a
= 0.9816$).  For smaller $\delta$ values the leeway factors derived
by these authors exceed $\alpha$ in \eqref{eq:par} for $k = 1 =
k_\mathrm{a}$.  Figure \ref{fig:u} also shows the $\alpha$ curve
obtained by \citet{Beron-etal-15}.  Note that it lies below that
one derived here, and it also very slowly tends to unity as $\delta\to
\infty$.  Because of this and the additional freedom in choosing
$k$ and $k_\mathrm{a}$, the new formula for $u$ has more margin
(leeway!) than its predecessor for modeling in an wider range of
conditions.

\subsection{Limitations and heuristic extensions}

The Maxey--Riley theory for inertial ocean dynamics proposed in
this paper has several limitations.  First is its restriction to
spherical particles (objects), which constrains its ability to
account for the motion of flotsam in general. Posing the general
problem of a rigid body of arbitrary shape moving in the flow of a
fluid is a very difficult task, which is beyond the scope of our
paper.  However, a simple heuristic fix, which can be expected to
be valid for sufficiently small objects, is to multiply $\tau$ in
\eqref{eq:par} by $K$, a \emph{shape factor} satisfying \cite{Ganser-93}
\begin{equation}
  K^{-1} = \frac{1}{3}\frac{a_\mathrm{n}}{a_\mathrm{v}} +
  \frac{2}{3}\frac{a_\mathrm{s}}{a_\mathrm{v}}.
  \label{eq:K}
\end{equation}
Here $a_\mathrm{n}$, $a_\mathrm{s}$, and $a_\mathrm{v}$ are the
radii of the sphere with equivalent projected area, surface area,
and equivalent volume, respectively, whose average provide an
appropriate choice for $a$.  A caveat is that $K$ is nonunique for
nonisometric shapes owing to orientation dependence.  If the
orientation is not known, \citet{Ganser-93} recommends to use
the average of the two extreme values of $K$.

Second, in deriving the Maxey--Riley set \eqref{eq:MR} we have
assumed a flat air--sea interface, ignoring the effects of the
Stokes drift arising from material orbits not being closed under a
wavy water surface \cite{Phillips-77}.   A first step toward
including these effects is at the level of the ocean component of
the carrying flow, $v$.  One option is to take $v$ as the output
from a coupled ocean--wave circulation model \cite{Breivik-etal-15}.
Another, less challenging option is to add \cite{Craik-82} to any
given representation of $v$ that ignores gravity wave effects, a
Stokes drift velocity $v_\mathrm{S}$.  To estimate $v_\mathrm{S}$
there several options depending on whether the directional wave
spectrum is known \cite{Jenkins-89} or not \cite{Webb-Fox-11,
Tamura-etal-12, Breivik-etal-16}.  The simplest rule is to make
$v_\mathrm{S}$ a small fraction of the air velocity $v_\mathrm{a}$
assuming that wind and waves are aligned and that the wave field
is in a steady state \cite{Wu-83}.

Finally, ignoring lateral gradients and temporal variations of the
density of the advecting fluid can be consequential, particularly
near frontal regions.  While the original Maxey--Riley set was
derived for the case of homogeneous carrying fluid density, heuristic
extensions to the inhomogeneous case have been proposed
\cite{Tanga-Provenzale-94}, which can be considered.  More
specifically, \citet{Tanga-Provenzale-94} considered the motion of
particles in a stable stratified fluid with buoyancy oscillating
around a reference density.  This translated into making parameter
$\delta$ in the original Maxey--Riley set a periodic function of
time by making $\rho_\mathrm{p}$ a periodic function of time
while $\rho$ is kept constant.  These heuristics may be modified to
investigate the situation in which an inertial particle with fixed
density $\rho_\mathrm{p}$ moves through an ambient fluid with
density $\rho$ changing in space and time.  This corresponds to
making $\delta$ a predefined arbitrary function of $x$ and $t$.  In
our case, $\rho$ is the density of the seawater. The air density
does not appear in the Maxey--Riley set \eqref{eq:MR}.  Indeed, the
only air parameter is the air viscosity, which can be kept safely
fixed.  The condition $\delta(x,t) \ge 1$ needed for the Maxey--Riley
set to remain valid should not be difficult to be satisfied for an
initially sufficiently buoyant particle.

\section{Behavior at limiting particle buoyancies and small-size
asymptotics}

\subsection{The neutrally buoyant case}

Setting $\delta = 1$, the Maxey--Riley set \eqref{eq:MR}
reduces to
\begin{equation}
  \dot v_\mathrm{p} + \left(f +
  \tfrac{1}{3}\omega\right)v_\mathrm{p}^\perp
  + \tau^{-1} v_\mathrm{p} =
  \frac{\D{v}}{\D{t}} + 
  \left(f + \tfrac{1}{3}\omega\right)v^\perp +
  \tau^{-1}v, 
  \label{eq:MR-neutral} 
\end{equation}
with
\begin{equation}
  \tau = \frac{a^2}{3\mu/\rho}.  
  \label{eq:tau-neutral}
\end{equation} 
The resulting set coincides \emph{exactly} with the Maxey--Riley
equation for neutrally buoyant particles as considered in \citet[Chapter
7]{Montabone-02}, which is the standard Maxey--Riley with Coriolis
and lift forces included, but with Faxen corrections and memory
term neglected  (cf.\ \citet{Cartwright-etal-10}, Section 4.1)
Evaluated at $\delta = 1$, the Maxey--Riley set for inertial surface
ocean dynamics derived by \citet{Beron-etal-16} has the same form
as \eqref{eq:MR-neutral} except for the terms produced by the lift
force, which was not included in that formulation.

Such dynamics are quite special: they coincide, irrespective of the
size of the particle (equivalently, the value of $\tau$), with those
of Lagrangian (seawater in the present case) particles if $v_\mathrm{p}
= v$ initially.  To see this, following \citet{Babiano-etal-00}
closely, we add and subtract $(\nabla v) v_\mathrm{p}$ to and from
the right-hand-side of \eqref{eq:MR-neutral} so it recasts as the
linear system
\begin{equation}
  \dot y = A y,\quad
  y:= v_\mathrm{p} - v,\quad 
  A:=-\left(\nabla v + \left(f + \tfrac{1}{3}\omega\right)J +
  \tau^{-1}\Id\right),
  \label{eq:z}
\end{equation} 
where $\dot v = \smash{\frac{\d{}}{\d{t}}}v = \partial_t v + (\nabla
v) v_\mathrm{p}$ is the total derivative of $v$ taken a long a
particle trajectory, satisfying $\dot x = v_\mathrm{p}$.
Clearly, the trivial solution $y = 0$ is invariant under the
dynamics.  In other words,
\begin{equation}
  \mathcal N := \{(x,t,v_\mathrm{p}) \mid v_\mathrm{p} =
  v(x,t),\ (x,t)\in D\times I\}
  \label{eq:N}
\end{equation}
represents an invariant manifold (modulo its boundary, which has
corners due to finiteness of $I$) that is unique as it does not
depend on the choice of $\tau$.

However, in the nonrotating case ($f = 0$) and ignoring the lift
force, the motion of neutrally buoyant particles of finite size is
known from numerical analysis \cite{Babiano-etal-00, Vilela-etal-06}
as well as laboratory experimentation \cite{Sapsis-etal-11} to
possibly deviate from that of Lagrangian particles.
\citet{Sapsis-Haller-08} rigorously addressed this problem by
deriving a sufficient condition for global attractivity of $\mathcal
N$ in that case as well as a necessary condition for local instability
of $\mathcal N$.  It turns out that, because $J = - J^\top$, the
same conditions as those obtained by \citet{Sapsis-Haller-08} are
found in the present geophysical setting with Coriolis and lift
forces (cf.\ Appendix \ref{ap:neutral} for details).  In other
words, these terms contribute to neither setting the attractivity
property of $\mathcal N$, nor controlling the growth of perturbations
off $\mathcal N$.

Specifically, let $S := \smash{\frac{1}{2}}(\nabla v + (\nabla
v)^\top)$ be the rate-of-strain tensor.  Then for $\mathcal N$ to
be globally attracting, i.e., for $v_\mathrm{p}$ to approach $v$
and hence neutrally buoyant finite-size particle motion to synchronize
with seawater particle motion in the long run in $D$, it is sufficient
that $S + \tau^{-1}\Id$ be positive definite for all $x \in D$ over
the time interval $I$, or, equivalently,
\begin{equation}
  \tau < \frac{2}{\sqrt{S_\mathrm{n}^2+S_\mathrm{s}^2} - \nabla\cdot
  v}, 
  \label{eq:Matt}
\end{equation}
where $S_\mathrm{n} := \partial_1v^1 - \partial_2v^2$ and $S_\mathrm{s}
:= \partial_2v^1 + \partial_1v^2$ respectively are normal and shear
strain components, for all $x \in D$ over the time interval $I$.
Clearly, for the latter to be realized over the finite-time interval
$I$, $v_\mathrm{p}$ must initially lie sufficiently close to $v$,
a restriction that is not required when $I = \mathbb{R}$ as in
\citet{Sapsis-Haller-08}.  In the geophysically relevant incompressible
case, \eqref{eq:Matt} reduces to $\tau < \smash{1/\sqrt{|\det S|}}$
for all $(x,t) \in D\times I$.  On the other hand, instantaneous
divergence away from $\mathcal N$ will take place where $S +
\tau^{-1}\Id$ is sign indefinite, or, equivalently, where \eqref{eq:Matt}
is violated.

\subsection{The maximally buoyant case}

The limit $\delta\to \infty$ is dynamically less sophisticated than
the $\delta = 1$ case of the previous section.  In this limit,
$\tau = 0$ and hence the Maxey--Riley set \eqref{eq:MR} reduces
to simply
\begin{equation}
  v_\mathrm{p} = v_\mathrm{a}.
  \label{eq:va}
\end{equation}
A maximally buoyant particle lies on the assumed flat surface ocean
and, irrespective of its size, its motion is synchronized at all
times with that of air particle (i.e., Lagrangian) motion.  In
this limit, \eqref{eq:MR} and the Maxey--Riley set derived by
\citet{Beron-etal-16} behave identically.

\subsection{Slow manifold reduction}

Because of the small particle size assumption, it is natural to
investigate the asymptotic behavior of the Maxey--Riley set
\eqref{eq:MR} as $\tau\to 0$, as it has been done for the Maxey--Riley
set in its standard fluid mechanics form \cite{Rubin-etal-95,
Burns-etal-99, Mograbi-Bar-06, Haller-Sapsis-08} and its earlier
adaptations for ocean dynamics \cite{Beron-etal-15, Beron-etal-16}.

To carry the above investigation formally, we first rescale space
and time by a characteristic length scale $L$ and characteristic
time scale $T = L/V$ where $V$ is a characteristic velocity.  Then
write the Maxey--Riley set \eqref{eq:MR} as an autonomous dynamical
system in the extended phase space $D\times I\times \mathbb R^2$
with variables $(x,\varphi,v_\mathrm{p})$, where $\varphi = t$,
namely,
\begin{align}
  \dot x 
  &= 
  v_\mathrm{p},\label{eq:MRsys-a}\\ 
  \dot\varphi
  &=
  1,\label{eq:MRsys-b}\\
  \tau \dot v_\mathrm{p}
  &=
  u - v_\mathrm{p} - 
  \tau \left(f + \tfrac{1}{3}R\omega\right)v_\mathrm{p}^\perp + 
  \tau R\frac{\D{v}}{\D{t}} + 
  \tau R\left(f + \tfrac{1}{3}\omega\right)v^\perp.\label{eq:MRsys-c}
\end{align}
All variables are here understood with no fear of confusion to be
dimensionless. In particular, the dimensionless $\tau$
parameter,
\begin{equation}
  \tau =
  \frac{(1-\frac{1}{6}\Phi)}{3\left(k^{-1}(1-\Psi) +
  \gamma k_\mathrm{a}^{-1}\Psi\right)\delta
  }\cdot \mathrm{St},
  \label{eq:eps}
\end{equation}
where $\mathrm{St} :=
\smash{\left(\frac{a}{L}\right)^2}\mathrm{Re}$ is a
Stokes number with $\mathrm{Re} := \smash{\frac{VL}{\mu/\rho}}$
the Reynolds number.  We assume
\begin{equation}
  \tau \ll 1.
\end{equation}

Now, from \eqref{eq:MRsys-a}--\eqref{eq:MRsys-b} it is clear that
$v_\mathrm{p}$ is a fast variable changing at $O(\tau^{-1})$ speed
while $(x,\varphi)$, changing at $O(1)$ speed, is a slow variable.
The coexistence of fast and slow variables in system
\eqref{eq:MRsys-a}--\eqref{eq:MRsys-b} makes it a singular perturbation
problem \cite{Fenichel-79, Jones-95}.  To see this, we rewrite
\eqref{eq:MRsys-a}--\eqref{eq:MRsys-b} using the fast timescale
\cite{Haller-Sapsis-08}
\begin{equation}
  \mathcal T := \frac{t - t_0}{\tau},
\end{equation}
where $t_0 \neq 0$, namely,
\begin{align}
  x' 
  &= 
  \tau v_\mathrm{p},\label{eq:MRsys2-a}\\ 
  \varphi'
  &=
  \tau,\label{eq:MRsys2-b}\\
  v_\mathrm{p}'
  &=
  u - v_\mathrm{p} - 
  \tau \left(f + \tfrac{1}{3}R\omega\right)v_\mathrm{p}^\perp + 
  \tau R\frac{\D{v}}{\D{t}} + 
  \tau R\left(f + \tfrac{1}{3}\omega\right)v^\perp,\label{eq:MRsys2-c}
\end{align}
where $' := \smash{\frac{\d{}}{\d{\mathcal T}}}$. The $\tau = 0$
limit of \eqref{eq:MRsys2-a}--\eqref{eq:MRsys2-c} has an invariant
normally hyperbolic manifold (modulo its boundary) $\mathcal S_0$
filled with fixed points, which globally attracts its solutions
exponentially fast in time.  This \emph{critical manifold} has a
graph representation:
\begin{equation}
  \mathcal S_0 = \{(x,\varphi,v_\mathrm{p}) \mid v_\mathrm{p} =
  u(x,\varphi),\ x\in D,\, \varphi\in I\}.
  \label{eq:S0}
\end{equation}
The motion of \eqref{eq:MRsys2-a}--\eqref{eq:MRsys2-c} at $\tau =
0$ is trivial: trajectories off $\mathcal S_0$ are attracted to it
and get stuck there.  By contrast, \eqref{eq:MRsys-a}--\eqref{eq:MRsys-c}
at $\tau = 0$ blows the motion on $\mathcal S_0$ to produce nontrivial
behavior on it, whereas the motion off $\mathcal S_0$ is not defined.
The idea of Fenichel's \cite{Fenichel-79, Jones-95} geometric
singular perturbation theory is to enable realization of the fast
and slow aspects of the motion simultaneously as follows.

Assume that $v$ and $v_\mathrm{a}$, and hence their $\delta$-weighted
average $u$, are $C^r$ smooth (i.e., $r$ times continuously
differentiable) in their arguments with $r > 1$.  Then when $0 <
\tau \ll 1$, there exists a unique (up to an
$O(\smash{\mathrm{e}^{-1/\tau}})$ error), locally invariant (i.e.,
with trajectories only possibly leaving through the boundary),
globally attracting manifold
\begin{equation}
  \mathcal S_\tau := \left\{(x,\varphi,v_\mathrm{p}) \mid v_\mathrm{p} =
  u_\tau(x,\varphi),\ (x,\varphi)\in D\times I\right\},
  \label{eq:Stau}
\end{equation}
where
\begin{equation}
  u_\tau(x,\varphi) = u(x,\varphi) + \sum\nolimits_1^r \tau^n
  u_n(x,\varphi) + O(\tau^{r+1}), 
  \label{eq:taylor}
\end{equation}
which is $C^r$ $O(\tau)$-close to $\mathcal S_0$ and $C^r$-diffeomorphic
to it.  The manifold $\mathcal S_\tau$ is called a \emph{slow
manifold} since the restriction of \eqref{eq:MRsys-a}--\eqref{eq:MRsys-c}
to $\mathcal S_\tau$ is a slowly varying system, namely,
\begin{equation}
  x' = \tau\left.v_\mathrm{p}\right|_{\mathcal S_\tau} = \tau u(x,t) +
  \sum\nolimits_1^r \tau^{n+1} u_n(x,t) + O(\tau^{r+2}).
\end{equation}
Moreover, this system controls the motion off $\mathcal S_\tau$ as
follows.  When $\tau = 0$, each point off $\mathcal S_0$ belongs
to the stable manifold of $\mathcal S_0$, which is foliated by its
distinct stable fibers (stable manifolds of points on $\mathcal
S_0$).  The stable manifold of $\mathcal S_0$ and its stable fibers
perturb along with $\mathcal S_0$.  Consequently, for $0 < \tau \ll
1$ each point off $\mathcal S_\tau$ is connected to a point on
$\mathcal S_\tau$ by a fiber in the sense that it follows a trajectory
that approaches its partner on $\mathcal S_\tau$ exponentially fast
in time.

The function defining $\mathcal S_\tau$ is found by plugging the
Taylor expansion in \eqref{eq:taylor} into
\eqref{eq:MRsys2-a}--\eqref{eq:MRsys2-c} and equating $\tau$-power
terms. This gives, following steps similar to those in Appendix
\ref{ap:lift},
\begin{align}
  u_1
  ={}&
  R\frac{\D{v}}{\D{t}} + R \left(f +
  \tfrac{1}{3}\omega\right) v^\perp - \frac{\D{u}}{\D{t}} -
  \left(f + \tfrac{1}{3}R\omega\right) u^\perp\label{eq:u1}\\
  u_n
  ={}&
  - \left(f + \tfrac{1}{3}R\omega\right)u_{n-1}^\perp\nonumber\\
  &- 
  \partial_t u_{n-1} - (\nabla u_{n-1})u - (\nabla u)u_{n-1}\nonumber\\ 
  &-
  \sum\nolimits_{m=1}^{n-2}(\nabla u_m)u_{n-m-1},\quad n \ge
  2,\label{eq:un}
\end{align}
which fully determine $\mathcal S_\tau$.  Switching back to the
original time scale, the leading-order contribution to the Maxey--Riley
system \eqref{eq:MR} on the slow manifold $\mathcal S_\tau$, in
dimensional variables, is
\begin{equation}
\boxed{
\dot{x} = v_\mathrm{p} \sim u +  \tau\bigg(
R\frac{\D{v}}{\D{t}} + R \left(f +
\tfrac{1}{3}\omega\right) v^\perp - \frac{\D{u}}{\D{t}} -
\left(f + \tfrac{1}{3}R\omega\right) u^\perp\bigg)
}
\label{eq:MRslow}
\end{equation}
as $\tau\to 0$. The reduced system \eqref{eq:MRslow} may be referred
to as the \emph{inertial equation} \footnote{The slow manifold
$\mathcal S_\tau$ and the Maxey--Riley equation restricted to
$\mathcal S_\tau$ formally satisfy the definition of \emph{inertial
manifold} and \emph{inertial equation}, respectively, developed in
the study of long-time-asymptotic behavior (attractors) of
infinite-dimensional dynamical systems \cite{Temam-90}.  In such
systems, actual attractors are hard to compute and are generally
not even manifolds.  The inertial manifold is easier to compute,
smooth, and contains the attractor.  It is unclear to us why these
constructs are called ``inertial,'' but this certainly is not related
to resistance of an object to a change in its velocity as meant
here.} following nomenclature employed in earlier work
\cite{Haller-Sapsis-08, Beron-etal-15}.

Several remarks are in order.  Firstly, the nondimensionalization
above makes sense in the $\delta$-range where $\alpha$ is small,
which is rather large (cf.\ Fig.\ \ref{fig:u}).  Indeed, since the
magnitude of $v$ is typically smaller than that of $v_\mathrm{a}$,
if $v$ is scaled using $V$, then $v_\mathrm{a}$ can be scaled using
$\alpha^{-1}V$ under the assumption that $\alpha$ is small enough.
This way $u$ will scale like $V$ as required.

Second, rapid changes in time of the carrying flow velocity,
represented by $u$, will lead to rapid changes on $\mathcal S_\tau$,
thereby hindering its efficacy in absorbing trajectories of the
Maxey--Riley equation over finite time \cite{Haller-Sapsis-08,
Haller-Sapsis-10}.  Yet appropriate redefinition of the slow manifold
involving history integrals of the fast time scale \cite{Roberts-08}
can compensate the effects of such rapid variations even if they
are stochastic.

Third, unlike the Maxey--Riley set \eqref{eq:MR}, its slow manifold
reduction \eqref{eq:MRslow} does not require specification of the
initial velocity, which is not known in general.  Also, \eqref{eq:MRslow}
does not include the term $v_\mathrm{p}/\tau$ present in \eqref{eq:MR}.
This term is known to produce numerical instability in long
backward-time integration, e.g., as required is source inversion
\cite{Olascoaga-etal-06, Miron-etal-19b}, unless specialized numerical
techniques \cite{Gear-Kevrekidis-03} are used.  Furthermore, according
to Theorem 2 of \citet{Sapsis-Haller-08}, the starting position
$x(t_0)$ of any solution $(x(t),v_\mathrm{p}(t))$ of \eqref{eq:MRslow}
may be recovered with $O(\tau)$ precision.

Fourth, representing a simpler model than the full Maxey--Riley set
\eqref{eq:MR}, the reduced equation \eqref{eq:MRslow} can provide
insight that is difficult---if not impossible---to be gained from
the analysis of the full system, as we show below.

Lastly, we note one difference with the slow manifold of the
Maxey--Riley set in its standard fluid mechanics setting with lift
force.  As we show in Appendix \ref{ap:lift}, the lift force makes
an $O(\tau^2)$ contribution to the slow manifold in that setting.
This is unlike the slow manifold in the present setting, in which
case the contribution is $O(\tau)$.

\section{Qualitative performance relative to observations}

\subsection{Trapping of flotsam inside mesoscale eddies}

Using in-situ measurements from sea campaign Expedition 7th Continent
in the North Atlantic subtropical gyre, data from satellite
observations and models, \citet{Brach-etal-18} recently provided
evidence that mesoscale anticyclonic eddies are more efficient at
trapping flotasm within than cyclonic ones.  Indeed, they found
microplastic concentrations nearly ten times higher in an anticyclonic
eddy surveyed than in a nearby cyclonic eddy.  This phenomenon
is predicted by the Maxey--Riley set proposed here.

Specifically, suppose that there are no winds ($v_\mathrm{a} = 0$)
and the ocean flow is quasigeostrophic.  To wit, $v =
\smash{\nabla^\perp}\psi + O(\mathrm{Ro}^2)$, $\smash{\partial_t}v
= O(\mathrm{Ro}^2)$, and $f = f_0 + O(\mathrm{Ro})$, where $\psi$
is a streamfuction (e.g., $\psi = g\smash{f_0^{-1}}\eta$) and
$\mathrm{Ro} = V/L|f_0| > 0$ small is the Rossby number
\cite{Pedlosky-87}.  Under these conditions, to the lowest order
in $\mathrm{Ro}$, the reduced Maxey--Riley set \eqref{eq:MRslow}
simplifies to
\begin{equation}
  \dot{x} = v_\mathrm{p} \sim (1-\alpha)\nabla^\perp\psi +  \tau
  (1-R-\alpha) f_0 \nabla\psi.
  \label{eq:MRslow-geo}
\end{equation}

As defined by \citet{Haller-etal-16}, a \emph{rotationally coherent
vortex} is a material region $U(t)$, $t\in [t_0,t_0+T]\subset I$,
enclosed by the outermost, sufficiently convex isoline of the
Lagrangian averaged vorticity deviation (LAVD) field enclosing a
local maximum. For the quasigeostrophic flow above, the LAVD is
given by
\begin{equation}
  \mathrm{LAVD}_{t_0}^t(x_0) := \int_{t_0}^t
  |\nabla^2\psi(F_{t_0}^s(x_0),s) - \overline{\nabla^2\psi}(s)|
  \d{s},
\end{equation}
where $F_{t_0}^t(x_0)$ is a trajectory of $\nabla^\perp\psi$ starting
from $x_0$ at $t_0$ and the overline represents an average on $D$.
As a consequence, the elements of the boundaries of such material
regions $U(t)$ complete the same total material rotation relative
to the mean material rotation of the whole fluid mass in the domain
$D$ that contains them.  This property of the boundaries is observed
\cite{Haller-etal-16} to restrict their filamentation to be mainly
tangential under advection from $t_0$ to $t_0+T$.

Assume that $D$ is large enough so $\overline{\nabla^2\psi}$ nearly
vanishes and bear in mind that $1-R-\alpha \ge
0$.  Then applying on \eqref{eq:MRslow-geo} Theorem of 2 of
\citet{Haller-etal-16}, which essentially is an application of
Liouville's theorem, one finds that a trajectory launched from a
nondegenerate maximum $x_0^*$ of $\mathrm{LAVD}_{t_0}^{t_0+T}(x_0)$
attracts or repel trajectories of \eqref{eq:MRslow-geo} depending
on the sign of
\begin{equation}
  f_0\nabla^2\psi(F_{t_0}^t(x_0^*),t),t)
  \label{eq:fw}
\end{equation}
over the time interval $[t_0,t_0+T]$. More precisely, a rotationally
coherent quasigeostrophic vortex $U(t)$ will contain an attractor
(resp., repeller) over $[t_0,t_0+T]$ staying $O(\tau)$-close to its
center $F_{t_0}^t(x_0^*)$ if \eqref{eq:fw} is negative (resp.,
positive).  In other words, cyclonic (resp., anticyclonic) mesoscale
such eddies disperse away (resp., concentrate within) inertial
particles floating at the surface of the ocean.  This result, which
holds in the presence of a sufficiently calm uniform wind, is
consistent with the observations reported by \citet{Brach-etal-18}.

The earlier oceanic implementations \cite{Beron-etal-15, Beron-etal-16}
of the Maxey--Riley formalism predict the behavior that is at odds
with the above result.  This may be a consequence of the heuristics
considered by \citet{Beron-etal-16} being too restrictive.  The
case of \citet{Beron-etal-15} is different because the only adaptation
made was the inclusion of the Coriolis force.  As considered, then,
the set is not in principle meant to be valid for a particle floating
at the sea surface, but rather for a particle immersed in a fluid
as in the standard formulation.  Indeed, that set does not seem
possible to be obtained as a limit of the set derived here except
for neutrally buoyant particles.

We finally note that \citet{Beron-etal-15} present observational
evidence of \emph{Sargassum} (a pelagic brown algae) accumulating
in a cold-core (i.e., cyclonic) Gulf Stream ring (eddy), which seems
at odds with the observations discussed by \citet{Brach-etal-18}
(cf.\ also \citet{Brooks-etal-19}).  An important difference with
microplastic particles is that \emph{Sargassum} presents in the
form of mats, which are better modeled as \emph{networks} of buoyant
particles than as individual particles.  Work in progress
\cite{Beron-19} is revealing that \emph{elastic chains} of
sufficiently small buoyant particles evolving under the Maxey--Riley
set derived here collect in cyclonic rotationally coherent
quasigeostrophic eddies provided that the chains are sufficiently
stiff.

\subsection{Great garbage patches}

The NOAA's Global Drifter Program (GDP) is an array of drifting
buoys used to measure the near surface ocean Lagrangian circulation
\cite{Lumpkin-Pazos-07}.  A GDP drifter follows the Surface Velocity
Program design \cite{Niiler-Paduan-95}, with a spherical float,
which includes a transmitter to relay data via satellite, tethered
to a holey sock drogue (sea anchor), centered at 15 m depth.
\citet{Beron-etal-16} noted that GDP drifters have lost their drogues
exhibit different time-asymptotic behavior than those that have not
along their lifetime.  More specifically, the undrogued drifters
tend in the long run to accumulate in the centers of the subtropical
gyres, most notably the Atlantic and Pacific subtropical gyres.  By
contrast drogued drifters tend to acquire more heterogeneous
distributions in the long term.  The regions where undrogued drifters
concentrate coincide with the regions where microplastics maximize
their densities as observations reveal \cite{Cozar-etal-14}.  In
particular, the region where flotsam accumulates in the North Pacific
is referred to as the Great Pacific Garbage Patch
\cite{Lebreton-etal-2018}.  We proceed to show that the Maxey--Riley
set proposed here is able to predict ``garbage patches'' consistent
with observed behavior, thereby allowing to interpret this behavior
as produced by inertial effects as suggested by \citet{Beron-etal-16}
using an early version of the set derived here.

\begin{figure}[t!]
  \centering
  \includegraphics[width=\textwidth]{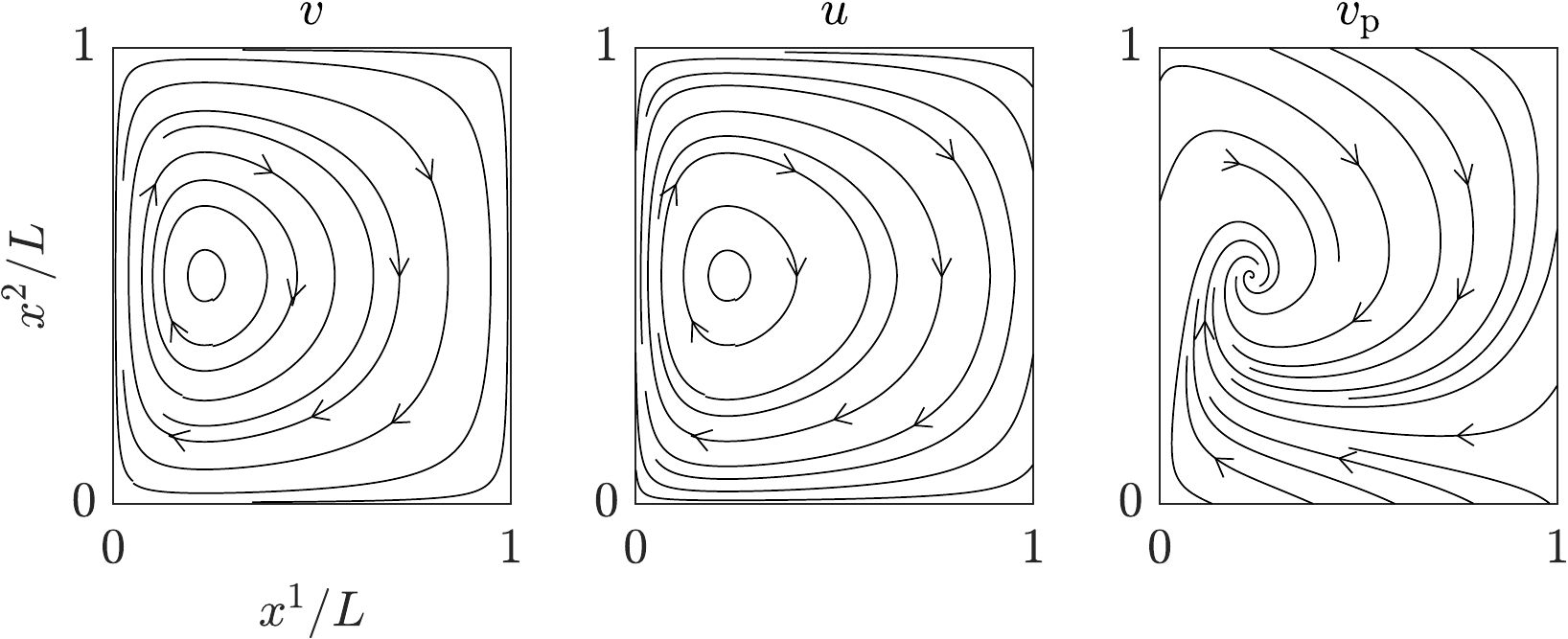}%
  \caption{Streamlines of the Stommel wind-driven circulation model
  velocity (left), the $\delta$-weighted velocity resulting from
  this velocity and the wind field that drives the Stommel gyre
  (middle), and dominant part of inertial particle velocity on the
  slow manifold of the Maxey--Riley set resulting from feeding
  the later with the aforementioned sewater and air velocities
  (right).}
  \label{fig:stommel}
\end{figure}

We first do this by considering as in \citet{Beron-etal-16} the
conceptual model of wind-driven circulation due to \citet{Stommel-48}.
The steady flow in such a barotropic model is quasigeostrophic,
i.e., $v = \nabla^\perp\psi(x) = O(\mathrm{Ro})$, and has an
anticyclonic basin-wide gyre in the northern hemisphere, so $\omega
= \nabla^2\psi \le 0$, driven by steady westerlies and trade winds,
namely, $v_\mathrm{a} = W(x^2)e_1$ with $W'(x^2) \ge 0$.  The
leading-order contribution to inertial particle velocity on the
slow manifold \eqref{eq:MRslow} takes the form
\begin{equation}
  v_\mathrm{p} = (1-\alpha)\nabla^\perp\psi + \alpha We_1 +
 \tau f_0 \big((1-R-\alpha) \nabla\psi + \alpha W e_2\big)
 \label{eq:stommel}
\end{equation}
with an $O(\mathrm{R_0}^2)$ error.  The divergence of this velocity,
\begin{equation}
  \nabla\cdot  v_\mathrm{p} = \tau f_0 \big(
  (1-R-\alpha)\nabla^2\psi - \alpha W'(x^2)\big).
  \label{eq:div}
\end{equation}
Recalling that $1-R(\delta) \ge \alpha(\delta) \ge 0$, it follows that
$\nabla\cdot v_\mathrm{p} \le 0$, which promotes attraction of
inertial particles toward the interior of the gyre in a manner akin
to undrogued drifters and plastic debris.

A precise localization of the attractor can be attained by inspecting
the streamfunction and the wind field.  A simple expression for the
streamfuction is \cite{Haidvogel-Bryan-92}
\begin{equation}
  \psi = \frac{\pi F}{H\beta}\Big(1-x^1/L -
  \mathrm{e}^{-\frac{x^1}{r/\beta}}\Big)\sin\frac{\pi x^2}{L},
\end{equation}
where $H$ is the (thermocline) depth, $r$ is the (bottom) friction
coefficient, $L$ here is the length of a square domain, and $F$ is the wind
stress (per unit density) amplitude, which sets the amplitude of
the wind field:
\begin{equation}
  W = \sign\left(x^2-\tfrac{1}{2}L\right)
  \sqrt{\frac{\rho F}{\rho_\mathrm{a}C_\mathrm{D}}} 
  \sqrt{\sign\left(\tfrac{1}{2}L - x^2\right) 
  \cos\frac{\pi x^2}{L}},
\end{equation}
where $C_\mathrm{D}$ is a (dimensionless) drag coefficient.  (We
note that $W$ is $C^\infty$ everywhere except at $x^2 = \frac{1}{2}L$,
a set of measure zero. Thus \eqref{eq:stommel} is a valid approximation
to \eqref{eq:MR} almost everywhere in the domain.) Figure
\ref{fig:stommel} shows streamlines of $v = \nabla^\perp\psi$ on
the left, $u = (1-\alpha)\nabla^\perp\psi + \alpha We_1$ in the
middle, and $v_\mathrm{p}$ given by \eqref{eq:stommel} on the right.
Parameters for the Stommel model are taken as in \citet{Stommel-48}
with $C_\mathrm{D} = 1.2 \times 10^{-3}$ (e.g., \citet{Large-Pond-81}).
Soft inertial parameters are chosen to represent undrogued GDP
drifters, namely, $\delta = 2$ and $a = 17.5$ cm. We have set also
$k = 1 = k_\mathrm{a}$.  The rest of the parameters are hard, typical
seawater and air values.  This gives $R = 0.6$ and $\tau = 0.0968$
d.  Variations of the soft parameters do not change the qualitative
aspects of the solution.  The streamlines of $v$ show a center,
displaced westward, resulting from the $\beta$ effect.  The streamlines
of $u$ are similar, with a center in precisely the same place.  This
is located at $(x^1,x^2) =
(-\smash{\frac{r}{\beta}}\log\smash{\frac{r}{\beta L}},\frac{1}{2}L)$.
The stability type of this equilibrium is changed to a \emph{stable
spiral} when the inertial velocity is $v_\mathrm{p}$ considered.
(Indeed, we have numerically verified that, at that point, $\nabla
v$ and $\nabla u$ both have complex conjugate pure imaginary
eigenvalues, while the complex conjugate eigenvalue pair of $\nabla
v_\mathrm{p}$ has a negative real part.) This thus shows explicitly
where inertial particles accumulate and further that finite-size
effects, produced by the term proportional to $\tau$ in $v_\mathrm{p}$,
are responsible for driving the accumulation.  A search-and-rescue
type model, i.e., one for which $\dot x = u$ neglecting those
effects, is not enough to realize it, as \citet{Beron-etal-16} noted
earlier.

\begin{figure}[t!]
  \centering%
  \includegraphics[width=\textwidth]{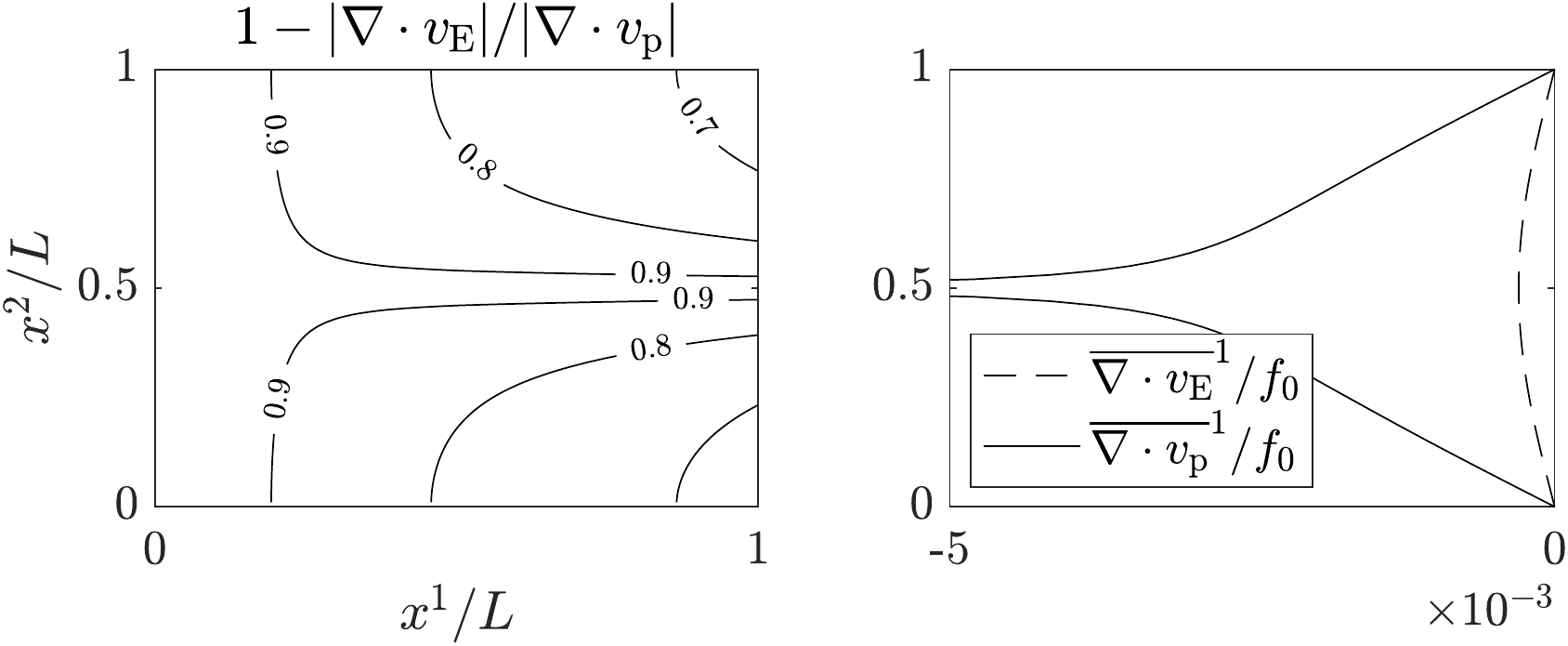}%
  \caption{For the Stommel model, relative difference of inertial
  and Ekman divergence magnitudes (left) and zonally averaged
  inertial and Ekman divergences normalized by the Coriolis parameter
  (right).}
  \label{fig:ekman}%
\end{figure}

We finalize the Stommel model analysis by comparing the divergence
of the inertial velocity \eqref{eq:div} with that one that would result
from the wind stress curl (Ekman divergence), namely,
\begin{equation}
  \nabla\cdot v_\mathrm{E} = -\frac{\pi F}{f_0HL}\sin\frac{\pi
  x^2}{L},
\end{equation}
which is nonpositive.  The comparison in presented in Figure
\ref{fig:ekman}.  Note that $|\nabla\cdot v_\mathrm{p}|$ dominates
over $|\nabla\cdot v_\mathrm{E}|$ in the domain (left panel) while
both are much smaller than $f_0$ (right panel), reason for which
the Ekman convergence does not enter in the Stommel model (it is a
higher-order effect in the Rossby number Ro).  It is important to
realize that the divergence $\smash{\overline{\nabla\cdot
v_\mathrm{p}}^1/f_0}$ at $x^2 = \smash{\frac{1}{2}}L$ is not a
deficiency of the Maxey--Riley description of inertial effects, but
rather a consequence of the convenient form of the wind stress
assumed by Stommel in his model, which leads to a divergence of the
associated wind there.

We now proceed to test the ability of the Maxey--Riley set derived
here to promote inertial particle concentration in the subtropical
gyres in a realistic setting following \citet{Beron-etal-16}. We
focus on the North Atlantic for simplicity as subtropical gyres in
the other oceans behave similarly.  The exception is the Indian
Ocean, where aggregation of inertial particles is not so evident,
suggesting that ocean and atmospheric conditions are peculiar there
\cite{vanderMheen-etal-19, Miron-etal-19b}.

Thus we feed the full spherical form of the Maxey--Riley set
\eqref{eq:MR-sph} with $v$ as given by surface ocean velocity from
the Global $1/12^\circ$ HYCOM (HYbrid-Coordinate Ocean Model) $+$
NCODA (Navy Coupled Ocean Data Assimilation) Ocean Reanalysis
\cite{Cummings-Smedstad-13}, and $v_\mathrm{a}$ as the wind velocity
from the National Centers for Environmental Prediction (NCEP) Climate
Forecast System Reanalysis (CFSR) employed to construct the wind
stress applied on the model. (To be more precise, the NCEP winds
are provided at 10 m, so we multiply them by one half following
\citet{Hsu-etal-94} to infer $v_\mathrm{a}$.) This way ocean currents
and winds are made dynamically consistent with each other.  

Specifically, we partition the North Atlantic domain into $5^{\circ}
\times 5^{\circ}$ longitude--latitude boxes and construct a matrix
of probabilities, $P$, of the drifters and the inertial particles
to transitioning, irrespective of the start time, among them over
a short time.  Such a time-independent $P$ represents a discrete
autonomous transfer operator which governs the evolution of tracer
probability densities, satisfying a stationary advection--diffusion
process, on a Markov chain defined on the boxes of the partition
\cite{Froyland-etal-14, Miron-etal-17, Miron-etal-19a, Miron-etal-19b,
Olascoaga-etal-18}.  Thus given an initial probability vector
$\mathbf f$, this is forward evolved under left multiplication by
$P$, namely, $\mathbf f_n = \mathbf fP^n$, $n = 1,2,\dotsc$.  This
way long-term evolution can be investigated in a probabilistic sense
without the need of long trajectory records, which may be generated
numerically but are not available from observations.  To construct
$P$ we set a transition time of 5 days, which is longer than the
Lagrangian decorrelation time scale, estimated to be of the order
of 1 day near the ocean surface \cite{LaCasce-08}, thereby
guaranteeing sufficiently negligible memory into the past that the
Markov assumption can be expected to hold well.

\begin{figure}[t!]
  \centering 
  \includegraphics[width=\textwidth]{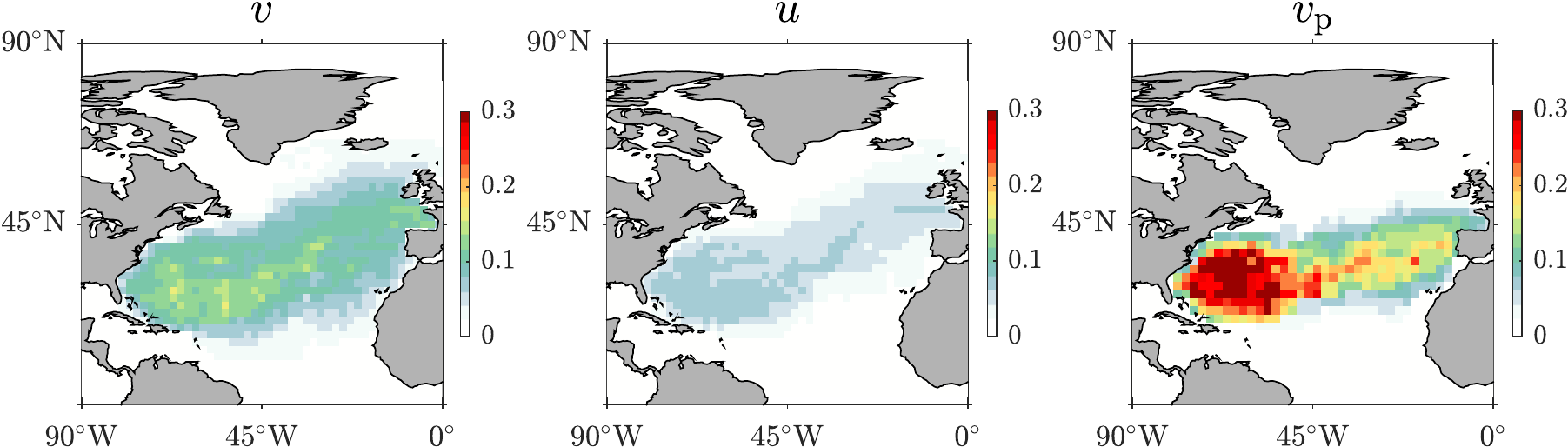}%
  \caption{Long-term distribution in the North Atlantic of an
  initially uniform probability density under action of an autonomous
  transfer operator constructed using short-run trajectories produced
  by surface HyCOM velocity (left), $\delta$-weighted velocity where
  seawater velocity is the HyCOM and the air velocity is the NCP
  wind used to force HyCOM (middle), and trajectories produced by
  the Maxey--Riley set derived in this paper fed with these seawater
  and air velocities (right). Soft inertial parameters choices are
  $\delta = 2$ and $a = 17.5$ cm, representing undrogued GDP drifters,
  and $k = 1 = k_\mathrm{a}$.  Densities are subjected to a fourth-root
  transformation.}
  \label{fig:hycom}%
\end{figure}

Figure \ref{fig:hycom} shows distributions in the North Atlantic
after 10 years of an initially uniform probability density evolving
under the action of transition matrix constructed using trajectories
produced by HyCOM surface ocean velocity output (left), trajectories
of the $\delta$-weighted velocity $u$ resulting from setting $v$
to be the HyCOM velocity and $v_\mathrm{a}$ to be the NCEP winds
used to force the model (middle), and trajectories produced by the
Maxey--Riley set fed with these velocities. (The full spherical
form \eqref{eq:MR-sph} of the set is employed in these calculations;
trajectories of $v$ and $u$ are computed by integrating the left
set in \eqref{eq:vfaf} with $v_\mathrm{f}$ replaced by $v$ and $u$,
respectively.)  Parameters were taken as above to represent undrogued
GDP drifters, $\delta = 2$ and $a = 17.5$ cm.  Note the good
qualitative agreement with the results based on the conceptual
wind-driven circulation model of Stommel of Figure \ref{fig:stommel}.
Note the density values, which are subjected to a fourth-root
transformation.  Inertial particles reveal accumulation toward the
center of the gyre, while seawater particles and particles evolving
under the $\delta$-weighted velocity $u$ do not.  Indeed, the
densities corresponding to the latter are low and distributed more
homogeneously over the gyre.  Garbage patches in the ocean tend to
localize in the center of the gyres consistent with the inertial
particles.  These reinforces the idea put forth by \citet{Beron-etal-16}
that inertial effects dominate the production of such patches. Ekman
transport convergence, the only mechanism acting in the absence of
inertial effects, does not control garbage accumulation despite
earlier \cite{Maximenko-etal-12} and recent \cite{vanderMheen-etal-19}
claims. Furthermore, ignoring finite-size effects results in strong
dispersion by view of the very low density values attained. This
questions the validity of the leeway modeling framework.  The results
based on the slow manifold reduction of the Maxey--Riley set are
indistinguishable from those based on the full set, which was
$v$-initialized with mean HyCOM velocity.  This provides support
for the validity of the slow manifold reduction.

\begin{figure}[t!]
  \centering 
  \includegraphics[width=\textwidth]{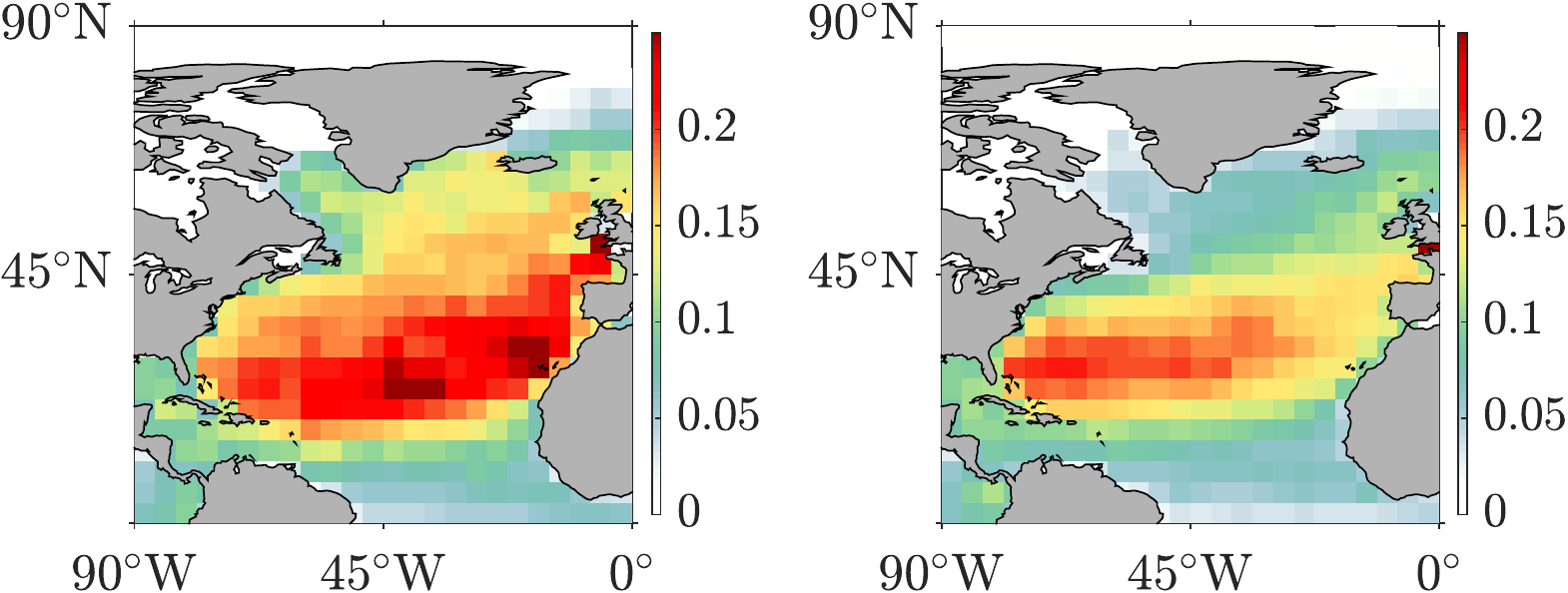}%
  \caption{Long-term distribution in the North Atlantic of an
  initially uniform probability density under action of an autonomous
  transfer operator constructed using short-run drogued (left) and
  undrogued (right) trajectories of satellite-tracked drifting buoys
  from the NOAA Global Drifter Program. Densities are subjected to
  a fourth-root transformation}
  \label{fig:svp}%
\end{figure}

Finally, Figure \ref{fig:svp} shows the distribution of an initially
uniform probability density after 10 years of forward evolution
under a discrete action of a transfer operator constructed using
drogued (left) and undrogued (right) drifter trajectories from the
GDP dataset.  Note how undrogued drifter density in the long run
tends to concentrate in the center of the gyre more evidently than
drogued drifter density.  Such a difference was not noted in previous
work \cite{vanSebille-etal-12, Maximenko-etal-12} which also used
probabilistic approaches to investigate long-term behavior.  Very
importantly, note that this behavior resembles quite well the
simulated behavior described above, providing a reality check for
it.  More specifically, the drogued drifters behave in a manner akin
to seawater particles.  The undrogued drifters, by contrast, behave
more like inertial particles, which represent a prototype for flotsam
in general as undrogued drifters and plastic debris present a similar
tendency to aggregate in the interior of the subtropical gyres.

\section{Concluding remarks}

In this paper we have proposed a theory for the motion finite-size
particles floating at the ocean surface based on the Maxey--Riley
set, the de-jure fluid dynamics framework for inertial particle
motion investigation.  The theory thus consist of a Maxey--Riley
set obtained by vertically averaging the various forces involved
in the original Maxey--Riley set, appropriately adapted to account
for planet's rotation and sphericity effects, across an assumed
small spherical particle that floats at a flat air--sea interface
and thus is subjected to the action ocean currents and winds.

The inertial particle velocity of the resulting  Maxey--Riley set
is shown to decay exponentially fast in time to a limit that is
$O(a^2)$-close, where $a$ is the particle radius, to an average of
the seawater and air velocities weighted by a function of the
seawater-to-particle density ratio.  This weighted average velocity
has a form which is similar to the so-called leeway velocity that
forms the basis for search-and-rescue modeling.  Such a leeway model
is not sufficient to explain the role of mesoscale eddies as traps
for marine debris or the formation of garbage patches in subtropical
gyres, which are phenomena dominated by finite-size effects.

The resulting Maxey--Riley set either outperforms or has potential
for outperforming an earlier proposed set \cite{Beron-etal-16} in
various aspects.  For instance, in the neutrally buoyant case,
inertial particle motion is synchronized with seawater (i.e.,
Lagrangian) particle motion under the same conditions as in the
original Maxey--Riley set without Coriolis and lift forces. Also,
the newly proposed set predicts concentration of particles inside
anticyclonic mesoscale eddies consistent with observations of marine
microplastic debris.  On the other hand, including lift force the
new Maxey--Riley set is expected to better represent particle
dispersion in the presence of fast submesoscale eddy motion.
Furthermore, the proposed heuristic shape corrections raise the
earlier set limitation to spherical particles. Finally, recommendations
were made for accounting for Stokes drift effects are expected to
improve the earlier set performance in the presence of waves, and
for incorporating the effects of inhomogeneities of the carrying
density field, which can be consequential near frontal regions.

We close by noting that a paper in preparation \cite{Olascoaga-etal-19}
reports on the results from a field experiment which involved the
deployment, in the Gulf Stream and other areas of the North Atlantic,
and subsequently tracking, using global satellite positioning, of
buoys of varied buoyancies, sizes, and shapes.  In that paper the
Maxey--Riley set derived here is shown capable of reproducing
individual trajectories with unexpected accuracy given the uncertainty
around the ocean current and wind representations, providing strong
support the validity of the set.

\begin{acknowledgments}
Clarifying comments on geometric singular perturbation theory by
Chris Jones and George Haller are sincerely appreciated.  We thank
Tony Roberts for calling our attention to relevant work
\cite{Gear-Kevrekidis-03, Roberts-08} that we had overlooked.  The
drifter data were collected by the NOAA Global Drifter Program
(http://\allowbreak www.\allowbreak aoml.\allowbreak noaa.\allowbreak
gov/\allowbreak phod/dac). The $1/12^\circ$ Global HYCOM$+$NCODA
Ocean Reanalysis was funded by the U.S.\ Navy and the Modeling and
Simulation Coordination Office.  Computer time was made available
by the DoD High Performance Computing Modernization Program. The
output and forcing are publicly available at http://hycom.org.  Our
work was supported by the University of Miami's Cooperative Institute
for Marine \& Atmospheric Studies (CIMAS).
\end{acknowledgments}

\appendix
\numberwithin{equation}{section}

\section{Inertial ocean dynamics on the sphere}\label{ap:sph}

Let $a_\odot$ be the mean radius of the Earth, and consider the
rescaled longitude ($\lambda $) and latitude ($\vartheta $)
coordinates, namely,
\begin{equation}
  x^1 = (\lambda - \lambda_0)\cdot a_\odot\cos\vartheta_0,\quad
  x^2 = (\vartheta - \vartheta_0)\cdot a_\odot, 
  \label{eq:lonlat}
\end{equation}
respectively, measured from an arbitrary location on the surface
of the planet.  Consider further the following geometric coefficients
\cite{Ripa-JPO-97b}:
\begin{equation}
  \gamma_\odot := \sec\vartheta_0\cos\vartheta,\quad 
  \tau_\odot := a_\odot^{-1}\tan\vartheta.
\end{equation}

The (horizontal) velocity of a fluid particle and its acceleration
as measured by a terrestrial observer are (cf.\ \citet{Ripa-JPO-97b,
Beron-03})
\begin{equation}
  v_\mathrm{f} = 
  \begin{pmatrix}
	 \gamma_\odot & 0\\
	 0 & 1
  \end{pmatrix}
  \dot x,\quad
  a_\mathrm{f} = \dot v_\mathrm{f} + (f + \tau_\odot
  v_\mathrm{f}^1)v_\mathrm{f}^\perp,
  \label{eq:vfaf}
\end{equation}
respectively, where $f = 2\Omega\sin\vartheta$.  It is important
to realize that this is not a mere change of coordinates from
Cartesian to spherical.  Very differently, this is a consequence
of the gravitational force, which, attracting the particle to the
nearest pole, is required to sustain a steady rotation, with angular
velocity $\pm\Omega$, relative to a fixed frame, at any point on
the planet's surface.  The terrestrial observer is then left with
\emph{just} the Coriolis force, in the absence of any other forces,
to describe motion on the surface of the Earth. A very enlightening
way to derive the formula for the acceleration in \eqref{eq:vfaf}
is from Hamilton's principle, with the Lagrangian as written by an
observer standing in a fixed frame, so the only force acting on the
particle (in the absence of any other forces) is the gravitational
one, and the coordinates employed by this observer related to those
rotating with the planet \eqref{eq:lonlat} (cf.\ \citet{Ripa-JPO-97b,
Beron-03}). This is in essence what Pierre Simon de Laplace (1749--1827)
did to derive his theory of tides and at the same time discover the
Coriolis force over a quarter of a century before Gaspard Gustave
de Coriolis (1792--1843) was born \cite{Ripa-RMF-95}. For a nice
account on the history of this, many times misunderstood, force,
cf.\ \citet{Ripa-FCE-97}.

By a similar token, the fluid's Eulerian acceleration takes the
form
\begin{equation}
  \frac{\D{v_\mathrm{f}}}{\D{t}} + (f + \tau_\odot
  v_\mathrm{f}^1)v_\mathrm{f}^\perp,
  \label{eq:euler}
\end{equation}
where
\begin{equation}
  \frac{\D{v_\mathrm{f}}}{\D{t}} = \partial_t v_\mathrm{f} + (\nabla
  v_\mathrm{f}) \dot x = \partial_t v_\mathrm{f} +
  (\gamma_\odot^{-1}\partial_1 v_\mathrm{f})v_\mathrm{f}^1 +
  (\partial_2 v_\mathrm{f})v_\mathrm{f}^2.
  \label{eq:DvDt}
\end{equation}

Equations \label{eq:va} hold for a particle of fluid, either seawater
or air, and also for an inertial particle.  The acceleration of the
inertial particle on the left-hand-side of (yet to be evaluated)
Maxey--Riley set \eqref{eq:mr} and in the added mass force \eqref{eq:AM}
is the $\beta$-plane form of $a_\mathrm{f}$ in \eqref{eq:vfaf} for
the case of an inertial particle, resulting from making $\gamma_\odot
= 1$, $\tau_\odot = 0$, and $f = f_0 + \beta x^2$, which, despite
its wide used, does not represent a consistent leading order in
$|x^2|/a_\odot \ll 1$ approximation \cite{Ripa-JPO-97b}.  In turn,
the fluid's Eulerian acceleration that appears in the flow force
\eqref{eq:FF} and the added mass term \eqref{eq:AM} is the $\beta$-plane
form of \eqref{eq:euler}--\eqref{eq:DvDt}.

With the above in mind, the Maxey--Riley set \eqref{eq:MR} on the
sphere then takes the form
\begin{equation}
  \dot v_\mathrm{p} + \left(f + \tau_\odot v_\mathrm{p}^1 +
  \tfrac{1}{3}R\omega\right)v_\mathrm{p}^\perp
  + \tau^{-1} v_\mathrm{p} =
  R\frac{\D{v}}{\D{t}} + 
  R\left(f + \tau_\odot v^1 + \tfrac{1}{3}\omega\right)v^\perp +
  \tau^{-1}u, 
  \label{eq:MR-sph}
\end{equation}
with $\frac{\D{}}{\D{t}}v$ given by \eqref{eq:DvDt} and 
\begin{equation}
  \omega = \gamma_\odot^{-1}\partial_1v^2 -
  \gamma_\odot^{-1}\partial_2(\gamma_\odot v^1)
  = \gamma_\odot^{-1}\partial_1v^2 - \partial_2v^1 + \tau_\odot
  v^1
\end{equation}
as it follows from its definition, $\omega
:= \smash{\lim_{\Delta x^1\Delta x^2\to 0}}
\smash{\frac{1}{\gamma_\odot\Delta x^1\Delta x^2}} \smash{\oint}
(\gamma_\odot v^1\d{x^1} + v^2\d{x^2})$, and noting that
$\gamma_\odot'(x^2)/\gamma_\odot(x^2) = -\tau_\odot(x^2)$

Applying the slow-manifold reduction on \eqref{eq:MR-sph} it
follows, to leading order on the slow manifold, that
\begin{equation}
  \dot{x} \sim u +  \tau\bigg( R\frac{\D{v}}{\D{t}}
  + R \left(f + \tau_\odot v^1 + \tfrac{1}{3}\omega\right) v^\perp
  - \frac{\D{u}}{\D{t}} - \left(f + \tau_\odot v^1 +
  \tfrac{1}{3}R\omega\right) u^\perp\bigg),
  \label{eq:MRslow-sph}
\end{equation}
where $\frac{\D{}}{\D{t}}u$ is as in \eqref{eq:DvDt} with $v_\mathrm{f}$
replaced by $u$.

\section{Attractivity and instability conditions for neutrally
buoyant particles}\label{ap:neutral}

To derive an attractivity condition for $\mathcal N$ in the present
geophysical setting ($f \neq 0$) with lift force, we follow
\citet{Sapsis-Haller-08} closely by first fixing a solution $(y,x)(t)$
to \eqref{eq:z}, which fixes $A(x(t),t)$. Then noting that $y^\top
Ay = y^\top(rA + (1-r)A^\top)y$ for any $r\in \mathbb{R}$, one
finds, using $r=\frac{1}{2}$, that $y^\top Ay \le \max\spec
\smash{\frac{A+A^\top}{2}}\cdot |y|^2$, which follows from real
symmetric $A+A^\top$ admitting an orthogonal diagonalization.  Now,
taking into account that $J = - J^\top$,
\begin{equation}
  \frac{A+A^\top}{2} = S+\tau^{-1} \Id
  \label{eq:ApAt}
\end{equation}
and hence
\begin{equation}
  \frac{1}{2}\frac{\d{}}{\d{t}}|y|^2 = - y^\top (S+\tau^{-1} \Id) y\le 
  -\min\spec(S+\tau^{-1} \Id)\cdot|y|^2.
  \label{eq:dy2dt}
\end{equation}
Integrating from $t = t_0$ to $t > t_0$,
\begin{equation}
  |y(t)|^2 \le |y(t_0)|^2\mathrm{e}^{\textstyle{-\int_{t_0}^t
  (S(x(s),s)+\tau^{-1} \Id)\d{s}}}.
  \label{eq:int}
\end{equation}
Then for $|y(t)|$ to decay from $|y(t_0)|$ as $t$ increases, it is
sufficient that the integrand in \eqref{eq:int} be positive for all
$x\in D$ over the time interval $I$, from which the global attractivity
condition \eqref{eq:Matt} follows.

As noted by \citet{Sapsis-Haller-08}, perturbations off $\mathcal
N$ which are initially sufficiently small will grow or decay depending
on the sign of the instantaneous stability indicator 
\begin{equation}
  \Lambda(x_0,t_0) = \lim_{t\to t_0} 
  \smash{\frac{2}{t - t_0}}\log||P_{t_0}^t||_2
\end{equation}
Here $P_{t_0}^t$ satisfies
\begin{equation}
  \dot P_{t_0}^t = A(x(t;t_0,x_0),t)P_{t_0}^t,\quad
  P_{t_0}^{t_0} = \Id,
\end{equation}
so $y(t;t_0,y_0,x_0) = P_{t_0}^ty_0$, to wit, $P_{t_0}^t$ represents
the fundamental matrix solution of \eqref{eq:z} for initial condition
$(y,x)(t_0) = (y_0,x_0)$. Taylor expanding $P_{t_0}^t$ one finds
\begin{equation}
  P_{t_0}^t = \Id + A_0\cdot (t-t_0) +  O((t-t_0)^2),
\end{equation}
where the shorthand notation $A_0 = A(t_0)$ is used, and hence
\begin{equation}
  (P_{t_0}^t)^\top P_{t_0}^t = \Id + (A_0 + A_0^\top)\cdot (t-t_0)
  + O((t-t_0)^2).
\end{equation}
Then
\begin{equation}
  (||P_{t_0}^t||_2)^2 = 1 - 2\min\spec(S_0+\tau^{-1}\Id)\cdot (t-t_0)
  + O((t-t_0)^2),
\end{equation}
where \eqref{eq:ApAt} was used.  Now, using
$\log\left(1+\smash{\sum_1^\infty} c_n\varepsilon^n\right) =
c_1\varepsilon + O(\varepsilon^2)$ for $\varepsilon
> 0$ small, one finds
\begin{equation}
  \Lambda(x_0,t_0) = -2\min\spec(S(x_0,t_0) + \tau^{-1}\Id).
\end{equation}
Replacing $(x_0,t_0)$ with $(x(t),t)$, it follows that instantaneous
divergence away from $\mathcal N$ will take place where $S +
\tau^{-1}\Id$ is sign indefinite, or, equivalently, where \eqref{eq:Matt}
is violated.

\section{Slow manifold reduction in the standard fluid mechanics
setting with lift force}\label{ap:lift}

The standard fluid mechanics Maxey--Riley equation with lift
force is given by (cf.\ \citet{Montabone-02}, Chapter 4)
\begin{equation}
  \dot v_\mathrm{p} + 
  \tfrac{1}{2}R\omega v_\mathrm{p}^\perp
  + \tau^{-1} v_\mathrm{p} =
  \tfrac{3}{2}R\frac{\D{v}}{\D{t}} + 
  \tfrac{1}{2}R\omega v^\perp +
  \tau^{-1}v,
  \label{eq:MRlift}
\end{equation}
where $v$ is any carrying flow velocity and
\begin{equation}
  \tau := \frac{2R}{9}\cdot \frac{a^2}{\mu/\rho},\quad
  R: = \frac{2\delta}{2+\delta}.
\end{equation}
In nondimensional variables with time rescaled as in \S 4, the above
equation in system form reads
\begin{align}
  x' 
  &= 
  \tau v_\mathrm{p},\\ 
  \varphi'
  &=
  \tau,\\
  v_\mathrm{p}'
  &= 
  v - v_\mathrm{p} -
  \tfrac{1}{2}\tau R\omega v_\mathrm{p}^\perp + 
  \tfrac{3}{2}\tau R\smash{\frac{\D{v}}{\D{t}}} + 
  \tfrac{1}{2}\tau R\omega v^\perp.
  \label{eq:MRsys3-c}
\end{align}

The critical manifold for the above system is
\begin{equation}
  \mathcal S_0 = \{(x,\varphi,v_\mathrm{p}) \mid v_\mathrm{p} =
  v(x,\varphi),\ x\in D,\, \varphi\in I\},
  \label{eq:S0}
\end{equation}
so the slow manifold takes the form:
\begin{equation}
  \mathcal S_\tau := \{(x,\varphi,v_\mathrm{p}) \mid v_\mathrm{p} =
  v(x,\varphi) + \sum\nolimits_1^r \tau^n v_n(x,\varphi) + O(\tau^{r+1}),\
  (x,\varphi)\in D\times I\}.  
  \label{eq:Stau-lift}
\end{equation}
Differentiating the equation defining $\mathcal S_\tau$ above with
respect to $\mathfrak t$,
\begin{align}
  v_\mathrm{p}' 
  ={}& 
  \left[(\nabla v)x' + \partial_\varphi v \varphi' +
  \sum\nolimits_1^r\tau^n\big((\nabla v_n)x' +
  \partial_\varphi v_n \varphi'\big) + O(\tau^{r+1})\right]_{\mathcal S_\tau}\nonumber\\
  ={}&
  \tau \frac{\D{v}}{\D{t}} + \sum\nolimits_{n=2}^r\tau^n
  \Big(\partial_t v_{n-1} + (\nabla v_{n-1})v + (\nabla
  v)v_{n-1}\nonumber\\ 
  &+
  \sum\nolimits_{m=1}^{n-2}(\nabla v_m)v_{n-m-1}\Big) +
  O(\tau^{r+2}).
  \label{eq:dSdt}
\end{align}
Then restricting \eqref{eq:MRsys3-c} to $\mathcal S_\tau$,
i.e.,
\begin{align}
  v_\mathrm{p}'
  ={}&
  \Big[
  v - v_\mathrm{p} -
  \tfrac{1}{2}\tau R\omega v_\mathrm{p}^\perp + 
  \tfrac{3}{2}\tau R\smash{\frac{\D{v}}{\D{t}}} + 
  \tfrac{1}{2}\tau R \omega v^\perp
  \Big]_{\mathcal S_\tau}\nonumber\\
  ={}&
  - \sum\nolimits_1^r \tau^n v_n -
  \tfrac{1}{2}\tau R\omega \left(v^\perp +
  \sum\nolimits_1^r \tau^n v_n^\perp\right) +   
  \tfrac{3}{2}\tau R\smash{\frac{\D{v}}{\D{t}}}\nonumber\\
  &+ 
  \tfrac{1}{2}\tau R \omega v^\perp +
  O(\tau^{r+2}),\label{eq:dvdt-2}
\end{align}
and equating equal $\tau$-power terms in \eqref{eq:dSdt} and
\eqref{eq:dvdt-2}, we obtain
\begin{align}
  v_1
  ={}&
  \left(\tfrac{3}{2}R-1\right)\frac{\D{v}}{\D{t}}\label{eq:v1}\\
  v_n
  ={}&
  - \tfrac{1}{2}R\omega v_{n-1}^\perp\nonumber\\
  &- 
  \partial_t v_{n-1} - (\nabla v_{n-1})v - (\nabla v)v_{n-1}\nonumber\\ 
  &-
  \sum\nolimits_{m=1}^{n-2}(\nabla v_m)v_{n-m-1},\quad n \ge
  2.\label{eq:vn}
\end{align}
The Maxey--Riley set \eqref{eq:MRlift} on the slow manifold
$\mathcal S_\tau$ reduces to
\begin{equation}
  \dot{x} = v_\mathrm{p} = v(x,t) + \sum\nolimits_1^r \tau^n
  v_n(x,t) + O(\tau^{r+1})
\end{equation}
with $v_n(x,t)$ as given in \eqref{eq:v1}--\eqref{eq:vn}. Note
that the lift force makes an $O(\tau^2)$ contribution to
$\mathcal S_\tau$.

\bibliography{fot}

\begin{thebibliography}{106}%
\makeatletter
\providecommand \@ifxundefined [1]{%
 \@ifx{#1\undefined}
}%
\providecommand \@ifnum [1]{%
 \ifnum #1\expandafter \@firstoftwo
 \else \expandafter \@secondoftwo
 \fi
}%
\providecommand \@ifx [1]{%
 \ifx #1\expandafter \@firstoftwo
 \else \expandafter \@secondoftwo
 \fi
}%
\providecommand \natexlab [1]{#1}%
\providecommand \enquote  [1]{``#1''}%
\providecommand \bibnamefont  [1]{#1}%
\providecommand \bibfnamefont [1]{#1}%
\providecommand \citenamefont [1]{#1}%
\providecommand \href@noop [0]{\@secondoftwo}%
\providecommand \href [0]{\begingroup \@sanitize@url \@href}%
\providecommand \@href[1]{\@@startlink{#1}\@@href}%
\providecommand \@@href[1]{\endgroup#1\@@endlink}%
\providecommand \@sanitize@url [0]{\catcode `\\12\catcode `\$12\catcode
  `\&12\catcode `\#12\catcode `\^12\catcode `\_12\catcode `\%12\relax}%
\providecommand \@@startlink[1]{}%
\providecommand \@@endlink[0]{}%
\providecommand \url  [0]{\begingroup\@sanitize@url \@url }%
\providecommand \@url [1]{\endgroup\@href {#1}{\urlprefix }}%
\providecommand \urlprefix  [0]{URL }%
\providecommand \Eprint [0]{\href }%
\providecommand \doibase [0]{http://dx.doi.org/}%
\providecommand \selectlanguage [0]{\@gobble}%
\providecommand \bibinfo  [0]{\@secondoftwo}%
\providecommand \bibfield  [0]{\@secondoftwo}%
\providecommand \translation [1]{[#1]}%
\providecommand \BibitemOpen [0]{}%
\providecommand \bibitemStop [0]{}%
\providecommand \bibitemNoStop [0]{.\EOS\space}%
\providecommand \EOS [0]{\spacefactor3000\relax}%
\providecommand \BibitemShut  [1]{\csname bibitem#1\endcsname}%
\let\auto@bib@innerbib\@empty
\bibitem [{\citenamefont {Stokes}(1851)}]{Stokes-51}%
  \BibitemOpen
  \bibfield  {author} {\bibinfo {author} {\bibfnamefont {G.~G.}\ \bibnamefont
  {Stokes}},\ }\bibfield  {title} {\enquote {\bibinfo {title} {{On the Effect
  of the Internal Friction of Fluids on the Motion of Pendulums}},}\
  }\href@noop {} {\bibfield  {journal} {\bibinfo  {journal} {Transactions of
  the Cambridge Philosophical Society}\ }\textbf {\bibinfo {volume} {9}},\
  \bibinfo {pages} {8} (\bibinfo {year} {1851})}\BibitemShut {NoStop}%
\bibitem [{\citenamefont {Basset}(1888)}]{Basset-88}%
  \BibitemOpen
  \bibfield  {author} {\bibinfo {author} {\bibfnamefont {A.~B.}\ \bibnamefont
  {Basset}},\ }\bibfield  {title} {\enquote {\bibinfo {title} {Treatise on
  hydrodynamics},}\ \ }(\bibinfo  {publisher} {Deighton Bell},\ \bibinfo
  {address} {London},\ \bibinfo {year} {1888})\ Chap.~\bibinfo {chapter} {22},
  pp.\ \bibinfo {pages} {285--297}\BibitemShut {NoStop}%
\bibitem [{\citenamefont {Boussinesq}(1885)}]{Boussinesq-85}%
  \BibitemOpen
  \bibfield  {author} {\bibinfo {author} {\bibfnamefont {J.~V.}\ \bibnamefont
  {Boussinesq}},\ }\bibfield  {title} {\enquote {\bibinfo {title} {Sur la
  r\'esistance qu\'oppose un fluide ind\'efini au repos, sans pesanteur, au
  mouvement vari\'e d\'une sph\'ere solide qu'il mouille sur toute sa surface,
  quand les vitesses restent bien continues et assez faibles pour que leurs
  carr\'es et produits soient n\'egligeables},}\ }\href@noop {} {\bibfield
  {journal} {\bibinfo  {journal} {Comptes Rendu de l'Academie des Sciences}\
  }\textbf {\bibinfo {volume} {100}},\ \bibinfo {pages} {935--937} (\bibinfo
  {year} {1885})}\BibitemShut {NoStop}%
\bibitem [{\citenamefont {Oseen}(1927)}]{Oseen-27}%
  \BibitemOpen
  \bibfield  {author} {\bibinfo {author} {\bibfnamefont {C.~W.}\ \bibnamefont
  {Oseen}},\ }\href@noop {} {\emph {\bibinfo {title} {Hydrodynamik}}}\
  (\bibinfo  {publisher} {Akademische Verlagsgesellschaft},\ \bibinfo {address}
  {Leipzig},\ \bibinfo {year} {1927})\BibitemShut {NoStop}%
\bibitem [{\citenamefont {Tchen}(1947)}]{Tchen-47}%
  \BibitemOpen
  \bibfield  {author} {\bibinfo {author} {\bibfnamefont {C.~M.}\ \bibnamefont
  {Tchen}},\ }\href@noop {} {Ph.D. thesis},\ \bibinfo  {school} {Delft,
  Martinus Nijhoff, The Hage} (\bibinfo {year} {1947})\BibitemShut {NoStop}%
\bibitem [{\citenamefont {Corrsin}\ and\ \citenamefont
  {Lumely}(1956)}]{Corrsin-Lumely-56}%
  \BibitemOpen
  \bibfield  {author} {\bibinfo {author} {\bibfnamefont {S.}~\bibnamefont
  {Corrsin}}\ and\ \bibinfo {author} {\bibfnamefont {J.}~\bibnamefont
  {Lumely}},\ }\href@noop {} {\bibfield  {journal} {\bibinfo  {journal} {Appl.
  Sci. Res. A}\ }\textbf {\bibinfo {volume} {6}},\ \bibinfo {pages} {114}
  (\bibinfo {year} {1956})}\BibitemShut {NoStop}%
\bibitem [{\citenamefont {Maxey}\ and\ \citenamefont
  {Riley}(1983)}]{Maxey-Riley-83}%
  \BibitemOpen
  \bibfield  {author} {\bibinfo {author} {\bibfnamefont {M.~R.}\ \bibnamefont
  {Maxey}}\ and\ \bibinfo {author} {\bibfnamefont {J.~J.}\ \bibnamefont
  {Riley}},\ }\bibfield  {title} {\enquote {\bibinfo {title} {Equation of
  motion for a small rigid sphere in a nonuniform flow},}\ }\href@noop {}
  {\bibfield  {journal} {\bibinfo  {journal} {Phys. Fluids}\ }\textbf {\bibinfo
  {volume} {26}},\ \bibinfo {pages} {883} (\bibinfo {year} {1983})}\BibitemShut
  {NoStop}%
\bibitem [{\citenamefont {Riley}(1971)}]{Riley-71}%
  \BibitemOpen
  \bibfield  {author} {\bibinfo {author} {\bibfnamefont {J.~J.}\ \bibnamefont
  {Riley}},\ }\href@noop {} {Ph.D. thesis},\ \bibinfo  {school} {The John
  Hopkins University, Baltimore, Maryland} (\bibinfo {year} {1971})\BibitemShut
  {NoStop}%
\bibitem [{\citenamefont {Gatignol}(1983)}]{Gatignol-83}%
  \BibitemOpen
  \bibfield  {author} {\bibinfo {author} {\bibfnamefont {R.}~\bibnamefont
  {Gatignol}},\ }\bibfield  {title} {\enquote {\bibinfo {title} {The faxen
  formulae for a rigid particle in an unsteady non-uniform stokes flow},}\
  }\href@noop {} {\bibfield  {journal} {\bibinfo  {journal} {J. Mec. Theor.
  Appl.}\ }\textbf {\bibinfo {volume} {1}},\ \bibinfo {pages} {143--160}
  (\bibinfo {year} {1983})}\BibitemShut {NoStop}%
\bibitem [{\citenamefont {Auton}, \citenamefont {Hunt},\ and\ \citenamefont
  {{Prud'homme}}(1988)}]{Auton-etal-88}%
  \BibitemOpen
  \bibfield  {author} {\bibinfo {author} {\bibfnamefont {T.~R.}\ \bibnamefont
  {Auton}}, \bibinfo {author} {\bibfnamefont {F.~C.~R.}\ \bibnamefont {Hunt}},
  \ and\ \bibinfo {author} {\bibfnamefont {M.}~\bibnamefont {{Prud'homme}}},\
  }\bibfield  {title} {\enquote {\bibinfo {title} {The force exerted on a body
  in inviscid unsteady non-uniform rotational flow},}\ }\href@noop {}
  {\bibfield  {journal} {\bibinfo  {journal} {J. Fluid. Mech.}\ }\textbf
  {\bibinfo {volume} {197}},\ \bibinfo {pages} {241} (\bibinfo {year}
  {1988})}\BibitemShut {NoStop}%
\bibitem [{\citenamefont {Michaelides}(1997)}]{Michaelides-97}%
  \BibitemOpen
  \bibfield  {author} {\bibinfo {author} {\bibfnamefont {E.~E.}\ \bibnamefont
  {Michaelides}},\ }\bibfield  {title} {\enquote {\bibinfo {title}
  {{Review---The transient equation of motion for particles, bubbles and
  droplets}},}\ }\href@noop {} {\bibfield  {journal} {\bibinfo  {journal}
  {ASME. J. Fluids Eng.}\ }\textbf {\bibinfo {volume} {119}},\ \bibinfo {pages}
  {233--247} (\bibinfo {year} {1997})}\BibitemShut {NoStop}%
\bibitem [{\citenamefont {Provenzale}(1999)}]{Provenzale-99}%
  \BibitemOpen
  \bibfield  {author} {\bibinfo {author} {\bibfnamefont {A.}~\bibnamefont
  {Provenzale}},\ }\bibfield  {title} {\enquote {\bibinfo {title} {Transport by
  coherent barotropic vortices},}\ }\href@noop {} {\bibfield  {journal}
  {\bibinfo  {journal} {Annu. Rev. Fluid Mech.}\ }\textbf {\bibinfo {volume}
  {31}},\ \bibinfo {pages} {55--93} (\bibinfo {year} {1999})}\BibitemShut
  {NoStop}%
\bibitem [{\citenamefont {Cartwright}\ \emph {et~al.}(2010)\citenamefont
  {Cartwright}, \citenamefont {Feudel}, \citenamefont {K\'arolyi},
  \citenamefont {{de Moura}}, \citenamefont {Piro},\ and\ \citenamefont
  {T\'el}}]{Cartwright-etal-10}%
  \BibitemOpen
  \bibfield  {author} {\bibinfo {author} {\bibfnamefont {J.~H.~E.}\
  \bibnamefont {Cartwright}}, \bibinfo {author} {\bibfnamefont
  {U.}~\bibnamefont {Feudel}}, \bibinfo {author} {\bibfnamefont
  {G.}~\bibnamefont {K\'arolyi}}, \bibinfo {author} {\bibfnamefont
  {A.}~\bibnamefont {{de Moura}}}, \bibinfo {author} {\bibfnamefont
  {O.}~\bibnamefont {Piro}}, \ and\ \bibinfo {author} {\bibfnamefont
  {T.}~\bibnamefont {T\'el}},\ }\bibfield  {title} {\enquote {\bibinfo {title}
  {Dynamics of finite-size particles in chaotic fluid flows},}\ }in\ \href@noop
  {} {\emph {\bibinfo {booktitle} {Nonlinear Dynamics and Chaos: Advances and
  Perspectives}}},\ \bibinfo {editor} {edited by\ \bibinfo {editor}
  {\bibnamefont {{M. Thiel et al.}}}}\ (\bibinfo  {publisher} {Springer-Verlag
  Berlin Heidelberg},\ \bibinfo {year} {2010})\ pp.\ \bibinfo {pages}
  {51--87}\BibitemShut {NoStop}%
\bibitem [{\citenamefont {Babiano}\ \emph {et~al.}(2000)\citenamefont
  {Babiano}, \citenamefont {Cartwright}, \citenamefont {Piro},\ and\
  \citenamefont {Provenzale}}]{Babiano-etal-00}%
  \BibitemOpen
  \bibfield  {author} {\bibinfo {author} {\bibfnamefont {A.}~\bibnamefont
  {Babiano}}, \bibinfo {author} {\bibfnamefont {J.~H.}\ \bibnamefont
  {Cartwright}}, \bibinfo {author} {\bibfnamefont {O.}~\bibnamefont {Piro}}, \
  and\ \bibinfo {author} {\bibfnamefont {A.}~\bibnamefont {Provenzale}},\
  }\bibfield  {title} {\enquote {\bibinfo {title} {Dynamics of a small
  neutrally buoyant sphere in a fluid and targeting in {H}amiltonian
  systems},}\ }\href@noop {} {\bibfield  {journal} {\bibinfo  {journal} {Phys.
  Rev. Lett.}\ }\textbf {\bibinfo {volume} {84}},\ \bibinfo {pages}
  {5,764--5,767} (\bibinfo {year} {2000})}\BibitemShut {NoStop}%
\bibitem [{\citenamefont {Vilela}, \citenamefont {{de Moura}},\ and\
  \citenamefont {Grebogi}(2006)}]{Vilela-etal-06}%
  \BibitemOpen
  \bibfield  {author} {\bibinfo {author} {\bibfnamefont {R.~D.}\ \bibnamefont
  {Vilela}}, \bibinfo {author} {\bibfnamefont {A.~P.~S.}\ \bibnamefont {{de
  Moura}}}, \ and\ \bibinfo {author} {\bibfnamefont {C.}~\bibnamefont
  {Grebogi}},\ }\bibfield  {title} {\enquote {\bibinfo {title} {Finite-size
  effects on open chaotic advection},}\ }\href@noop {} {\bibfield  {journal}
  {\bibinfo  {journal} {Phys. Rev. E}\ }\textbf {\bibinfo {volume} {73}},\
  \bibinfo {pages} {026302} (\bibinfo {year} {2006})}\BibitemShut {NoStop}%
\bibitem [{\citenamefont {Breivik}\ \emph {et~al.}(2013)\citenamefont
  {Breivik}, \citenamefont {Allen}, \citenamefont {Maisondieu},\ and\
  \citenamefont {Olagnon}}]{Breivik-etal-13}%
  \BibitemOpen
  \bibfield  {author} {\bibinfo {author} {\bibfnamefont {{\O}.}~\bibnamefont
  {Breivik}}, \bibinfo {author} {\bibfnamefont {A.~A.}\ \bibnamefont {Allen}},
  \bibinfo {author} {\bibfnamefont {C.}~\bibnamefont {Maisondieu}}, \ and\
  \bibinfo {author} {\bibfnamefont {M.}~\bibnamefont {Olagnon}},\ }\bibfield
  {title} {\enquote {\bibinfo {title} {Advances in search and rescue at sea},}\
  }\href {\doibase 10.1007/s10236-012-0581-1} {\bibfield  {journal} {\bibinfo
  {journal} {Ocean Dynamics}\ }\textbf {\bibinfo {volume} {63}},\ \bibinfo
  {pages} {83--88} (\bibinfo {year} {2013})}\BibitemShut {NoStop}%
\bibitem [{\citenamefont {Bellomo}\ \emph {et~al.}(2015)\citenamefont
  {Bellomo}, \citenamefont {Griffa}, \citenamefont {Cosoli}, \citenamefont
  {Falco}, \citenamefont {Gerin}, \citenamefont {Iermano}, \citenamefont
  {Kalampokis}, \citenamefont {Kokkini}, \citenamefont {Lana}, \citenamefont
  {Magaldi}, \citenamefont {Mamoutos}, \citenamefont {Mantovani}, \citenamefont
  {Marmain}, \citenamefont {Potiris}, \citenamefont {Sayol}, \citenamefont
  {Barbin}, \citenamefont {Berta}, \citenamefont {Borghini}, \citenamefont
  {Bussani}, \citenamefont {Corgnati}, \citenamefont {Dagneaux}, \citenamefont
  {Gaggelli}, \citenamefont {Guterman}, \citenamefont {Mallarino},
  \citenamefont {Mazzoldi}, \citenamefont {Molcard}, \citenamefont {Orfila},
  \citenamefont {Poulain}, \citenamefont {Quentin}, \citenamefont
  {Tintor{\'e}}, \citenamefont {Uttieri}, \citenamefont {Vetrano},
  \citenamefont {Zambianchi},\ and\ \citenamefont
  {Zervakis}}]{Bellomo-etal-15}%
  \BibitemOpen
  \bibfield  {author} {\bibinfo {author} {\bibfnamefont {L.}~\bibnamefont
  {Bellomo}}, \bibinfo {author} {\bibfnamefont {A.}~\bibnamefont {Griffa}},
  \bibinfo {author} {\bibfnamefont {S.}~\bibnamefont {Cosoli}}, \bibinfo
  {author} {\bibfnamefont {P.}~\bibnamefont {Falco}}, \bibinfo {author}
  {\bibfnamefont {R.}~\bibnamefont {Gerin}}, \bibinfo {author} {\bibfnamefont
  {I.}~\bibnamefont {Iermano}}, \bibinfo {author} {\bibfnamefont
  {A.}~\bibnamefont {Kalampokis}}, \bibinfo {author} {\bibfnamefont
  {Z.}~\bibnamefont {Kokkini}}, \bibinfo {author} {\bibfnamefont
  {A.}~\bibnamefont {Lana}}, \bibinfo {author} {\bibfnamefont {M.}~\bibnamefont
  {Magaldi}}, \bibinfo {author} {\bibfnamefont {I.}~\bibnamefont {Mamoutos}},
  \bibinfo {author} {\bibfnamefont {C.}~\bibnamefont {Mantovani}}, \bibinfo
  {author} {\bibfnamefont {J.}~\bibnamefont {Marmain}}, \bibinfo {author}
  {\bibfnamefont {E.}~\bibnamefont {Potiris}}, \bibinfo {author} {\bibfnamefont
  {J.}~\bibnamefont {Sayol}}, \bibinfo {author} {\bibfnamefont
  {Y.}~\bibnamefont {Barbin}}, \bibinfo {author} {\bibfnamefont
  {M.}~\bibnamefont {Berta}}, \bibinfo {author} {\bibfnamefont
  {M.}~\bibnamefont {Borghini}}, \bibinfo {author} {\bibfnamefont
  {A.}~\bibnamefont {Bussani}}, \bibinfo {author} {\bibfnamefont
  {L.}~\bibnamefont {Corgnati}}, \bibinfo {author} {\bibfnamefont
  {Q.}~\bibnamefont {Dagneaux}}, \bibinfo {author} {\bibfnamefont
  {J.}~\bibnamefont {Gaggelli}}, \bibinfo {author} {\bibfnamefont
  {P.}~\bibnamefont {Guterman}}, \bibinfo {author} {\bibfnamefont
  {D.}~\bibnamefont {Mallarino}}, \bibinfo {author} {\bibfnamefont
  {A.}~\bibnamefont {Mazzoldi}}, \bibinfo {author} {\bibfnamefont
  {A.}~\bibnamefont {Molcard}}, \bibinfo {author} {\bibfnamefont
  {A.}~\bibnamefont {Orfila}}, \bibinfo {author} {\bibfnamefont {P.-M.}\
  \bibnamefont {Poulain}}, \bibinfo {author} {\bibfnamefont {C.}~\bibnamefont
  {Quentin}}, \bibinfo {author} {\bibfnamefont {J.}~\bibnamefont
  {Tintor{\'e}}}, \bibinfo {author} {\bibfnamefont {M.}~\bibnamefont
  {Uttieri}}, \bibinfo {author} {\bibfnamefont {A.}~\bibnamefont {Vetrano}},
  \bibinfo {author} {\bibfnamefont {E.}~\bibnamefont {Zambianchi}}, \ and\
  \bibinfo {author} {\bibfnamefont {V.}~\bibnamefont {Zervakis}},\ }\bibfield
  {title} {\enquote {\bibinfo {title} {Toward an integrated hf radar network in
  the mediterranean sea to improve search and rescue and oil spill response:
  the tosca project experience},}\ }\href@noop {} {\bibfield  {journal}
  {\bibinfo  {journal} {Journal of Operational Oceanography}\ }\textbf
  {\bibinfo {volume} {8}},\ \bibinfo {pages} {95--107} (\bibinfo {year}
  {2015})}\BibitemShut {NoStop}%
\bibitem [{\citenamefont {Gower}\ and\ \citenamefont
  {King}(2008)}]{Gower-King-08}%
  \BibitemOpen
  \bibfield  {author} {\bibinfo {author} {\bibfnamefont {J.}~\bibnamefont
  {Gower}}\ and\ \bibinfo {author} {\bibfnamefont {S.}~\bibnamefont {King}},\
  }\href@noop {} {\enquote {\bibinfo {title} {{Satellite images show the
  Movement of floating \emph{Sargassum} in the Gulf of Mexico and Atlantic
  Ocean}},}\ }\bibinfo {howpublished} {Available from Nature Precedings
  (http://\allowbreak hdl.\allowbreak handle.\allowbreak net/\allowbreak
  10101/\allowbreak npre.\allowbreak 2008.\allowbreak 1894.\allowbreak 1)}
  (\bibinfo {year} {2008})\BibitemShut {NoStop}%
\bibitem [{\citenamefont {Brooks}, \citenamefont {Coles},\ and\ \citenamefont
  {Coles}(2019)}]{Brooks-etal-19}%
  \BibitemOpen
  \bibfield  {author} {\bibinfo {author} {\bibfnamefont {M.~T.}\ \bibnamefont
  {Brooks}}, \bibinfo {author} {\bibfnamefont {V.~J.}\ \bibnamefont {Coles}}, \
  and\ \bibinfo {author} {\bibfnamefont {W.~C.}\ \bibnamefont {Coles}},\
  }\bibfield  {title} {\enquote {\bibinfo {title} {Inertia influences pelagic
  {\emph{sargassum}} advection and distribution},}\ }\href@noop {} {\bibfield
  {journal} {\bibinfo  {journal} {Geophysical Research Letters}\ }\textbf
  {\bibinfo {volume} {46}},\ \bibinfo {pages} {2610--2618} (\bibinfo {year}
  {2019})}\BibitemShut {NoStop}%
\bibitem [{\citenamefont {Law}\ \emph {et~al.}(2010)\citenamefont {Law},
  \citenamefont {Mor\'et-Ferguson}, \citenamefont {Maximenko}, \citenamefont
  {Proskurowski}, \citenamefont {Peacock}, \citenamefont {Hafner},\ and\
  \citenamefont {Reddy}}]{Law-etal-10}%
  \BibitemOpen
  \bibfield  {author} {\bibinfo {author} {\bibfnamefont {K.~L.}\ \bibnamefont
  {Law}}, \bibinfo {author} {\bibfnamefont {S.}~\bibnamefont
  {Mor\'et-Ferguson}}, \bibinfo {author} {\bibfnamefont {N.~A.}\ \bibnamefont
  {Maximenko}}, \bibinfo {author} {\bibfnamefont {G.}~\bibnamefont
  {Proskurowski}}, \bibinfo {author} {\bibfnamefont {E.~E.}\ \bibnamefont
  {Peacock}}, \bibinfo {author} {\bibfnamefont {J.}~\bibnamefont {Hafner}}, \
  and\ \bibinfo {author} {\bibfnamefont {C.~M.}\ \bibnamefont {Reddy}},\
  }\bibfield  {title} {\enquote {\bibinfo {title} {{Plastic accumulation in the
  North Atlantic subtropical gyre}},}\ }\href@noop {} {\bibfield  {journal}
  {\bibinfo  {journal} {Science}\ }\textbf {\bibinfo {volume} {329}},\ \bibinfo
  {pages} {1185--1188} (\bibinfo {year} {2010})}\BibitemShut {NoStop}%
\bibitem [{\citenamefont {Cozar}\ \emph {et~al.}(2014)\citenamefont {Cozar},
  \citenamefont {Echevarria}, \citenamefont {Gonzalez-Gordillo}, \citenamefont
  {Irigoien}, \citenamefont {Ubeda}, \citenamefont {Hernandez-Leon},
  \citenamefont {Palma}, \citenamefont {Navarro}, \citenamefont {Garcia-de
  Lomas}, \citenamefont {andrea}, \citenamefont {Fernandez-de Puelles},\ and\
  \citenamefont {Duarte}}]{Cozar-etal-14}%
  \BibitemOpen
  \bibfield  {author} {\bibinfo {author} {\bibfnamefont {A.}~\bibnamefont
  {Cozar}}, \bibinfo {author} {\bibfnamefont {F.}~\bibnamefont {Echevarria}},
  \bibinfo {author} {\bibfnamefont {J.~I.}\ \bibnamefont {Gonzalez-Gordillo}},
  \bibinfo {author} {\bibfnamefont {X.}~\bibnamefont {Irigoien}}, \bibinfo
  {author} {\bibfnamefont {B.}~\bibnamefont {Ubeda}}, \bibinfo {author}
  {\bibfnamefont {S.}~\bibnamefont {Hernandez-Leon}}, \bibinfo {author}
  {\bibfnamefont {A.~T.}\ \bibnamefont {Palma}}, \bibinfo {author}
  {\bibfnamefont {S.}~\bibnamefont {Navarro}}, \bibinfo {author} {\bibfnamefont
  {J.}~\bibnamefont {Garcia-de Lomas}}, \bibinfo {author} {\bibfnamefont
  {R.}~\bibnamefont {andrea}}, \bibinfo {author} {\bibfnamefont {M.~L.}\
  \bibnamefont {Fernandez-de Puelles}}, \ and\ \bibinfo {author} {\bibfnamefont
  {C.~M.}\ \bibnamefont {Duarte}},\ }\bibfield  {title} {\enquote {\bibinfo
  {title} {Plastic debris in the open ocean},}\ }\href {\doibase
  10.1073/pnas.1314705111} {\bibfield  {journal} {\bibinfo  {journal} {Proc.
  Nat. Acad. Sci. USA}\ }\textbf {\bibinfo {volume} {111}},\ \bibinfo {pages}
  {10239--10244} (\bibinfo {year} {2014})}\BibitemShut {NoStop}%
\bibitem [{\citenamefont {Trinanes}\ \emph {et~al.}(2016)\citenamefont
  {Trinanes}, \citenamefont {Olascoaga}, \citenamefont {Goni}, \citenamefont
  {Maximenko}, \citenamefont {Griffin},\ and\ \citenamefont
  {Hafner}}]{Trinanes-etal-16}%
  \BibitemOpen
  \bibfield  {author} {\bibinfo {author} {\bibfnamefont {J.~A.}\ \bibnamefont
  {Trinanes}}, \bibinfo {author} {\bibfnamefont {M.~J.}\ \bibnamefont
  {Olascoaga}}, \bibinfo {author} {\bibfnamefont {G.~J.}\ \bibnamefont {Goni}},
  \bibinfo {author} {\bibfnamefont {N.~A.}\ \bibnamefont {Maximenko}}, \bibinfo
  {author} {\bibfnamefont {D.~A.}\ \bibnamefont {Griffin}}, \ and\ \bibinfo
  {author} {\bibfnamefont {J.}~\bibnamefont {Hafner}},\ }\bibfield  {title}
  {\enquote {\bibinfo {title} {{Analysis of flight MH370 potential debris
  trajectories using ocean observations and numerical model results}},}\
  }\href@noop {} {\bibfield  {journal} {\bibinfo  {journal} {Journal of
  Operational Oceanography}\ }\textbf {\bibinfo {volume} {9}},\ \bibinfo
  {pages} {126--138} (\bibinfo {year} {2016})}\BibitemShut {NoStop}%
\bibitem [{\citenamefont {Miron}\ \emph
  {et~al.}(2019{\natexlab{a}})\citenamefont {Miron}, \citenamefont
  {Beron-Vera}, \citenamefont {Olascoaga},\ and\ \citenamefont
  {Koltai}}]{Miron-etal-19b}%
  \BibitemOpen
  \bibfield  {author} {\bibinfo {author} {\bibfnamefont {P.}~\bibnamefont
  {Miron}}, \bibinfo {author} {\bibfnamefont {F.~J.}\ \bibnamefont
  {Beron-Vera}}, \bibinfo {author} {\bibfnamefont {M.~J.}\ \bibnamefont
  {Olascoaga}}, \ and\ \bibinfo {author} {\bibfnamefont {P.}~\bibnamefont
  {Koltai}},\ }\bibfield  {title} {\enquote {\bibinfo {title}
  {{Markov-chain-inspired search for MH370}},}\ }\href@noop {} {\bibfield
  {journal} {\bibinfo  {journal} {Chaos: An Interdisciplinary Journal of
  Nonlinear Science}\ }\textbf {\bibinfo {volume} {29}},\ \bibinfo {pages}
  {041105} (\bibinfo {year} {2019}{\natexlab{a}})}\BibitemShut {NoStop}%
\bibitem [{\citenamefont {Rypina}\ \emph {et~al.}(2013)\citenamefont {Rypina},
  \citenamefont {Jayne}, \citenamefont {Yoshida}, \citenamefont {Macdonald},
  \citenamefont {Douglas},\ and\ \citenamefont {Buesseler}}]{Rypina-etal-13a}%
  \BibitemOpen
  \bibfield  {author} {\bibinfo {author} {\bibfnamefont {I.}~\bibnamefont
  {Rypina}}, \bibinfo {author} {\bibfnamefont {S.~R.}\ \bibnamefont {Jayne}},
  \bibinfo {author} {\bibfnamefont {S.}~\bibnamefont {Yoshida}}, \bibinfo
  {author} {\bibfnamefont {A.~M.}\ \bibnamefont {Macdonald}}, \bibinfo {author}
  {\bibfnamefont {E.}~\bibnamefont {Douglas}}, \ and\ \bibinfo {author}
  {\bibfnamefont {K.}~\bibnamefont {Buesseler}},\ }\bibfield  {title} {\enquote
  {\bibinfo {title} {{Short-term dispersal of Fukushima-derived radionuclides
  off Japan: modeling efforts and model-data intercomparison}},}\ }\href@noop
  {} {\bibfield  {journal} {\bibinfo  {journal} {Biogeosciences}\ }\textbf
  {\bibinfo {volume} {10}},\ \bibinfo {pages} {4973--4990} (\bibinfo {year}
  {2013})}\BibitemShut {NoStop}%
\bibitem [{\citenamefont {Matthews}\ \emph {et~al.}(2017)\citenamefont
  {Matthews}, \citenamefont {Ostrovsky}, \citenamefont {Yoshikawa},
  \citenamefont {Komori},\ and\ \citenamefont {Tamura}}]{Matthews-etal-17}%
  \BibitemOpen
  \bibfield  {author} {\bibinfo {author} {\bibfnamefont {J.~P.}\ \bibnamefont
  {Matthews}}, \bibinfo {author} {\bibfnamefont {L.}~\bibnamefont {Ostrovsky}},
  \bibinfo {author} {\bibfnamefont {Y.}~\bibnamefont {Yoshikawa}}, \bibinfo
  {author} {\bibfnamefont {S.}~\bibnamefont {Komori}}, \ and\ \bibinfo {author}
  {\bibfnamefont {H.}~\bibnamefont {Tamura}},\ }\bibfield  {title} {\enquote
  {\bibinfo {title} {{Dynamics and early post-tsunami evolution of floating
  marine debris near Fukushima Daiichi}},}\ }\href@noop {} {\bibfield
  {journal} {\bibinfo  {journal} {Nature Geosci.}\ }\textbf {\bibinfo {volume}
  {10}},\ \bibinfo {pages} {598--603} (\bibinfo {year} {2017})}\BibitemShut
  {NoStop}%
\bibitem [{\citenamefont {Szanyi}, \citenamefont {Lukovich},\ and\
  \citenamefont {Barber}(2016)}]{Szanyi-etal-16}%
  \BibitemOpen
  \bibfield  {author} {\bibinfo {author} {\bibfnamefont {S.}~\bibnamefont
  {Szanyi}}, \bibinfo {author} {\bibfnamefont {J.~V.}\ \bibnamefont
  {Lukovich}}, \ and\ \bibinfo {author} {\bibfnamefont {D.~G.}\ \bibnamefont
  {Barber}},\ }\bibfield  {title} {\enquote {\bibinfo {title} {Lagrangian
  analysis of sea-ice dynamics in the arctic ocean},}\ }\href@noop {}
  {\bibfield  {journal} {\bibinfo  {journal} {Polar Research}\ }\textbf
  {\bibinfo {volume} {35}},\ \bibinfo {pages} {30778} (\bibinfo {year}
  {2016})}\BibitemShut {NoStop}%
\bibitem [{\citenamefont {Nielsen}(1994)}]{Nielsen-94}%
  \BibitemOpen
  \bibfield  {author} {\bibinfo {author} {\bibfnamefont {P.}~\bibnamefont
  {Nielsen}},\ }\bibfield  {title} {\enquote {\bibinfo {title} {Suspended
  sediment particle motion in coastal flows},}\ }\href@noop {} {\bibfield
  {journal} {\bibinfo  {journal} {Coastal Engineering Proceedings}\ }\textbf
  {\bibinfo {volume} {1}},\ \bibinfo {pages} {2406--2416} (\bibinfo {year}
  {1994})}\BibitemShut {NoStop}%
\bibitem [{\citenamefont {Reigada}\ \emph {et~al.}(2003)\citenamefont
  {Reigada}, \citenamefont {Hillary}, \citenamefont {Bees}, \citenamefont
  {Sancho},\ and\ \citenamefont {Sagues}}]{Reigada-etal-03}%
  \BibitemOpen
  \bibfield  {author} {\bibinfo {author} {\bibfnamefont {R.}~\bibnamefont
  {Reigada}}, \bibinfo {author} {\bibfnamefont {R.~M.}\ \bibnamefont
  {Hillary}}, \bibinfo {author} {\bibfnamefont {M.~A.}\ \bibnamefont {Bees}},
  \bibinfo {author} {\bibfnamefont {J.~M.}\ \bibnamefont {Sancho}}, \ and\
  \bibinfo {author} {\bibfnamefont {F.}~\bibnamefont {Sagues}},\ }\bibfield
  {title} {\enquote {\bibinfo {title} {Plankton blooms induced by turbulent
  flows},}\ }\href@noop {} {\bibfield  {journal} {\bibinfo  {journal} {Proc. R.
  Soc. B: Biological Sciences}\ }\textbf {\bibinfo {volume} {270}},\ \bibinfo
  {pages} {875--880} (\bibinfo {year} {2003})}\BibitemShut {NoStop}%
\bibitem [{\citenamefont {Peng}\ and\ \citenamefont
  {Dabiri}(2009)}]{Peng-Dabiri-09}%
  \BibitemOpen
  \bibfield  {author} {\bibinfo {author} {\bibfnamefont {J.}~\bibnamefont
  {Peng}}\ and\ \bibinfo {author} {\bibfnamefont {J.~O.}\ \bibnamefont
  {Dabiri}},\ }\bibfield  {title} {\enquote {\bibinfo {title} {{Transport of
  inertial particles by Lagrangian Coherent Structures: application to
  predator-prey interaction in jellyfish feeding}},}\ }\href@noop {} {\bibfield
   {journal} {\bibinfo  {journal} {J. Fluid Mech.}\ }\textbf {\bibinfo {volume}
  {623}},\ \bibinfo {pages} {75--84} (\bibinfo {year} {2009})}\BibitemShut
  {NoStop}%
\bibitem [{\citenamefont {Monroy}\ \emph {et~al.}(2016)\citenamefont {Monroy},
  \citenamefont {Hern\'andez-Garc\'{\i}a}, \citenamefont {Rossi},\ and\
  \citenamefont {L\'opez}}]{Monroy-etal-16}%
  \BibitemOpen
  \bibfield  {author} {\bibinfo {author} {\bibfnamefont {P.}~\bibnamefont
  {Monroy}}, \bibinfo {author} {\bibfnamefont {E.}~\bibnamefont
  {Hern\'andez-Garc\'{\i}a}}, \bibinfo {author} {\bibfnamefont
  {V.}~\bibnamefont {Rossi}}, \ and\ \bibinfo {author} {\bibfnamefont
  {C.}~\bibnamefont {L\'opez}},\ }\bibfield  {title} {\enquote {\bibinfo
  {title} {Modeling the dynamical sinking of biogenic particles in oceanic
  flow},}\ }\href@noop {} {\  (\bibinfo {year} {2016})}\BibitemShut {NoStop}%
\bibitem [{\citenamefont {Beron-Vera}\ \emph {et~al.}(2015)\citenamefont
  {Beron-Vera}, \citenamefont {Olascoaga}, \citenamefont {Haller},
  \citenamefont {Farazmand}, \citenamefont {{Tri\~nanes}},\ and\ \citenamefont
  {Wang}}]{Beron-etal-15}%
  \BibitemOpen
  \bibfield  {author} {\bibinfo {author} {\bibfnamefont {F.~J.}\ \bibnamefont
  {Beron-Vera}}, \bibinfo {author} {\bibfnamefont {M.~J.}\ \bibnamefont
  {Olascoaga}}, \bibinfo {author} {\bibfnamefont {G.}~\bibnamefont {Haller}},
  \bibinfo {author} {\bibfnamefont {M.}~\bibnamefont {Farazmand}}, \bibinfo
  {author} {\bibfnamefont {J.}~\bibnamefont {{Tri\~nanes}}}, \ and\ \bibinfo
  {author} {\bibfnamefont {Y.}~\bibnamefont {Wang}},\ }\bibfield  {title}
  {\enquote {\bibinfo {title} {{Dissipative inertial transport patterns near
  coherent Lagrangian eddies in the ocean}},}\ }\href@noop {} {\bibfield
  {journal} {\bibinfo  {journal} {Chaos}\ }\textbf {\bibinfo {volume} {25}},\
  \bibinfo {pages} {087412} (\bibinfo {year} {2015})}\BibitemShut {NoStop}%
\bibitem [{\citenamefont {Haller}\ and\ \citenamefont
  {Beron-Vera}(2013)}]{Haller-Beron-13}%
  \BibitemOpen
  \bibfield  {author} {\bibinfo {author} {\bibfnamefont {G.}~\bibnamefont
  {Haller}}\ and\ \bibinfo {author} {\bibfnamefont {F.~J.}\ \bibnamefont
  {Beron-Vera}},\ }\bibfield  {title} {\enquote {\bibinfo {title} {{Coherent
  Lagrangian vortices: The black holes of turbulence}},}\ }\href {\doibase
  10.1017/jfm.2013.391} {\bibfield  {journal} {\bibinfo  {journal} {J. Fluid
  Mech.}\ }\textbf {\bibinfo {volume} {731}},\ \bibinfo {pages} {R4} (\bibinfo
  {year} {2013})}\BibitemShut {NoStop}%
\bibitem [{\citenamefont {Haller}\ and\ \citenamefont
  {Beron-Vera}(2014)}]{Haller-Beron-14}%
  \BibitemOpen
  \bibfield  {author} {\bibinfo {author} {\bibfnamefont {G.}~\bibnamefont
  {Haller}}\ and\ \bibinfo {author} {\bibfnamefont {F.~J.}\ \bibnamefont
  {Beron-Vera}},\ }\bibfield  {title} {\enquote {\bibinfo {title} {{Addendum to
  `Coherent Lagrangian vortices: The black holes of turbulence'}},}\
  }\href@noop {} {\bibfield  {journal} {\bibinfo  {journal} {J. Fluid Mech.}\
  }\textbf {\bibinfo {volume} {755}},\ \bibinfo {pages} {R3} (\bibinfo {year}
  {2014})}\BibitemShut {NoStop}%
\bibitem [{\citenamefont {Haller}\ \emph {et~al.}(2016)\citenamefont {Haller},
  \citenamefont {Hadjighasem}, \citenamefont {Farazmand},\ and\ \citenamefont
  {Huhn}}]{Haller-etal-16}%
  \BibitemOpen
  \bibfield  {author} {\bibinfo {author} {\bibfnamefont {G.}~\bibnamefont
  {Haller}}, \bibinfo {author} {\bibfnamefont {A.}~\bibnamefont {Hadjighasem}},
  \bibinfo {author} {\bibfnamefont {M.}~\bibnamefont {Farazmand}}, \ and\
  \bibinfo {author} {\bibfnamefont {F.}~\bibnamefont {Huhn}},\ }\bibfield
  {title} {\enquote {\bibinfo {title} {Defining coherent vortices objectively
  from the vorticity},}\ }\href@noop {} {\bibfield  {journal} {\bibinfo
  {journal} {J. Fluid Mech.}\ }\textbf {\bibinfo {volume} {795}},\ \bibinfo
  {pages} {136--173} (\bibinfo {year} {2016})}\BibitemShut {NoStop}%
\bibitem [{\citenamefont {Beron-Vera}, \citenamefont {Olascoaga},\ and\
  \citenamefont {Lumpkin}(2016)}]{Beron-etal-16}%
  \BibitemOpen
  \bibfield  {author} {\bibinfo {author} {\bibfnamefont {F.~J.}\ \bibnamefont
  {Beron-Vera}}, \bibinfo {author} {\bibfnamefont {M.~J.}\ \bibnamefont
  {Olascoaga}}, \ and\ \bibinfo {author} {\bibfnamefont {R.}~\bibnamefont
  {Lumpkin}},\ }\bibfield  {title} {\enquote {\bibinfo {title} {Inertia-induced
  accumulation of flotsam in the subtropical gyres},}\ }\href@noop {}
  {\bibfield  {journal} {\bibinfo  {journal} {Geophys. Res. Lett.}\ }\textbf
  {\bibinfo {volume} {43}},\ \bibinfo {pages} {12228--12233} (\bibinfo {year}
  {2016})}\BibitemShut {NoStop}%
\bibitem [{\citenamefont {Maximenko}\ and\ \citenamefont
  {Niiler}(2006)}]{Maximenko-Niiler-06}%
  \BibitemOpen
  \bibfield  {author} {\bibinfo {author} {\bibfnamefont {N.~A.}\ \bibnamefont
  {Maximenko}}\ and\ \bibinfo {author} {\bibfnamefont {P.~P.}\ \bibnamefont
  {Niiler}},\ }\bibfield  {title} {\enquote {\bibinfo {title} {Mean surface
  circulation of the global ocean inferred from satellite altimeter and drifter
  data},}\ }in\ \href@noop {} {\emph {\bibinfo {booktitle} {15 years of
  Progress in Radar Altimetry}}}\ (\bibinfo  {publisher} {ESA Publication
  SP-614},\ \bibinfo {year} {2006})\BibitemShut {NoStop}%
\bibitem [{\citenamefont {Brach}\ \emph {et~al.}(2018)\citenamefont {Brach},
  \citenamefont {Deixonne}, \citenamefont {Bernard}, \citenamefont {Durand},
  \citenamefont {Desjean}, \citenamefont {Perez}, \citenamefont {van Sebille},\
  and\ \citenamefont {ter Halle}}]{Brach-etal-18}%
  \BibitemOpen
  \bibfield  {author} {\bibinfo {author} {\bibfnamefont {L.}~\bibnamefont
  {Brach}}, \bibinfo {author} {\bibfnamefont {P.}~\bibnamefont {Deixonne}},
  \bibinfo {author} {\bibfnamefont {M.-F.}\ \bibnamefont {Bernard}}, \bibinfo
  {author} {\bibfnamefont {E.}~\bibnamefont {Durand}}, \bibinfo {author}
  {\bibfnamefont {M.-C.}\ \bibnamefont {Desjean}}, \bibinfo {author}
  {\bibfnamefont {E.}~\bibnamefont {Perez}}, \bibinfo {author} {\bibfnamefont
  {E.}~\bibnamefont {van Sebille}}, \ and\ \bibinfo {author} {\bibfnamefont
  {A.}~\bibnamefont {ter Halle}},\ }\bibfield  {title} {\enquote {\bibinfo
  {title} {Anticyclonic eddies increase accumulation of microplastic in the
  north atlantic subtropical gyre},}\ }\href {\doibase
  https://doi.org/10.1016/j.marpolbul.2017.10.077} {\bibfield  {journal}
  {\bibinfo  {journal} {Marine Pollution Bulletin}\ }\textbf {\bibinfo {volume}
  {126}},\ \bibinfo {pages} {191--196} (\bibinfo {year} {2018})}\BibitemShut
  {NoStop}%
\bibitem [{\citenamefont {Chelton}\ \emph {et~al.}(2011)\citenamefont
  {Chelton}, \citenamefont {Gaube}, \citenamefont {Schlax}, \citenamefont
  {Early},\ and\ \citenamefont {Samelson}}]{Chelton-etal-11b}%
  \BibitemOpen
  \bibfield  {author} {\bibinfo {author} {\bibfnamefont {D.~B.}\ \bibnamefont
  {Chelton}}, \bibinfo {author} {\bibfnamefont {P.}~\bibnamefont {Gaube}},
  \bibinfo {author} {\bibfnamefont {M.~G.}\ \bibnamefont {Schlax}}, \bibinfo
  {author} {\bibfnamefont {J.~J.}\ \bibnamefont {Early}}, \ and\ \bibinfo
  {author} {\bibfnamefont {R.~M.}\ \bibnamefont {Samelson}},\ }\bibfield
  {title} {\enquote {\bibinfo {title} {The influence of nonlinear mesoscale
  eddies on near-surface oceanic chlorophyll},}\ }\href@noop {} {\bibfield
  {journal} {\bibinfo  {journal} {Science}\ }\textbf {\bibinfo {volume}
  {334}},\ \bibinfo {pages} {328--332} (\bibinfo {year} {2011})}\BibitemShut
  {NoStop}%
\bibitem [{\citenamefont {Olascoaga}\ \emph {et~al.}(2019)\citenamefont
  {Olascoaga}, \citenamefont {Beron-Vera}, \citenamefont {P.~Miron},
  \citenamefont {Trinanes}, \citenamefont {Lumpkin},\ and\ \citenamefont
  {Goni}}]{Olascoaga-etal-19}%
  \BibitemOpen
  \bibfield  {author} {\bibinfo {author} {\bibfnamefont {M.~J.}\ \bibnamefont
  {Olascoaga}}, \bibinfo {author} {\bibfnamefont {F.~J.}\ \bibnamefont
  {Beron-Vera}}, \bibinfo {author} {\bibfnamefont {U.~R.}\ \bibnamefont
  {P.~Miron}}, \bibinfo {author} {\bibfnamefont {J.}~\bibnamefont {Trinanes}},
  \bibinfo {author} {\bibfnamefont {R.}~\bibnamefont {Lumpkin}}, \ and\
  \bibinfo {author} {\bibfnamefont {G.~J.}\ \bibnamefont {Goni}},\ }\href@noop
  {} {\enquote {\bibinfo {title} {Observation and quantification of inertial
  effects on the drift of floating objects at the ocean surface},}\ }\bibinfo
  {howpublished} {In prepaparation.} (\bibinfo {year} {2019})\BibitemShut
  {NoStop}%
\bibitem [{Note1()}]{Note1}%
  \BibitemOpen
  \bibinfo {note} {While having a positive discriminant, the cubic polynomial
  in \protect \textup {\hbox {\mathsurround \z@ \protect \normalfont
  (\ignorespaces \ref {eq:cubic}\unskip \@@italiccorr )}} is irreducible over
  the reals. Thus while its three roots are real, they require complex numbers
  to be expressed in radicals \cite {Wantzel-43}.}\BibitemShut {Stop}%
\bibitem [{\citenamefont {Auton}(1987)}]{Auton-87}%
  \BibitemOpen
  \bibfield  {author} {\bibinfo {author} {\bibfnamefont {T.~R.}\ \bibnamefont
  {Auton}},\ }\bibfield  {title} {\enquote {\bibinfo {title} {The lift force on
  a spherical body in a rotational flow},}\ }\href@noop {} {\bibfield
  {journal} {\bibinfo  {journal} {Journal of Fluid Mechanics}\ }\textbf
  {\bibinfo {volume} {183}},\ \bibinfo {pages} {199--218} (\bibinfo {year}
  {1987})}\BibitemShut {NoStop}%
\bibitem [{\citenamefont {Beron-Vera}\ \emph {et~al.}(2018)\citenamefont
  {Beron-Vera}, \citenamefont {Hadjighasem}, \citenamefont {Xia}, \citenamefont
  {Olascoaga},\ and\ \citenamefont {Haller}}]{Beron-etal-18a}%
  \BibitemOpen
  \bibfield  {author} {\bibinfo {author} {\bibfnamefont {F.~J.}\ \bibnamefont
  {Beron-Vera}}, \bibinfo {author} {\bibfnamefont {A.}~\bibnamefont
  {Hadjighasem}}, \bibinfo {author} {\bibfnamefont {Q.}~\bibnamefont {Xia}},
  \bibinfo {author} {\bibfnamefont {M.~J.}\ \bibnamefont {Olascoaga}}, \ and\
  \bibinfo {author} {\bibfnamefont {G.}~\bibnamefont {Haller}},\ }\bibfield
  {title} {\enquote {\bibinfo {title} {{Coherent Lagrangian swirls among
  submesoscale motions}},}\ }\href {\doibase 10.1073/pnas.1701392115}
  {\bibfield  {journal} {\bibinfo  {journal} {Proc. Natl. Acad. Sci. U.S.A.}\
  }\textbf {\bibinfo {volume} {Mar 2018}},\ \bibinfo {pages} {201701392}
  (\bibinfo {year} {2018})}\BibitemShut {NoStop}%
\bibitem [{\citenamefont {Montabone}(2002)}]{Montabone-02}%
  \BibitemOpen
  \bibfield  {author} {\bibinfo {author} {\bibfnamefont {L.}~\bibnamefont
  {Montabone}},\ }\emph {\bibinfo {title} {Vortex Dynamics and Particle
  Transport in Barotropic Turbulence}},\ \href@noop {} {Ph.D. thesis},\
  \bibinfo  {school} {University of Genoa}, \bibinfo {address} {Italy}
  (\bibinfo {year} {2002})\BibitemShut {NoStop}%
\bibitem [{\citenamefont {Henderson}, \citenamefont {Gwynllyw},\ and\
  \citenamefont {Barenghi}(2007)}]{Henderson-etal-07}%
  \BibitemOpen
  \bibfield  {author} {\bibinfo {author} {\bibfnamefont {K.~L.}\ \bibnamefont
  {Henderson}}, \bibinfo {author} {\bibfnamefont {D.~R.}\ \bibnamefont
  {Gwynllyw}}, \ and\ \bibinfo {author} {\bibfnamefont {C.~F.}\ \bibnamefont
  {Barenghi}},\ }\bibfield  {title} {\enquote {\bibinfo {title} {{Particle
  tracking in Taylor--Couette flow}},}\ }\href@noop {} {\bibfield  {journal}
  {\bibinfo  {journal} {European Journal of Mechanics - B/Fluids}\ }\textbf
  {\bibinfo {volume} {26}},\ \bibinfo {pages} {738 -- 748} (\bibinfo {year}
  {2007})}\BibitemShut {NoStop}%
\bibitem [{\citenamefont {Sapsis}\ \emph {et~al.}(2011)\citenamefont {Sapsis},
  \citenamefont {Ouellette}, \citenamefont {Gollub},\ and\ \citenamefont
  {Haller}}]{Sapsis-etal-11}%
  \BibitemOpen
  \bibfield  {author} {\bibinfo {author} {\bibfnamefont {T.~P.}\ \bibnamefont
  {Sapsis}}, \bibinfo {author} {\bibfnamefont {N.~T.}\ \bibnamefont
  {Ouellette}}, \bibinfo {author} {\bibfnamefont {J.~P.}\ \bibnamefont
  {Gollub}}, \ and\ \bibinfo {author} {\bibfnamefont {G.}~\bibnamefont
  {Haller}},\ }\bibfield  {title} {\enquote {\bibinfo {title} {Neutrally
  buoyant particle dynamics in fluid flows: Comparison of experiments with
  lagrangian stochastic models},}\ }\href@noop {} {\bibfield  {journal}
  {\bibinfo  {journal} {Physics of Fluids}\ }\textbf {\bibinfo {volume} {23}},\
  \bibinfo {pages} {093304} (\bibinfo {year} {2011})}\BibitemShut {NoStop}%
\bibitem [{\citenamefont {Farazmand}\ and\ \citenamefont
  {Haller}(2015)}]{Farazmand-Haller-15}%
  \BibitemOpen
  \bibfield  {author} {\bibinfo {author} {\bibfnamefont {M.}~\bibnamefont
  {Farazmand}}\ and\ \bibinfo {author} {\bibfnamefont {G.}~\bibnamefont
  {Haller}},\ }\bibfield  {title} {\enquote {\bibinfo {title} {{The
  Maxey--Riley equation: Existence, uniqueness and regularity of solutions}},}\
  }\href@noop {} {\bibfield  {journal} {\bibinfo  {journal} {Nonlinear
  Analysis: Real World Applications}\ }\textbf {\bibinfo {volume} {22}},\
  \bibinfo {pages} {98--106} (\bibinfo {year} {2015})}\BibitemShut {NoStop}%
\bibitem [{\citenamefont {Langlois}, \citenamefont {Farazmand},\ and\
  \citenamefont {Haller}(2015)}]{Langlois-etal-15}%
  \BibitemOpen
  \bibfield  {author} {\bibinfo {author} {\bibfnamefont {G.~P.}\ \bibnamefont
  {Langlois}}, \bibinfo {author} {\bibfnamefont {M.}~\bibnamefont {Farazmand}},
  \ and\ \bibinfo {author} {\bibfnamefont {G.}~\bibnamefont {Haller}},\
  }\bibfield  {title} {\enquote {\bibinfo {title} {Asymptotic dynamics of
  inertial particles with memory},}\ }\href {\doibase
  10.1007/s00332-015-9250-0} {\bibfield  {journal} {\bibinfo  {journal}
  {Journal of Nonlinear Science}\ }\textbf {\bibinfo {volume} {25}},\ \bibinfo
  {pages} {1225--1255} (\bibinfo {year} {2015})}\BibitemShut {NoStop}%
\bibitem [{\citenamefont {Daitche}\ and\ \citenamefont
  {T\'el}(2011)}]{Daitche-Tel-11}%
  \BibitemOpen
  \bibfield  {author} {\bibinfo {author} {\bibfnamefont {A.}~\bibnamefont
  {Daitche}}\ and\ \bibinfo {author} {\bibfnamefont {T.}~\bibnamefont
  {T\'el}},\ }\bibfield  {title} {\enquote {\bibinfo {title} {Memory effects
  are relevant for chaotic advection of inertial particles},}\ }\href@noop {}
  {\bibfield  {journal} {\bibinfo  {journal} {Phys. Rev. Lett.}\ }\textbf
  {\bibinfo {volume} {107}},\ \bibinfo {pages} {244501} (\bibinfo {year}
  {2011})}\BibitemShut {NoStop}%
\bibitem [{\citenamefont {Daitche}\ and\ \citenamefont
  {T\'el}(2014)}]{Daitche-Tel-14}%
  \BibitemOpen
  \bibfield  {author} {\bibinfo {author} {\bibfnamefont {A.}~\bibnamefont
  {Daitche}}\ and\ \bibinfo {author} {\bibfnamefont {T.}~\bibnamefont
  {T\'el}},\ }\bibfield  {title} {\enquote {\bibinfo {title} {Memory effects in
  chaotic advection of inertial particles},}\ }\href@noop {} {\bibfield
  {journal} {\bibinfo  {journal} {New Journal of Physics}\ }\textbf {\bibinfo
  {volume} {16}},\ \bibinfo {pages} {073008} (\bibinfo {year}
  {2014})}\BibitemShut {NoStop}%
\bibitem [{\citenamefont {Sudharsan}, \citenamefont {Brunton},\ and\
  \citenamefont {Riley}(2016)}]{Sudharsan-etal-16}%
  \BibitemOpen
  \bibfield  {author} {\bibinfo {author} {\bibfnamefont {M.}~\bibnamefont
  {Sudharsan}}, \bibinfo {author} {\bibfnamefont {S.~L.}\ \bibnamefont
  {Brunton}}, \ and\ \bibinfo {author} {\bibfnamefont {J.~J.}\ \bibnamefont
  {Riley}},\ }\bibfield  {title} {\enquote {\bibinfo {title} {Lagrangian
  coherent structures and inertial particle dynamics},}\ }\href@noop {}
  {\bibfield  {journal} {\bibinfo  {journal} {Phys. Rev. E}\ }\textbf {\bibinfo
  {volume} {93}},\ \bibinfo {pages} {033108} (\bibinfo {year}
  {2016})}\BibitemShut {NoStop}%
\bibitem [{Note2()}]{Note2}%
  \BibitemOpen
  \bibinfo {note} {In an earlier geophysical adaptation of the Maxey--Riley
  equation \cite {Provenzale-99}, the centrifugal force was included as well,
  but this is actually balanced out by the gravitational force on the
  horizontal plane.}\BibitemShut {Stop}%
\bibitem [{\citenamefont {Pedlosky}(1987)}]{Pedlosky-87}%
  \BibitemOpen
  \bibfield  {author} {\bibinfo {author} {\bibfnamefont {J.}~\bibnamefont
  {Pedlosky}},\ }\href@noop {} {\emph {\bibinfo {title} {Geophysical {F}luid
  {D}ynamics}}},\ \bibinfo {edition} {2nd}\ ed.\ (\bibinfo  {publisher}
  {Springer},\ \bibinfo {year} {1987})\ p.\ \bibinfo {pages} {624
  pp.}\BibitemShut {Stop}%
\bibitem [{\citenamefont {Ripa}(1997{\natexlab{a}})}]{Ripa-JPO-97b}%
  \BibitemOpen
  \bibfield  {author} {\bibinfo {author} {\bibfnamefont {P.}~\bibnamefont
  {Ripa}},\ }\bibfield  {title} {\enquote {\bibinfo {title} {``{I}nertial''
  oscillations and the $\beta $-plane approximation({s})},}\ }\href@noop {}
  {\bibfield  {journal} {\bibinfo  {journal} {J. Phys. Oceanogr.}\ }\textbf
  {\bibinfo {volume} {27}},\ \bibinfo {pages} {633--647} (\bibinfo {year}
  {1997}{\natexlab{a}})}\BibitemShut {NoStop}%
\bibitem [{\citenamefont {R\"{o}hrs}\ \emph {et~al.}(2012)\citenamefont
  {R\"{o}hrs}, \citenamefont {Christensen}, \citenamefont {Hole}, \citenamefont
  {Brostr\"{o}m}, \citenamefont {Drivdal},\ and\ \citenamefont
  {Sundby}}]{Rohrs-etal-12}%
  \BibitemOpen
  \bibfield  {author} {\bibinfo {author} {\bibfnamefont {J.}~\bibnamefont
  {R\"{o}hrs}}, \bibinfo {author} {\bibfnamefont {K.~H.}\ \bibnamefont
  {Christensen}}, \bibinfo {author} {\bibfnamefont {L.~R.}\ \bibnamefont
  {Hole}}, \bibinfo {author} {\bibfnamefont {G.}~\bibnamefont {Brostr\"{o}m}},
  \bibinfo {author} {\bibfnamefont {M.}~\bibnamefont {Drivdal}}, \ and\
  \bibinfo {author} {\bibfnamefont {S.}~\bibnamefont {Sundby}},\ }\bibfield
  {title} {\enquote {\bibinfo {title} {Observation-based evaluation of surface
  wave effects on currents and trajectory forecasts},}\ }\href@noop {}
  {\bibfield  {journal} {\bibinfo  {journal} {Ocean Dyn.}\ }\textbf {\bibinfo
  {volume} {62}},\ \bibinfo {pages} {1519--1533} (\bibinfo {year}
  {2012})}\BibitemShut {NoStop}%
\bibitem [{\citenamefont {Nesterov}(2018)}]{Nesterov-18}%
  \BibitemOpen
  \bibfield  {author} {\bibinfo {author} {\bibfnamefont {O.}~\bibnamefont
  {Nesterov}},\ }\bibfield  {title} {\enquote {\bibinfo {title} {{Consideration
  of various aspects in a drift study of MH370 debris}},}\ }\href@noop {}
  {\bibfield  {journal} {\bibinfo  {journal} {Ocean Sci.}\ }\textbf {\bibinfo
  {volume} {14}},\ \bibinfo {pages} {387--402} (\bibinfo {year}
  {2018})}\BibitemShut {NoStop}%
\bibitem [{\citenamefont {Sozza}\ \emph {et~al.}(2016)\citenamefont {Sozza},
  \citenamefont {{De Lillo}}, \citenamefont {Musacchio},\ and\ \citenamefont
  {Boffetta}}]{Sozza-etal-16}%
  \BibitemOpen
  \bibfield  {author} {\bibinfo {author} {\bibfnamefont {A.}~\bibnamefont
  {Sozza}}, \bibinfo {author} {\bibfnamefont {F.}~\bibnamefont {{De Lillo}}},
  \bibinfo {author} {\bibfnamefont {S.}~\bibnamefont {Musacchio}}, \ and\
  \bibinfo {author} {\bibfnamefont {G.}~\bibnamefont {Boffetta}},\ }\bibfield
  {title} {\enquote {\bibinfo {title} {Large-scale confinement and small-scale
  clustering of floating particles in stratified turbulence},}\ }\href@noop {}
  {\bibfield  {journal} {\bibinfo  {journal} {Physical Review Fluids}\ }\textbf
  {\bibinfo {volume} {1}},\ \bibinfo {pages} {052401(R)} (\bibinfo {year}
  {2016})}\BibitemShut {NoStop}%
\bibitem [{\citenamefont {Breivik}\ and\ \citenamefont
  {Allen}(2008)}]{Breivik-Allen-08}%
  \BibitemOpen
  \bibfield  {author} {\bibinfo {author} {\bibfnamefont {{\O}.}~\bibnamefont
  {Breivik}}\ and\ \bibinfo {author} {\bibfnamefont {A.}~\bibnamefont
  {Allen}},\ }\bibfield  {title} {\enquote {\bibinfo {title} {{An operational
  search and rescue model for the Norwegian Sea and the North Sea}},}\
  }\href@noop {} {\bibfield  {journal} {\bibinfo  {journal} {J. Marine Syst.}\
  }\textbf {\bibinfo {volume} {69}},\ \bibinfo {pages} {99--113} (\bibinfo
  {year} {2008})}\BibitemShut {NoStop}%
\bibitem [{\citenamefont {Duhec}\ \emph {et~al.}(2015)\citenamefont {Duhec},
  \citenamefont {Jeanne}, \citenamefont {Maximenko},\ and\ \citenamefont
  {Hafner}}]{Duhec-etal-15}%
  \BibitemOpen
  \bibfield  {author} {\bibinfo {author} {\bibfnamefont {A.~V.}\ \bibnamefont
  {Duhec}}, \bibinfo {author} {\bibfnamefont {R.~F.}\ \bibnamefont {Jeanne}},
  \bibinfo {author} {\bibfnamefont {N.}~\bibnamefont {Maximenko}}, \ and\
  \bibinfo {author} {\bibfnamefont {J.}~\bibnamefont {Hafner}},\ }\bibfield
  {title} {\enquote {\bibinfo {title} {{Composition and potential origin of
  marine debris stranded in the Western Indian Ocean on remote Alphonse Island,
  Seychelles}},}\ }\href {\doibase 10.1016/j.marpolbul.2015.05.042} {\bibfield
  {journal} {\bibinfo  {journal} {Mar. Poll. Bull.}\ }\textbf {\bibinfo
  {volume} {96}},\ \bibinfo {pages} {76--86} (\bibinfo {year}
  {2015})}\BibitemShut {NoStop}%
\bibitem [{\citenamefont {Allshouse}\ \emph {et~al.}(2017)\citenamefont
  {Allshouse}, \citenamefont {Ivey}, \citenamefont {Lowe}, \citenamefont
  {Jones}, \citenamefont {Beegle-krause}, \citenamefont {Xu},\ and\
  \citenamefont {Peacock}}]{Allshouse-etal-17}%
  \BibitemOpen
  \bibfield  {author} {\bibinfo {author} {\bibfnamefont {M.~R.}\ \bibnamefont
  {Allshouse}}, \bibinfo {author} {\bibfnamefont {G.~N.}\ \bibnamefont {Ivey}},
  \bibinfo {author} {\bibfnamefont {R.~J.}\ \bibnamefont {Lowe}}, \bibinfo
  {author} {\bibfnamefont {N.~L.}\ \bibnamefont {Jones}}, \bibinfo {author}
  {\bibfnamefont {C.}~\bibnamefont {Beegle-krause}}, \bibinfo {author}
  {\bibfnamefont {J.}~\bibnamefont {Xu}}, \ and\ \bibinfo {author}
  {\bibfnamefont {T.}~\bibnamefont {Peacock}},\ }\bibfield  {title}
  {{\selectlanguage {English}\enquote {\bibinfo {title} {Impact of windage on
  ocean surface lagrangian coherent structures},}\ }}\href@noop {} {\bibfield
  {journal} {\bibinfo  {journal} {Environmental Fluid Mechanics}\ }\textbf
  {\bibinfo {volume} {17}},\ \bibinfo {pages} {473--483} (\bibinfo {year}
  {2017})}\BibitemShut {NoStop}%
\bibitem [{\citenamefont {Geyer}(1989)}]{Geyer-89}%
  \BibitemOpen
  \bibfield  {author} {\bibinfo {author} {\bibfnamefont {W.~R.}\ \bibnamefont
  {Geyer}},\ }\bibfield  {title} {\enquote {\bibinfo {title} {Field calibration
  of mixed-layer drifters},}\ }\href@noop {} {\bibfield  {journal} {\bibinfo
  {journal} {Journal of Atmospheric and Oceanic Technology}\ }\textbf {\bibinfo
  {volume} {6}},\ \bibinfo {pages} {333--342} (\bibinfo {year} {1989})},\
  \Eprint
  {http://arxiv.org/abs/https://doi.org/10.1175/1520-0426(1989)006<0333:FCOMLD>2.0.CO;2}
  {https://doi.org/10.1175/1520-0426(1989)006<0333:FCOMLD>2.0.CO;2}
  \BibitemShut {NoStop}%
\bibitem [{\citenamefont {Kundu}, \citenamefont {Cohen},\ and\ \citenamefont
  {Dowling}(2012)}]{Kundu-etal-12}%
  \BibitemOpen
  \bibfield  {author} {\bibinfo {author} {\bibfnamefont {P.~K.}\ \bibnamefont
  {Kundu}}, \bibinfo {author} {\bibfnamefont {I.~M.}\ \bibnamefont {Cohen}}, \
  and\ \bibinfo {author} {\bibfnamefont {D.~R.}\ \bibnamefont {Dowling}},\
  }\href {https://books.google.com/books?id=iUo\_4tsHQYUC} {\emph {\bibinfo
  {title} {Fluid Mechanics}}},\ \bibinfo {edition} {5th}\ ed.\ (\bibinfo
  {publisher} {Academic Press},\ \bibinfo {year} {2012})\ p.\ \bibinfo {pages}
  {891}\BibitemShut {NoStop}%
\bibitem [{\citenamefont {Daniel}\ \emph {et~al.}(2002)\citenamefont {Daniel},
  \citenamefont {Jan}, \citenamefont {Cabioc'h}, \citenamefont {Landau},\ and\
  \citenamefont {Loiseau}}]{Daniel-etal-02}%
  \BibitemOpen
  \bibfield  {author} {\bibinfo {author} {\bibfnamefont {P.}~\bibnamefont
  {Daniel}}, \bibinfo {author} {\bibfnamefont {G.}~\bibnamefont {Jan}},
  \bibinfo {author} {\bibfnamefont {F.}~\bibnamefont {Cabioc'h}}, \bibinfo
  {author} {\bibfnamefont {Y.}~\bibnamefont {Landau}}, \ and\ \bibinfo {author}
  {\bibfnamefont {E.}~\bibnamefont {Loiseau}},\ }\bibfield  {title} {\enquote
  {\bibinfo {title} {Drift modeling of cargo containers},}\ }\href@noop {}
  {\bibfield  {journal} {\bibinfo  {journal} {Spill Science \& Technology
  Bulletin}\ }\textbf {\bibinfo {volume} {7}},\ \bibinfo {pages} {279 -- 288}
  (\bibinfo {year} {2002})}\BibitemShut {NoStop}%
\bibitem [{\citenamefont {Ganser}(1993)}]{Ganser-93}%
  \BibitemOpen
  \bibfield  {author} {\bibinfo {author} {\bibfnamefont {G.~H.}\ \bibnamefont
  {Ganser}},\ }\bibfield  {title} {\enquote {\bibinfo {title} {A rational
  approach to drag prediction of spherical and nonspherical particles},}\
  }\href@noop {} {\bibfield  {journal} {\bibinfo  {journal} {Powder Tecnology}\
  }\textbf {\bibinfo {volume} {77}},\ \bibinfo {pages} {143--152} (\bibinfo
  {year} {1993})}\BibitemShut {NoStop}%
\bibitem [{\citenamefont {Phillips}(1997)}]{Phillips-77}%
  \BibitemOpen
  \bibfield  {author} {\bibinfo {author} {\bibfnamefont {O.~M.}\ \bibnamefont
  {Phillips}},\ }\href@noop {} {\emph {\bibinfo {title} {Dynamics of the Upper
  Ocean}}}\ (\bibinfo  {publisher} {Cambridge University Press},\ \bibinfo
  {year} {1997})\BibitemShut {NoStop}%
\bibitem [{\citenamefont {Breivik}\ \emph {et~al.}(2015)\citenamefont
  {Breivik}, \citenamefont {Mogensen}, \citenamefont {Bidlot}, \citenamefont
  {Balmaseda},\ and\ \citenamefont {Janssen}}]{Breivik-etal-15}%
  \BibitemOpen
  \bibfield  {author} {\bibinfo {author} {\bibfnamefont {{\O}.}~\bibnamefont
  {Breivik}}, \bibinfo {author} {\bibfnamefont {K.}~\bibnamefont {Mogensen}},
  \bibinfo {author} {\bibfnamefont {J.-R.}\ \bibnamefont {Bidlot}}, \bibinfo
  {author} {\bibfnamefont {M.~A.}\ \bibnamefont {Balmaseda}}, \ and\ \bibinfo
  {author} {\bibfnamefont {P.~A. E.~M.}\ \bibnamefont {Janssen}},\ }\bibfield
  {title} {\enquote {\bibinfo {title} {Surface wave effects in the nemo ocean
  model: Forced and coupled experiments},}\ }\href {\doibase
  10.1002/2014JC010565} {\bibfield  {journal} {\bibinfo  {journal} {J. Geophys.
  Res.}\ }\textbf {\bibinfo {volume} {120}},\ \bibinfo {pages} {2973--2992}
  (\bibinfo {year} {2015})}\BibitemShut {NoStop}%
\bibitem [{\citenamefont {Craik}(1982)}]{Craik-82}%
  \BibitemOpen
  \bibfield  {author} {\bibinfo {author} {\bibfnamefont {A.~D.~D.}\
  \bibnamefont {Craik}},\ }\bibfield  {title} {\enquote {\bibinfo {title} {The
  drift velocity of water waves},}\ }\href@noop {} {\bibfield  {journal}
  {\bibinfo  {journal} {J. Fluid Mech.}\ }\textbf {\bibinfo {volume} {116}},\
  \bibinfo {pages} {187--205} (\bibinfo {year} {1982})}\BibitemShut {NoStop}%
\bibitem [{\citenamefont {Jenkins}(1989)}]{Jenkins-89}%
  \BibitemOpen
  \bibfield  {author} {\bibinfo {author} {\bibfnamefont {A.~D.}\ \bibnamefont
  {Jenkins}},\ }\bibfield  {title} {\enquote {\bibinfo {title} {The use of a
  wave prediction model for driving a near-surface current model},}\
  }\href@noop {} {\bibfield  {journal} {\bibinfo  {journal} {Ocean Dyn.}\
  }\textbf {\bibinfo {volume} {42}},\ \bibinfo {pages} {133--149} (\bibinfo
  {year} {1989})}\BibitemShut {NoStop}%
\bibitem [{\citenamefont {Webb}\ and\ \citenamefont
  {Fox-Kemper}(2011)}]{Webb-Fox-11}%
  \BibitemOpen
  \bibfield  {author} {\bibinfo {author} {\bibfnamefont {A.}~\bibnamefont
  {Webb}}\ and\ \bibinfo {author} {\bibfnamefont {B.}~\bibnamefont
  {Fox-Kemper}},\ }\bibfield  {title} {\enquote {\bibinfo {title} {{Wave
  spectral moments and Stokes drift estimation}},}\ }\href@noop {} {\bibfield
  {journal} {\bibinfo  {journal} {Ocean Modell.}\ }\textbf {\bibinfo {volume}
  {40}},\ \bibinfo {pages} {273--288} (\bibinfo {year} {2011})}\BibitemShut
  {NoStop}%
\bibitem [{\citenamefont {Tamura}, \citenamefont {Miyazawa},\ and\
  \citenamefont {Oey}(2012)}]{Tamura-etal-12}%
  \BibitemOpen
  \bibfield  {author} {\bibinfo {author} {\bibfnamefont {H.}~\bibnamefont
  {Tamura}}, \bibinfo {author} {\bibfnamefont {Y.}~\bibnamefont {Miyazawa}}, \
  and\ \bibinfo {author} {\bibfnamefont {L.-Y.}\ \bibnamefont {Oey}},\
  }\bibfield  {title} {\enquote {\bibinfo {title} {{The Stokes drift and wave
  induced-mass flux in the North Pacific}},}\ }\href@noop {} {\bibfield
  {journal} {\bibinfo  {journal} {J. Geophys. Res.}\ }\textbf {\bibinfo
  {volume} {117}},\ \bibinfo {pages} {C08021} (\bibinfo {year}
  {2012})}\BibitemShut {NoStop}%
\bibitem [{\citenamefont {Breivik}, \citenamefont {Bidlot},\ and\ \citenamefont
  {Janssen}(2016)}]{Breivik-etal-16}%
  \BibitemOpen
  \bibfield  {author} {\bibinfo {author} {\bibfnamefont {{\O}.}~\bibnamefont
  {Breivik}}, \bibinfo {author} {\bibfnamefont {J.-R.}\ \bibnamefont {Bidlot}},
  \ and\ \bibinfo {author} {\bibfnamefont {P.~A.}\ \bibnamefont {Janssen}},\
  }\bibfield  {title} {\enquote {\bibinfo {title} {{A Stokes drift
  approximation based on the Phillips spectrum}},}\ }\href@noop {} {\bibfield
  {journal} {\bibinfo  {journal} {Ocean Modelling}\ }\textbf {\bibinfo {volume}
  {100}},\ \bibinfo {pages} {49 -- 56} (\bibinfo {year} {2016})}\BibitemShut
  {NoStop}%
\bibitem [{\citenamefont {Wu}(1983)}]{Wu-83}%
  \BibitemOpen
  \bibfield  {author} {\bibinfo {author} {\bibfnamefont {J.}~\bibnamefont
  {Wu}},\ }\bibfield  {title} {\enquote {\bibinfo {title} {Sea-surface drift
  currents induced by wind and waves},}\ }\href {\doibase
  10.1175/1520-0485(1983)013<1441:SSDCIB>2.0.CO;2} {\bibfield  {journal}
  {\bibinfo  {journal} {J. Phys. Oceanogr.}\ }\textbf {\bibinfo {volume}
  {13}},\ \bibinfo {pages} {1441--1451} (\bibinfo {year} {1983})},\ \Eprint
  {http://arxiv.org/abs/https://doi.org/10.1175/1520-0485(1983)013<1441:SSDCIB>2.0.CO;2}
  {https://doi.org/10.1175/1520-0485(1983)013<1441:SSDCIB>2.0.CO;2}
  \BibitemShut {NoStop}%
\bibitem [{\citenamefont {Tanga}\ and\ \citenamefont
  {Provenzale}(1994)}]{Tanga-Provenzale-94}%
  \BibitemOpen
  \bibfield  {author} {\bibinfo {author} {\bibfnamefont {P.}~\bibnamefont
  {Tanga}}\ and\ \bibinfo {author} {\bibfnamefont {A.}~\bibnamefont
  {Provenzale}},\ }\bibfield  {title} {\enquote {\bibinfo {title} {Dynamics of
  advected tracers with varying buoyancy},}\ }\href@noop {} {\bibfield
  {journal} {\bibinfo  {journal} {Physica D}\ }\textbf {\bibinfo {volume}
  {76}},\ \bibinfo {pages} {202--215} (\bibinfo {year} {1994})}\BibitemShut
  {NoStop}%
\bibitem [{\citenamefont {Sapsis}\ and\ \citenamefont
  {Haller}(2008)}]{Sapsis-Haller-08}%
  \BibitemOpen
  \bibfield  {author} {\bibinfo {author} {\bibfnamefont {T.}~\bibnamefont
  {Sapsis}}\ and\ \bibinfo {author} {\bibfnamefont {G.}~\bibnamefont
  {Haller}},\ }\bibfield  {title} {\enquote {\bibinfo {title} {Instabilities in
  the dynamics of neutrally buoyant particles},}\ }\href@noop {} {\bibfield
  {journal} {\bibinfo  {journal} {Physics of Fluids}\ }\textbf {\bibinfo
  {volume} {20}},\ \bibinfo {pages} {017102} (\bibinfo {year}
  {2008})}\BibitemShut {NoStop}%
\bibitem [{\citenamefont {Rubin}, \citenamefont {Jones},\ and\ \citenamefont
  {Maxey}(1995)}]{Rubin-etal-95}%
  \BibitemOpen
  \bibfield  {author} {\bibinfo {author} {\bibfnamefont {J.}~\bibnamefont
  {Rubin}}, \bibinfo {author} {\bibfnamefont {C.~K.~R.~T.}\ \bibnamefont
  {Jones}}, \ and\ \bibinfo {author} {\bibfnamefont {M.}~\bibnamefont
  {Maxey}},\ }\bibfield  {title} {\enquote {\bibinfo {title} {Settling and
  asymptotic motion of aerosol particles in a cellular flow field},}\ }\href
  {http://dx.doi.org/10.1007/BF01275644} {\bibfield  {journal} {\bibinfo
  {journal} {J. Nonlin. Sci.}\ }\textbf {\bibinfo {volume} {5}},\ \bibinfo
  {pages} {337--358} (\bibinfo {year} {1995})},\ \bibinfo {note}
  {10.1007/BF01275644}\BibitemShut {NoStop}%
\bibitem [{\citenamefont {Burns}, \citenamefont {Davis},\ and\ \citenamefont
  {Moore}(1999)}]{Burns-etal-99}%
  \BibitemOpen
  \bibfield  {author} {\bibinfo {author} {\bibfnamefont {T.~J.}\ \bibnamefont
  {Burns}}, \bibinfo {author} {\bibfnamefont {R.~W.}\ \bibnamefont {Davis}}, \
  and\ \bibinfo {author} {\bibfnamefont {E.~F.}\ \bibnamefont {Moore}},\
  }\bibfield  {title} {\enquote {\bibinfo {title} {A perturbation study of
  particle dynamics in a plane wake flow},}\ }\href@noop {} {\bibfield
  {journal} {\bibinfo  {journal} {J. Fluid Mech.}\ }\textbf {\bibinfo {volume}
  {384}},\ \bibinfo {pages} {1--26} (\bibinfo {year} {1999})}\BibitemShut
  {NoStop}%
\bibitem [{\citenamefont {Mograbi}\ and\ \citenamefont
  {Bar-Ziv}(2006)}]{Mograbi-Bar-06}%
  \BibitemOpen
  \bibfield  {author} {\bibinfo {author} {\bibfnamefont {E.}~\bibnamefont
  {Mograbi}}\ and\ \bibinfo {author} {\bibfnamefont {E.}~\bibnamefont
  {Bar-Ziv}},\ }\bibfield  {title} {\enquote {\bibinfo {title} {{On the
  asymptotic solution of the MaxeyÐRiley equation}},}\ }\href@noop {}
  {\bibfield  {journal} {\bibinfo  {journal} {Phys. Fluids}\ }\textbf {\bibinfo
  {volume} {18}},\ \bibinfo {pages} {051704} (\bibinfo {year}
  {2006})}\BibitemShut {NoStop}%
\bibitem [{\citenamefont {Haller}\ and\ \citenamefont
  {Sapsis}(2008)}]{Haller-Sapsis-08}%
  \BibitemOpen
  \bibfield  {author} {\bibinfo {author} {\bibfnamefont {G.}~\bibnamefont
  {Haller}}\ and\ \bibinfo {author} {\bibfnamefont {T.}~\bibnamefont
  {Sapsis}},\ }\bibfield  {title} {\enquote {\bibinfo {title} {Where do
  inertial particles go in fluid flows?}}\ }\href@noop {} {\bibfield  {journal}
  {\bibinfo  {journal} {Physica D}\ }\textbf {\bibinfo {volume} {237}},\
  \bibinfo {pages} {573--583} (\bibinfo {year} {2008})}\BibitemShut {NoStop}%
\bibitem [{\citenamefont {Fenichel}(1979)}]{Fenichel-79}%
  \BibitemOpen
  \bibfield  {author} {\bibinfo {author} {\bibfnamefont {N.}~\bibnamefont
  {Fenichel}},\ }\bibfield  {title} {\enquote {\bibinfo {title} {Geometric
  singular perturbation theory for ordinary differential equations},}\
  }\href@noop {} {\bibfield  {journal} {\bibinfo  {journal} {J. Differential
  Equations}\ }\textbf {\bibinfo {volume} {31}},\ \bibinfo {pages} {51--98}
  (\bibinfo {year} {1979})}\BibitemShut {NoStop}%
\bibitem [{\citenamefont {Jones}(1995)}]{Jones-95}%
  \BibitemOpen
  \bibfield  {author} {\bibinfo {author} {\bibfnamefont {C.~K.~R.~T.}\
  \bibnamefont {Jones}},\ }\enquote {\bibinfo {title} {{Dynamical Systems,
  Lecture Notes in Mathematics}},}\ \ (\bibinfo  {publisher}
  {Springer-Verlag},\ \bibinfo {address} {Berlin},\ \bibinfo {year} {1995})\
  Chap.\ \bibinfo {chapter} {Geometric Singular Perturbation Theory}, pp.\
  \bibinfo {pages} {44--118}\BibitemShut {NoStop}%
\bibitem [{Note3()}]{Note3}%
  \BibitemOpen
  \bibinfo {note} {The slow manifold $\protect \mathcal S_\tau $ and the
  Maxey--Riley equation restricted to $\protect \mathcal S_\tau $ formally
  satisfy the definition of \protect \emph {inertial manifold} and \protect
  \emph {inertial equation}, respectively, developed in the study of
  long-time-asymptotic behavior (attractors) of infinite-dimensional dynamical
  systems \cite {Temam-90}. In such systems, actual attractors are hard to
  compute and are generally not even manifolds. The inertial manifold is easier
  to compute, smooth, and contains the attractor. It is unclear to us why these
  constructs are called ``inertial,'' but this certainly is not related to
  resistance of an object to a change in its velocity as meant
  here.}\BibitemShut {Stop}%
\bibitem [{\citenamefont {Haller}\ and\ \citenamefont
  {Sapsis}(2010)}]{Haller-Sapsis-10}%
  \BibitemOpen
  \bibfield  {author} {\bibinfo {author} {\bibfnamefont {G.}~\bibnamefont
  {Haller}}\ and\ \bibinfo {author} {\bibfnamefont {T.}~\bibnamefont
  {Sapsis}},\ }\bibfield  {title} {\enquote {\bibinfo {title} {Localized
  instability and attraction along invariant manifolds},}\ }\href@noop {}
  {\bibfield  {journal} {\bibinfo  {journal} {Siam J. Applied Dynamical
  Systems}\ }\textbf {\bibinfo {volume} {9}},\ \bibinfo {pages} {611--633}
  (\bibinfo {year} {2010})}\BibitemShut {NoStop}%
\bibitem [{\citenamefont {Roberts}(2008)}]{Roberts-08}%
  \BibitemOpen
  \bibfield  {author} {\bibinfo {author} {\bibfnamefont {A.~J.}\ \bibnamefont
  {Roberts}},\ }\bibfield  {title} {\enquote {\bibinfo {title} {Normal form
  transforms separate slow and fast modes in stochastic dynamical systems},}\
  }\href@noop {} {\bibfield  {journal} {\bibinfo  {journal} {Physica A}\
  }\textbf {\bibinfo {volume} {387}},\ \bibinfo {pages} {12Ð38} (\bibinfo
  {year} {2008})}\BibitemShut {NoStop}%
\bibitem [{\citenamefont {Olascoaga}\ \emph {et~al.}(2016)\citenamefont
  {Olascoaga}, \citenamefont {Rypina}, \citenamefont {Brown}, \citenamefont
  {Beron-Vera}, \citenamefont {Ko\c{c}ak}, \citenamefont {Brand}, \citenamefont
  {Halliwell},\ and\ \citenamefont {Shay}}]{Olascoaga-etal-06}%
  \BibitemOpen
  \bibfield  {author} {\bibinfo {author} {\bibfnamefont {M.~J.}\ \bibnamefont
  {Olascoaga}}, \bibinfo {author} {\bibfnamefont {I.~I.}\ \bibnamefont
  {Rypina}}, \bibinfo {author} {\bibfnamefont {M.~G.}\ \bibnamefont {Brown}},
  \bibinfo {author} {\bibfnamefont {F.~J.}\ \bibnamefont {Beron-Vera}},
  \bibinfo {author} {\bibfnamefont {H.}~\bibnamefont {Ko\c{c}ak}}, \bibinfo
  {author} {\bibfnamefont {L.~E.}\ \bibnamefont {Brand}}, \bibinfo {author}
  {\bibfnamefont {G.~R.}\ \bibnamefont {Halliwell}}, \ and\ \bibinfo {author}
  {\bibfnamefont {L.~K.}\ \bibnamefont {Shay}},\ }\bibfield  {title} {\enquote
  {\bibinfo {title} {{Persistent transport barrier on the West Florida
  Shelf}},}\ }\href {\doibase 10.1029/2006GL027800} {\bibfield  {journal}
  {\bibinfo  {journal} {Geophysical Research Letters}\ }\textbf {\bibinfo
  {volume} {33}},\ \bibinfo {pages} {L22603} (\bibinfo {year} {2016})},\
  \Eprint
  {http://arxiv.org/abs/https://agupubs.onlinelibrary.wiley.com/doi/pdf/10.1029/2006GL027800}
  {https://agupubs.onlinelibrary.wiley.com/doi/pdf/10.1029/2006GL027800}
  \BibitemShut {NoStop}%
\bibitem [{\citenamefont {Gear}\ and\ \citenamefont
  {Kevrekidis}(2003)}]{Gear-Kevrekidis-03}%
  \BibitemOpen
  \bibfield  {author} {\bibinfo {author} {\bibfnamefont {C.}~\bibnamefont
  {Gear}}\ and\ \bibinfo {author} {\bibfnamefont {I.~G.}\ \bibnamefont
  {Kevrekidis}},\ }\bibfield  {title} {\enquote {\bibinfo {title} {Computing in
  the past with forward integration},}\ }\href@noop {} {\bibfield  {journal}
  {\bibinfo  {journal} {Phys. Lett. A}\ }\textbf {\bibinfo {volume} {321}},\
  \bibinfo {pages} {335Ð343} (\bibinfo {year} {2003})}\BibitemShut {NoStop}%
\bibitem [{\citenamefont {Beron-Vera}(2019)}]{Beron-19}%
  \BibitemOpen
  \bibfield  {author} {\bibinfo {author} {\bibfnamefont {F.~J.}\ \bibnamefont
  {Beron-Vera}},\ }\href@noop {} {\enquote {\bibinfo {title} {Preferential
  sampling of elastic inertial chains in the ocean},}\ }\bibinfo {howpublished}
  {Preprint} (\bibinfo {year} {2019})\BibitemShut {NoStop}%
\bibitem [{\citenamefont {Lumpkin}\ and\ \citenamefont
  {Pazos}(2007)}]{Lumpkin-Pazos-07}%
  \BibitemOpen
  \bibfield  {author} {\bibinfo {author} {\bibfnamefont {R.}~\bibnamefont
  {Lumpkin}}\ and\ \bibinfo {author} {\bibfnamefont {M.}~\bibnamefont
  {Pazos}},\ }\bibfield  {title} {\enquote {\bibinfo {title} {{Measuring
  surface currents with Surface Velocity Program drifters: the instrument, its
  data and some recent results}},}\ }in\ \href@noop {} {\emph {\bibinfo
  {booktitle} {Lagrangian Analysis and Prediction of Coastal and Ocean
  Dynamics}}},\ \bibinfo {editor} {edited by\ \bibinfo {editor} {\bibfnamefont
  {A.}~\bibnamefont {Griffa}}, \bibinfo {editor} {\bibfnamefont {A.~D.}\
  \bibnamefont {Kirwan}}, \bibinfo {editor} {\bibfnamefont {A.}~\bibnamefont
  {Mariano}}, \bibinfo {editor} {\bibfnamefont {T.}~\bibnamefont
  {\"{O}zg\"{o}kmen}}, \ and\ \bibinfo {editor} {\bibfnamefont
  {T.}~\bibnamefont {Rossby}}}\ (\bibinfo  {publisher} {Cambridge University
  Press},\ \bibinfo {year} {2007})\ Chap.~\bibinfo {chapter} {2}, pp.\ \bibinfo
  {pages} {39--67}\BibitemShut {NoStop}%
\bibitem [{\citenamefont {Niiler}\ and\ \citenamefont
  {Paduan}(1995)}]{Niiler-Paduan-95}%
  \BibitemOpen
  \bibfield  {author} {\bibinfo {author} {\bibfnamefont {P.~P.}\ \bibnamefont
  {Niiler}}\ and\ \bibinfo {author} {\bibfnamefont {J.~D.}\ \bibnamefont
  {Paduan}},\ }\bibfield  {title} {\enquote {\bibinfo {title} {{Wind-driven
  Motions in the northeastern Pacific as measured by Lagrangian drifters}},}\
  }\href@noop {} {\bibfield  {journal} {\bibinfo  {journal} {J. Phys.
  Oceanogr.}\ }\textbf {\bibinfo {volume} {25}},\ \bibinfo {pages} {2819--2830}
  (\bibinfo {year} {1995})}\BibitemShut {NoStop}%
\bibitem [{\citenamefont {Lebreton}\ \emph {et~al.}(2018)\citenamefont
  {Lebreton}, \citenamefont {Slat}, \citenamefont {Ferrari}, \citenamefont
  {Sainte-Rose}, \citenamefont {Aitken}, \citenamefont {Marthouse},
  \citenamefont {Hajbane}, \citenamefont {Cunsolo}, \citenamefont {Schwarz},
  \citenamefont {Levivier}, \citenamefont {Noble}, \citenamefont {Debeljak},
  \citenamefont {Maral}, \citenamefont {Schoeneich-Argent}, \citenamefont
  {Brambini},\ and\ \citenamefont {Reisser}}]{Lebreton-etal-2018}%
  \BibitemOpen
  \bibfield  {author} {\bibinfo {author} {\bibfnamefont {L.}~\bibnamefont
  {Lebreton}}, \bibinfo {author} {\bibfnamefont {B.}~\bibnamefont {Slat}},
  \bibinfo {author} {\bibfnamefont {F.}~\bibnamefont {Ferrari}}, \bibinfo
  {author} {\bibfnamefont {B.}~\bibnamefont {Sainte-Rose}}, \bibinfo {author}
  {\bibfnamefont {J.}~\bibnamefont {Aitken}}, \bibinfo {author} {\bibfnamefont
  {R.}~\bibnamefont {Marthouse}}, \bibinfo {author} {\bibfnamefont
  {S.}~\bibnamefont {Hajbane}}, \bibinfo {author} {\bibfnamefont
  {S.}~\bibnamefont {Cunsolo}}, \bibinfo {author} {\bibfnamefont
  {A.}~\bibnamefont {Schwarz}}, \bibinfo {author} {\bibfnamefont
  {A.}~\bibnamefont {Levivier}}, \bibinfo {author} {\bibfnamefont
  {K.}~\bibnamefont {Noble}}, \bibinfo {author} {\bibfnamefont
  {P.}~\bibnamefont {Debeljak}}, \bibinfo {author} {\bibfnamefont
  {H.}~\bibnamefont {Maral}}, \bibinfo {author} {\bibfnamefont
  {R.}~\bibnamefont {Schoeneich-Argent}}, \bibinfo {author} {\bibfnamefont
  {R.}~\bibnamefont {Brambini}}, \ and\ \bibinfo {author} {\bibfnamefont
  {J.}~\bibnamefont {Reisser}},\ }\bibfield  {title} {\enquote {\bibinfo
  {title} {Evidence that the great pacific garbage patch is rapidly
  accumulating plastic},}\ }\href@noop {} {\bibfield  {journal} {\bibinfo
  {journal} {Scientific Reports}\ }\textbf {\bibinfo {volume} {8}},\ \bibinfo
  {pages} {4666} (\bibinfo {year} {2018})}\BibitemShut {NoStop}%
\bibitem [{\citenamefont {Stommel}(1948)}]{Stommel-48}%
  \BibitemOpen
  \bibfield  {author} {\bibinfo {author} {\bibfnamefont {H.}~\bibnamefont
  {Stommel}},\ }\bibfield  {title} {\enquote {\bibinfo {title} {The westward
  intensification of wind-driven ocean currents},}\ }\href@noop {} {\bibfield
  {journal} {\bibinfo  {journal} {Trans. AGU}\ }\textbf {\bibinfo {volume}
  {29}},\ \bibinfo {pages} {202--206} (\bibinfo {year} {1948})}\BibitemShut
  {NoStop}%
\bibitem [{\citenamefont {Haidvogel}\ and\ \citenamefont
  {Bryan}(1992)}]{Haidvogel-Bryan-92}%
  \BibitemOpen
  \bibfield  {author} {\bibinfo {author} {\bibfnamefont {D.~B.}\ \bibnamefont
  {Haidvogel}}\ and\ \bibinfo {author} {\bibfnamefont {F.}~\bibnamefont
  {Bryan}},\ }\enquote {\bibinfo {title} {Climate system modeling},}\ \
  (\bibinfo  {publisher} {Oxford Press},\ \bibinfo {year} {1992})\ Chap.\
  \bibinfo {chapter} {Ocean general circulation modeling}, pp.\ \bibinfo
  {pages} {371--412}\BibitemShut {NoStop}%
\bibitem [{\citenamefont {Large}\ and\ \citenamefont
  {Pond}(1981)}]{Large-Pond-81}%
  \BibitemOpen
  \bibfield  {author} {\bibinfo {author} {\bibfnamefont {W.~G.}\ \bibnamefont
  {Large}}\ and\ \bibinfo {author} {\bibfnamefont {S.}~\bibnamefont {Pond}},\
  }\bibfield  {title} {\enquote {\bibinfo {title} {Open ocean momentum flux
  measurements in moderate to strong winds},}\ }\href@noop {} {\bibfield
  {journal} {\bibinfo  {journal} {Journal of Physical Oceanography}\ }\textbf
  {\bibinfo {volume} {11}},\ \bibinfo {pages} {324--336} (\bibinfo {year}
  {1981})}\BibitemShut {NoStop}%
\bibitem [{\citenamefont {van~der Mheen}, \citenamefont {Pattiaratchi},\ and\
  \citenamefont {van Sebille}(2019)}]{vanderMheen-etal-19}%
  \BibitemOpen
  \bibfield  {author} {\bibinfo {author} {\bibfnamefont {M.}~\bibnamefont
  {van~der Mheen}}, \bibinfo {author} {\bibfnamefont {C.}~\bibnamefont
  {Pattiaratchi}}, \ and\ \bibinfo {author} {\bibfnamefont {E.}~\bibnamefont
  {van Sebille}},\ }\bibfield  {title} {\enquote {\bibinfo {title} {Role of
  indian ocean dynamics on accumulation of buoyant debris},}\ }\href {\doibase
  10.1029/2018JC014806} {\bibfield  {journal} {\bibinfo  {journal} {Journal of
  Geophysical Research: Oceans}\ }\textbf {\bibinfo {volume} {124}},\ \bibinfo
  {pages} {doi:10.1029/2018JC014806} (\bibinfo {year} {2019})},\ \Eprint
  {http://arxiv.org/abs/https://agupubs.pericles-prod.literatumonline.com/doi/pdf/10.1029/2018JC014806}
  {https://agupubs.pericles-prod.literatumonline.com/doi/pdf/10.1029/2018JC014806}
  \BibitemShut {NoStop}%
\bibitem [{\citenamefont {Cummings}\ and\ \citenamefont
  {Smedstad}(2013)}]{Cummings-Smedstad-13}%
  \BibitemOpen
  \bibfield  {author} {\bibinfo {author} {\bibfnamefont {J.~A.}\ \bibnamefont
  {Cummings}}\ and\ \bibinfo {author} {\bibfnamefont {O.~M.}\ \bibnamefont
  {Smedstad}},\ }\bibfield  {title} {\enquote {\bibinfo {title} {Variational
  data analysis for the global ocean},}\ }in\ \href {\doibase
  10.1007/978-3-642-35088-7-13} {\emph {\bibinfo {booktitle} {Data Assimilation
  for Atmospheric, Oceanic and Hydrologic Applications}}},\ Vol.~\bibinfo
  {volume} {2},\ \bibinfo {editor} {edited by\ \bibinfo {editor} {\bibfnamefont
  {S.~K.}\ \bibnamefont {Park}}\ and\ \bibinfo {editor} {\bibfnamefont
  {L.}~\bibnamefont {Xu}}}\ (\bibinfo  {publisher} {Springer-Verlag Berlin
  Heidelberg},\ \bibinfo {year} {2013})\ Chap.~\bibinfo {chapter}
  {13}\BibitemShut {NoStop}%
\bibitem [{\citenamefont {Hsu}, \citenamefont {Meindl},\ and\ \citenamefont
  {Gilhousen}(1994)}]{Hsu-etal-94}%
  \BibitemOpen
  \bibfield  {author} {\bibinfo {author} {\bibfnamefont {S.~A.}\ \bibnamefont
  {Hsu}}, \bibinfo {author} {\bibfnamefont {E.~A.}\ \bibnamefont {Meindl}}, \
  and\ \bibinfo {author} {\bibfnamefont {D.~B.}\ \bibnamefont {Gilhousen}},\
  }\bibfield  {title} {\enquote {\bibinfo {title} {Determining the power-law
  wind-profile exponet under near-neutral stability conditions at sea},}\
  }\href@noop {} {\bibfield  {journal} {\bibinfo  {journal} {J. App. Met.}\
  }\textbf {\bibinfo {volume} {33}},\ \bibinfo {pages} {757--756} (\bibinfo
  {year} {1994})}\BibitemShut {NoStop}%
\bibitem [{\citenamefont {Froyland}, \citenamefont {Stuart},\ and\
  \citenamefont {{van Sebille}}(2014)}]{Froyland-etal-14}%
  \BibitemOpen
  \bibfield  {author} {\bibinfo {author} {\bibfnamefont {G.}~\bibnamefont
  {Froyland}}, \bibinfo {author} {\bibfnamefont {R.~M.}\ \bibnamefont
  {Stuart}}, \ and\ \bibinfo {author} {\bibfnamefont {E.}~\bibnamefont {{van
  Sebille}}},\ }\bibfield  {title} {\enquote {\bibinfo {title} {How
  well-connected is the surface of the global ocean?}}\ }\href@noop {}
  {\bibfield  {journal} {\bibinfo  {journal} {Chaos}\ }\textbf {\bibinfo
  {volume} {24}},\ \bibinfo {pages} {033126} (\bibinfo {year}
  {2014})}\BibitemShut {NoStop}%
\bibitem [{\citenamefont {Miron}\ \emph {et~al.}(2017)\citenamefont {Miron},
  \citenamefont {Beron-Vera}, \citenamefont {Olascoaga}, \citenamefont
  {Sheinbaum}, \citenamefont {P\'erez-Brunius},\ and\ \citenamefont
  {Froyland}}]{Miron-etal-17}%
  \BibitemOpen
  \bibfield  {author} {\bibinfo {author} {\bibfnamefont {P.}~\bibnamefont
  {Miron}}, \bibinfo {author} {\bibfnamefont {F.~J.}\ \bibnamefont
  {Beron-Vera}}, \bibinfo {author} {\bibfnamefont {M.~J.}\ \bibnamefont
  {Olascoaga}}, \bibinfo {author} {\bibfnamefont {J.}~\bibnamefont
  {Sheinbaum}}, \bibinfo {author} {\bibfnamefont {P.}~\bibnamefont
  {P\'erez-Brunius}}, \ and\ \bibinfo {author} {\bibfnamefont {G.}~\bibnamefont
  {Froyland}},\ }\bibfield  {title} {\enquote {\bibinfo {title} {{Lagrangian
  dynamical geography of the Gulf of Mexico}},}\ }\href {\doibase
  10.1038/s41598-017-07177-w} {\bibfield  {journal} {\bibinfo  {journal}
  {Scientific Reports}\ }\textbf {\bibinfo {volume} {7}},\ \bibinfo {pages}
  {7021} (\bibinfo {year} {2017})}\BibitemShut {NoStop}%
\bibitem [{\citenamefont {Miron}\ \emph
  {et~al.}(2019{\natexlab{b}})\citenamefont {Miron}, \citenamefont
  {Beron-Vera}, \citenamefont {Olascoaga}, \citenamefont {Froyland},
  \citenamefont {P\'erez-Brunius},\ and\ \citenamefont
  {Sheinbaum}}]{Miron-etal-19a}%
  \BibitemOpen
  \bibfield  {author} {\bibinfo {author} {\bibfnamefont {P.}~\bibnamefont
  {Miron}}, \bibinfo {author} {\bibfnamefont {F.~J.}\ \bibnamefont
  {Beron-Vera}}, \bibinfo {author} {\bibfnamefont {M.~J.}\ \bibnamefont
  {Olascoaga}}, \bibinfo {author} {\bibfnamefont {G.}~\bibnamefont {Froyland}},
  \bibinfo {author} {\bibfnamefont {P.}~\bibnamefont {P\'erez-Brunius}}, \ and\
  \bibinfo {author} {\bibfnamefont {J.}~\bibnamefont {Sheinbaum}},\ }\bibfield
  {title} {\enquote {\bibinfo {title} {{Lagrangian geography of the deep Gulf
  of Mexico}},}\ }\href@noop {} {\bibfield  {journal} {\bibinfo  {journal} {J.
  Phys. Oceanogr.}\ }\textbf {\bibinfo {volume} {49}},\ \bibinfo {pages}
  {269--290} (\bibinfo {year} {2019}{\natexlab{b}})}\BibitemShut {NoStop}%
\bibitem [{\citenamefont {Olascoaga}\ \emph {et~al.}(2018)\citenamefont
  {Olascoaga}, \citenamefont {Miron}, \citenamefont {Paris}, \citenamefont
  {P\'erez-Brunius}, \citenamefont {P\'erez-Portela}, \citenamefont {Smith},\
  and\ \citenamefont {Vaz}}]{Olascoaga-etal-18}%
  \BibitemOpen
  \bibfield  {author} {\bibinfo {author} {\bibfnamefont {M.~J.}\ \bibnamefont
  {Olascoaga}}, \bibinfo {author} {\bibfnamefont {P.}~\bibnamefont {Miron}},
  \bibinfo {author} {\bibfnamefont {C.}~\bibnamefont {Paris}}, \bibinfo
  {author} {\bibfnamefont {P.}~\bibnamefont {P\'erez-Brunius}}, \bibinfo
  {author} {\bibfnamefont {R.}~\bibnamefont {P\'erez-Portela}}, \bibinfo
  {author} {\bibfnamefont {R.~H.}\ \bibnamefont {Smith}}, \ and\ \bibinfo
  {author} {\bibfnamefont {A.}~\bibnamefont {Vaz}},\ }\bibfield  {title}
  {\enquote {\bibinfo {title} {{Connectivity of Pulley Ridge with remote
  locations as inferred from satellite-tracked drifter trajectories}},}\
  }\href@noop {} {\bibfield  {journal} {\bibinfo  {journal} {Journal of
  Geophysical Research}\ }\textbf {\bibinfo {volume} {123}},\ \bibinfo {pages}
  {5742--5750} (\bibinfo {year} {2018})}\BibitemShut {NoStop}%
\bibitem [{\citenamefont {LaCasce}(2008)}]{LaCasce-08}%
  \BibitemOpen
  \bibfield  {author} {\bibinfo {author} {\bibfnamefont {J.~H.}\ \bibnamefont
  {LaCasce}},\ }\bibfield  {title} {\enquote {\bibinfo {title} {{Statistics
  from Lagrangian observations}},}\ }\href@noop {} {\bibfield  {journal}
  {\bibinfo  {journal} {Progr. Oceanogr.}\ }\textbf {\bibinfo {volume} {77}},\
  \bibinfo {pages} {1--29} (\bibinfo {year} {2008})}\BibitemShut {NoStop}%
\bibitem [{\citenamefont {Maximenko}, \citenamefont {Hafner},\ and\
  \citenamefont {Niiler}(2012)}]{Maximenko-etal-12}%
  \BibitemOpen
  \bibfield  {author} {\bibinfo {author} {\bibfnamefont {A.~N.}\ \bibnamefont
  {Maximenko}}, \bibinfo {author} {\bibfnamefont {J.}~\bibnamefont {Hafner}}, \
  and\ \bibinfo {author} {\bibfnamefont {P.}~\bibnamefont {Niiler}},\
  }\bibfield  {title} {\enquote {\bibinfo {title} {{Pathways of marine debris
  derived from trajectories of Lagrangian drifters}},}\ }\href@noop {}
  {\bibfield  {journal} {\bibinfo  {journal} {Mar. Pollut. Bull.}\ }\textbf
  {\bibinfo {volume} {65}},\ \bibinfo {pages} {51--62} (\bibinfo {year}
  {2012})}\BibitemShut {NoStop}%
\bibitem [{\citenamefont {{van Sebille}}, \citenamefont {England},\ and\
  \citenamefont {Froyland}(2012)}]{vanSebille-etal-12}%
  \BibitemOpen
  \bibfield  {author} {\bibinfo {author} {\bibfnamefont {E.}~\bibnamefont {{van
  Sebille}}}, \bibinfo {author} {\bibfnamefont {E.~H.}\ \bibnamefont
  {England}}, \ and\ \bibinfo {author} {\bibfnamefont {G.}~\bibnamefont
  {Froyland}},\ }\bibfield  {title} {\enquote {\bibinfo {title} {Origin,
  dynamics and evolution of ocean garbage patches from observed surface
  drifters},}\ }\href@noop {} {\bibfield  {journal} {\bibinfo  {journal}
  {Environ. Res. Lett.}\ }\textbf {\bibinfo {volume} {7}},\ \bibinfo {pages}
  {044040} (\bibinfo {year} {2012})}\BibitemShut {NoStop}%
\bibitem [{\citenamefont {Beron-Vera}(2003)}]{Beron-03}%
  \BibitemOpen
  \bibfield  {author} {\bibinfo {author} {\bibfnamefont {F.~J.}\ \bibnamefont
  {Beron-Vera}},\ }\bibfield  {title} {\enquote {\bibinfo {title}
  {Constrained-{H}amiltonian shallow-water dynamics on the sphere},}\ }in\
  \href@noop {} {\emph {\bibinfo {booktitle} {Nonlinear Processes in
  Geophysical Fluid Dynamics: A Tribute to the Scientific Work of Pedro
  Ripa}}},\ \bibinfo {editor} {edited by\ \bibinfo {editor} {\bibfnamefont
  {O.~U.}\ \bibnamefont {Velasco-{F}uentes}}, \bibinfo {editor} {\bibfnamefont
  {J.}~\bibnamefont {Sheinbuam}}, \ and\ \bibinfo {editor} {\bibfnamefont
  {J.}~\bibnamefont {Ochoa}}}\ (\bibinfo  {publisher} {Kluwer},\ \bibinfo
  {year} {2003})\ pp.\ \bibinfo {pages} {29--51}\BibitemShut {NoStop}%
\bibitem [{\citenamefont {Ripa}(1995)}]{Ripa-RMF-95}%
  \BibitemOpen
  \bibfield  {author} {\bibinfo {author} {\bibfnamefont {P.}~\bibnamefont
  {Ripa}},\ }\bibfield  {title} {\enquote {\bibinfo {title} {Ca\'{\i}da libre y
  la figura de la {T}ierra},}\ }\href@noop {} {\bibfield  {journal} {\bibinfo
  {journal} {Rev. Mex. F\'{\i}s.}\ }\textbf {\bibinfo {volume} {41}},\ \bibinfo
  {pages} {106--127} (\bibinfo {year} {1995})}\BibitemShut {NoStop}%
\bibitem [{\citenamefont {Ripa}(1997{\natexlab{b}})}]{Ripa-FCE-97}%
  \BibitemOpen
  \bibfield  {author} {\bibinfo {author} {\bibfnamefont {P.}~\bibnamefont
  {Ripa}},\ }\href@noop {} {\emph {\bibinfo {title} {La {i}ncre\'{\i}ble
  {h}istoria de la {m}alentendida {f}uerza de {C}oriolis (The Incredible Story
  of the Misunderstood {C}oriolis Force)}}}\ (\bibinfo  {publisher} {Fondo de
  Cultura Econ\'omica},\ \bibinfo {year} {1997})\BibitemShut {NoStop}%
\bibitem [{\citenamefont {Wantzel}(1843)}]{Wantzel-43}%
  \BibitemOpen
  \bibfield  {author} {\bibinfo {author} {\bibfnamefont {L.}~\bibnamefont
  {Wantzel}},\ }\bibfield  {title} {\enquote {\bibinfo {title} {Classification
  des nombres incommensurables d'origine alg\'ebrique},}\ }\href@noop {}
  {\bibfield  {journal} {\bibinfo  {journal} {Nouvelles Annales de
  Math\'ematiques: Journal des Candidats aux \'Ecoles Polytechnique et
  Normale}\ }\textbf {\bibinfo {volume} {2}},\ \bibinfo {pages} {117--127}
  (\bibinfo {year} {1843})}\BibitemShut {NoStop}%
\bibitem [{\citenamefont {Temam}(1990)}]{Temam-90}%
  \BibitemOpen
  \bibfield  {author} {\bibinfo {author} {\bibfnamefont {R.}~\bibnamefont
  {Temam}},\ }\bibfield  {title} {\enquote {\bibinfo {title} {Inertial
  manifolds},}\ }\href@noop {} {\bibfield  {journal} {\bibinfo  {journal} {The
  Mathematical Intelligencer}\ }\textbf {\bibinfo {volume} {12}},\ \bibinfo
  {pages} {68--74} (\bibinfo {year} {1990})}\BibitemShut {NoStop}%
\end{thebibliography}%

\end{document}